\documentclass{MyArticle}

\begin{document}

\allowdisplaybreaks

\begin{center}
\begin{minipage}{.9\textwidth}
	\begin{center}
	{\large \textbf{UV/IR mixing as an artifact of non-covariant quantisation}}
	
	\vspace{20pt}
	
	{\large Kilian Hersent}\footnote{%
		\ \href{mailto:khersent@protonmail.com}{khersent@protonmail.com}
	},
	{\large Flavio Mercati}\footnote{%
		\ \href{mailto:fmercati@ubu.es}{fmercati@ubu.es}	
	}
	
	\vspace{7pt}	
	
	{Departamento de F\'{i}sica, Universidad de Burgos, 09001 Burgos, Spain.}
	
	\vspace{15pt}

		{\textbf{Abstract}}
	\end{center}    
We study the path integral quantisation of a scalar field on a generic noncommutative deformation of Minkowski space, built as a quantum homogeneous space of a deformed Poincar\'{e} group. We show that the procedure depends on the choice of noncommutative functional derivative, and we isolate two natural choices, distinguished by the space in which the Leibniz rule remains undeformed. The first, which carries the undeformed statistics, reproduces the standard scheme and yields $n$-point functions that break the deformed Poincar\'{e} covariance and exhibit the UV/IR mixing of \cite{Minwalla_2000}. The second, which adapts the functional calculus to the braided statistics of the fields in the spirit of \cite{Oeckl_2001}, yields covariant $n$-point functions free of this mixing. We trace both the covariance breaking and the mixing to a single source, the intertwining of external and loop momenta in the non-planar contributions, and conclude that, in the models considered, the UV/IR mixing of \cite{Minwalla_2000} is an artifact of a quantisation that breaks the deformed symmetry rather than a feature of noncommutativity itself. We further disentangle covariance, fixed by the quantisation scheme, from finiteness, fixed independently by the propagator, and illustrate the formalism on the $T$-Minkowski models, the Euclidean three-dimensional quantum gravity model, and the quantum two-sphere.
\end{minipage}
\end{center}

\newpage

\tableofcontents

\newpage

\section{Introduction}
\label{sec:intro}

\paragraph{}
Endowing spacetime with noncommutative coordinates is a long-standing idea, introduced by Snyder \cite{Snyder_1947} in the hope of taming the ultraviolet (UV) divergences of quantum field theory (QFT). It was revived by string theory \cite{Szabo_2003}, where a noncommutative field theory arises as the effective description of open-string dynamics in a $B$-field background, and it has since become \cite{Doplicher_1995}, and remains \cite{Addazi_2022}, a natural candidate for the effective description of quantum gravity in some semiclassical regime. This expectation is realised explicitly in three dimensions: integrating out the gravitational degrees of freedom in the Ponzano--Regge model coupled to matter yields a $\kappa$-deformed noncommutative field theory \cite{Freidel_2006}. Seen this way, Snyder's original motivation reappears as the programme of \emph{noncommutative regularisation} \cite{Majid_1990, Oeckl_2000}: a covariant short-distance regulator built into the geometry itself, in contrast with lattice regulators, which break the spacetime symmetries.

\paragraph{}
These hopes were tempered by the work of Minwalla, Van Raamsdonk and Seiberg \cite{Minwalla_2000}, which uncovered the UV/IR mixing effect: in noncommutative field theory the ultraviolet divergences not only persist but feed back into the infrared, making even a scalar field non-renormalisable. The effect was later found in gauge theories \cite{Matusis_2000} and in deformations of Minkowski other than Moyal \cite{Grosse_2006, Dimitrijevic_2018}. Several remedies were proposed \cite{Grosse_2005, Gurau_2009, Mirza_2006, Jafar_2006, Schenkel_2010}, mostly tailored to Moyal and difficult to generalise, while leaving the conceptual status of the mixing unsettled \cite{Hersent_2024a}. From the opposite standpoint, the effect has also been taken as a resource: as a window on the hierarchy problem, through an infrared scale of ultraviolet origin \cite{Craig_2020}, or as a mechanism for emergent gravity in $U(n)$ gauge theory \cite{Steinacker_2007}.

\paragraph{}
Two lines of evidence nonetheless suggest that noncommutative field theories can be well behaved. First, as emphasised by one of the authors \cite{Hersent_2024a}, the divergences are governed primarily by the integral of the propagator over momentum space: when the noncommutative geometry makes this integral converge, the ultraviolet divergence and its infrared echo are removed together. Second, and more strikingly, the early work of Oeckl \cite{Oeckl_2001} showed that a \emph{braided} quantisation of the fields behaves quite differently from the scheme of \cite{Minwalla_2000}: it admits a regularisation free of UV/IR mixing on $SU_q(2)$, while on Moyal a braided quantisation reproduces an \emph{undeformed} QFT \cite{Oeckl_2000}. The gap between these two outcomes, one finite and free of mixing, the other equivalent to the commutative theory, neither exhibiting the pathology of \cite{Minwalla_2000}, has to our knowledge never been confronted with the standard noncommutative result. Closing it is one of the aims of this paper.

\paragraph{}
The question of how to quantise noncommutative fields covariantly has regained attention through two recent developments. The first is the braided Batalin--Vilkovisky quantisation of fields on Moyal, formulated in the language of $L_\infty$-algebras \cite{Dimitrijevic_2023, Bogdanovic_2024, Dimitrijevic_2024}: there the Moyal $\mathfrak{R}$-matrix is used to implement the action of the BV Laplacian on products of fields, a construction that resonates with the braided derivations of Oeckl \cite{Oeckl_2001} and that has been argued to be free of UV/IR mixing. The second is the canonical quantisation of the $T$-Minkowski models, carried out in part by one of the authors \cite{Fabiano_2025}: by placing covariance under $T$-Poincar\'{e} at the centre of the construction, it extends the undeformed result of \cite{Oeckl_2000} to all unimodular $T$-deformations and, conversely, shows that the usual quantisation scheme (that of \cite{Minwalla_2000}) already breaks $T$-Poincar\'{e} covariance at the level of the free theory, precisely on the non-planar contributions.

\paragraph{}
In this paper we trace these phenomena to a single choice made in the quantisation: the notion of noncommutative functional derivative. We identify two natural notions, distinguished by the space in which the Leibniz rule remains undeformed. The \emph{$p$-Leibniz} derivative carries the undeformed (bosonic) statistics and reproduces the standard scheme of \cite{Minwalla_2000}; we show that it yields $n$-point functions that are neither quantum Poincar\'{e} covariant nor free of UV/IR mixing, and that the two defects share a single origin, the intertwining of external and loop momenta in the non-planar contributions. The \emph{$x$-Leibniz} derivative, inspired by Oeckl's braided quantisation \cite{Oeckl_2001}, instead adapts the functional calculus to the braided algebra of fields; it yields covariant $n$-point functions in which external momenta never intertwine, and which are accordingly free of this mixing. We are thus led to conclude that, in the models we consider, the UV/IR mixing of \cite{Minwalla_2000} is an artifact of a quantisation that breaks the deformed Poincar\'{e} symmetry, rather than a feature of noncommutativity itself. Along the way we make precise a distinction that is often blurred: covariance is fixed by the quantisation scheme, the braiding and the Leibniz rule, whereas the finiteness of the theory is fixed independently by the propagator, through the convergence of its momentum integral. To connect with \cite{Fabiano_2025}, we show in Appendix \ref{apx:cq} that the same dichotomy appears in canonical quantisation, where it becomes the choice between an undeformed and a braided oscillator algebra.

\paragraph{}
The paper is organised as follows. Section \ref{sec:Md} sets up the noncommutative Minkowski space as the dual of a quantum-group deformation of the translations, kept general enough to cover all deformations of Minkowski obtained by quantum-group methods to date. Section \ref{sec:mps} introduces the braiding and the braided multi-particle algebra, which give a covariant meaning to the notion of $n$-point function; most of the supporting computations are deferred to Appendix \ref{apx:mps}. Section \ref{sec:ncpi} contains our main result, the path integral quantisation of a scalar field with quartic interaction for both the $p$- and $x$-Leibniz derivatives. Section \ref{sec:ex} gathers three illustrative examples: the $T$-Minkowski models, the Euclidean three-dimensional (Ponzano--Regge) quantum gravity model, and the scalar field on the quantum two-sphere. Notation is introduced as needed; we use the Einstein summation convention and natural units, and we set off the more technical computations in greyed boxes, so that the main line of argument can be followed without them.

\paragraph{Conventions: invariance versus coinvariance.}
Quantum Poincar\'{e} symmetry can be encoded in two dual ways. One may act with the quantum Poincar\'{e} \emph{algebra} $\Poin_\ell$ via the module action $\actl$ and call $F$ \emph{invariant} when $X \actl F = \varepsilon(X) F$ for all $X \in \Poin_\ell$; or coact with the quantum Poincar\'{e} \emph{group} $\Poing_\ell$ via the comodule coaction $\coactl$ and call $F$ \emph{coinvariant} when $\coactl F = 1 \otimes F$. As $\Malg_\ell$ is a quantum homogeneous space, action and coaction are mutually dual and the two notions coincide: a quantity is $\Poin_\ell$-invariant iff it is $\Poing_\ell$-coinvariant, and a braiding is covariant for $\actl$, \eqref{eq:mps_braid_cov}, iff it is covariant for $\coactl$, \eqref{eq:mps_braid_cov_coac}. We do not commit to a single formulation. By default we phrase covariance through the action $\actl$; but where the coaction is more transparent or eases comparison with the literature -- most notably for the quantum $2$-sphere in Section~\ref{subsec:ex_SUq2}, where coinvariance $\coactl F = 1\otimes F$ is the natural language of \cite{Majid_1995, Oeckl_2001} -- we use the comodule formulation instead. No result depends on the choice.

\section{The noncommutative Minkowski spacetime}
\label{sec:Md}
\paragraph{}
In this section, we construct our noncommutative space, starting by building its symmetries. Since this work does not need a specific deformation, we keep this construction quite formal and only introduce the mathematical pieces we need. More detailed reviews on the physical and mathematical motivations for the definition of noncommutative Minkowski spacetimes can be found for example in \cite{Szabo_2003, Mercati_2023, Hersent_2023}.

\paragraph{}
We consider a quantum group deformation of the Poincar\'{e} algebra, noted $\Poin_\ell$, generated by translations $\{P_\mu\}_{\mu = 0, \ldots, d}$ and spacetime rotations $\{M_{\mu\nu}\}_{\mu,\nu=0,\ldots,d}$. Here, $\ell$ denotes the deformation parameter. We do not specify any choice of deformation since it is not needed for our discussion. Still, many examples exist, like $\theta$-Poincaré \cite{Aschieri_2005}, $\varrho$-Poincaré \cite{Dimitrijevic_2018} (or more generally T-Poincaré \cite{Mercati_2023}), $\kappa$-Poincaré \cite{Lukierski_1991, Majid_1994}, \textit{etc}... We make the hypothesis that the translations form a Hopf subalgebra of $\Poin_\ell$, noted $\Tran_\ell$. This hypothesis is actually satisfied by all the previously cited deformations of Poincaré. Similarly to the authors of \cite{Majid_1994}, we define the noncommutative Minkowski spacetime, noted $\Malg_\ell$, as the Hopf algebra dual to $\Tran_\ell$. It consists of the deformed algebra of functions over the Minkowski space which are multiplied thanks to a noncommutative product, often called the star-product. From this construction, the product of $\Malg_\ell$ is fully determined by the coalgebra structure of $\Tran_\ell$ and so by the chosen deformation of Poincaré. $\Malg_\ell$ is generated, as an algebra, by the deformed coordinate functions $\{x^\mu\}_{\mu=0, \ldots, d}$, dual elements of the $P_\mu$'s. The duality between $\Tran_\ell$ and $\Malg_\ell$ further allows us to define a $\Tran_\ell$-module algebra structure on $\Malg_\ell$ (see for example Section 1.3.5 of \cite{Klimyk_1997}). Explicitly, one can define an action $\actl : \Tran_\ell \otimes \Malg_\ell \to \Malg_\ell$ of the deformed translations on the algebra of functions on the noncommutative spacetime: $P_\mu \actl f$, for any $f \in \Malg_\ell$.

\paragraph{}
The full dual Hopf algebra of $\Poin_\ell$ consists of the quantum deformation of the (functions on the) Poincar\'{e} group, noted $\Poing_\ell$. Since $\Poin_\ell$ has a finite basis (consisting of the $P_\mu$'s and $M_{\mu\nu}$'s), its dual is a Hopf algebra generated by a finite number of elements consisting of translations $\{a^\mu\}_{\mu=0,\ldots,d}$ and Lorentz transformations $\{\tensor{\Lambda}{^{\mu}_\nu}\}_{\mu,\nu=0,\ldots,d}$. From the action $\actl$, one can define\footnote{
    Since the action $\actl$ is \textit{a priori} only defined on the deformed translations, this coaction cannot be defined by a mere dualisation. An additional structure of $\Poin_\ell$ is actually required, like a bicrossproduct (see \cite{Majid_1994}) or by defining $\Malg_\ell$ as a quantum homogeneous space (that is a subalgebra of coinvariants for the coproduct of $\Poing_\ell$, see for example \cite{Mercati_2023}). The details of the construction of $\coactl$ is again not relevant for our discussion, so one can simply make the hypothesis that such a coaction exists.
}
a coaction $\coactl : \Malg_\ell \to \Poing_\ell \otimes \Malg_\ell$. Once again, explicit examples of (co)action exist, like all $T$-Minkowski models \cite{Mercati_2024}, or the $\kappa$-Minkowski \cite{Poulain_2018} models.

\paragraph{}
Let us consider a plane wave of the form $e^{i p x} \in \Malg_\ell$, for $p \in \Mink{d}$, which is defined at least formally. Such plane waves corresponds to eigenvectors of the $P_\mu$'s, with associated eigenvalue $p_\mu$, \ie
\begin{align}
    \label{eq:Md_Pev}
	P_\mu \actl e^{i p x} = p_\mu e^{i p x}.
\end{align}
The definition of such a plane wave suffers from an ordering ambiguity, coming from the fact that the $x^\mu$'s are not commuting: see for example \cite{Kosinski_2000, Mercati_2018, Hersent_2024a}. However, a change in the ordering of the exponential can be associated to a specific change of basis of $\Tran_\ell$, and within this new basis \eqref{eq:Md_Pev} holds. This was proven in the specific case of $\kappa$-Minkowski in \cite{Mercati_2018}.

\paragraph{}
One can associate to the $p_\mu$'s a new law that emerges from the coproduct of $\Tran_\ell$. Indeed, by considering the product of two plane waves, one can write, using the definition of the action $\actl$,
\begin{align*}
	P_\mu \actl (e^{i p x} e^{i q x})
	&= \big(P_{\mu(1)} \actl e^{i p x} \big) \big( P_{\mu(2)} \actl e^{i q x} \big)
	= p_{\mu(1)} q_{\mu(2)} (e^{i p x} e^{i q x})
	:= (p \dplus q)_\mu (e^{i p x} e^{i q x})
	= P_\mu \actl e^{i (p \dplus q) x},
\end{align*}
where $\Delta(P_\mu) = P_{\mu(1)} \otimes P_{\mu(2)}$ is the Sweedler notations for the coproduct. The $\dplus$ is the noncommutative addition of momenta, with unit $0$ and inverse law $\dminus$. It consists in the law of the group of momenta that we call $(G_\ell, \dplus)$. The law $\dplus$ is noncommutative if and only if the spacetime coordinates $x^\mu$ do not commute among one another, as indeed both premises are equivalent to $\Tran_\ell$ being non-cocommutative. The explicit expressions of this $\dplus$ law for $\theta$, $\varrho$ and $\kappa$ deformations are displayed in \cite{Hersent_2024a}.

We further make the hypothesis that the group $G_\ell$ is locally compact. In the specific cases where the $x^\mu$'s form a Lie algebra, \ie $[x, x] \propto x$, $G_\ell$ corresponds to the Lie group of this Lie algebra, which is automatically locally compact. This is the case for all $T$ and $\kappa$ deformations for example. The local compactness of $G_\ell$  allow us to define a Haar measure and so to perform integration over the momentum space. We refer to \cite{Hersent_2024a, Mercati_2024} for more details, and to textbooks like \cite{Deitmar_2014} for the mathematical aspects of harmonic analysis. In what follows, we work with the left invariant Haar measure\footnote{
    One could work equivalently with the right invariant one.
}
noted $\Haar{p}$. We do not require that $G_\ell$ is unimodular \cite{Deitmar_2014}, \ie the modular function $\triangle : G_\ell \to \spReal$ is not necessarily identically $1$. In order to avoid confusion in notations with the coproduct, we will rather consider the inverse modular function, noted $\mathscr{I}(p) = \triangle(\dminus p) = 1 / \triangle(p)$ for any $p \in G_\ell$.

\paragraph{}
As a final hypothesis, we suppose that the integral over Minkowski space $\Mink{d}$ can be exported to its deformed version. Explicitly, there exists a weight $\int : \Malg_\ell \to \Cpx$ which corresponds to the spacetime integral in the commutative limit. We define a Fourier transform and its inverse as
\begin{align}
    \label{eq:Md_Ft}
	\phi(p)
	&= \frac{1}{(2\pi)^{d+1}} \int \tdl{x} e^{i (\dminus p) x} \phi(x), &
	\phi(x)
	&= \int \Haar{p} e^{i p x} \phi(p),
\end{align}
with $\phi \in \Malg_\ell$. We also use $\phi$ to denote its Fourier transform. 

We build a Dirac delta function, named $\delta$, in the momentum space as the function that satisfies $\delta(p) = 0$ for any $p \neq 0$ and $\int \Haar{p} \delta(p) = 1$. The properties of this delta function have been derived in \cite{Hersent_2024a, Agostini_2002}, and we compile some of them here without proof. From the above definition, one can show that
\begin{align}
    \label{eq:Md_mom_delta_def}
	\delta(p)
	&= \frac{1}{(2\pi)^{d+1}} \int \tdl{x} e^{i p x}.
\end{align}
Due to the noncommutativity of $\dplus$, one has that $\delta(p \dplus q) \neq \delta(q \dplus p)$. However, thanks to a change of variable, a deformed cyclicity property emerges as
\begin{align}
    \label{eq:Md_dcyc}
	\delta(p \dplus q)
	&= \mathscr{I}(q) \delta(q \dplus p).
\end{align}
In addition, one obtains from the definition of $\delta$ that
\begin{subequations}
\begin{align}
	\int \Haar{p} f(p) \, \delta(p \dplus q)
	&= \mathscr{I}(q) f(\dminus q), \\
	\int \Haar{p} f(p) \, \delta(q \dplus p)
	&= f(\dminus q).
\end{align}
\end{subequations}
Therefore, one should be careful with respect to which variable (left or right) the delta function is integrated over, the ``noncommutativity'' of the two expressions being captured by the modular function. By performing some change of variable, one can deduce
\begin{align}
	\int \Haar{p} f(p) \delta(p \dminus q)
	&= \int \Haar{p} f(p) \delta(q \dminus p)
	= \mathscr{I}(\dminus q) f(q), \\
	\delta(\dminus p)
	&= \delta(p).
\end{align}

The cyclicity property of the delta function \eqref{eq:Md_dcyc} can be used to obtain a cyclicity property of the spacetime integral:
\begin{align}
	\label{eq:Md_int_tcyc}
	\int \tdl{x} \phi_1(x) \phi_2(x)
	&= \int \tdl{x} (\mathscr{I} \actl \phi_2)(x) \phi_1(x),
\end{align}
for any $\phi_1,\phi_2 \in \Malg_\ell$. The element $\mathscr{I} \actl : \Malg_\ell \to \Malg_\ell$ is the operator that corresponds to the inverse Fourier transform of a multiplication by $\mathscr{I}$.

\paragraph{}
Finally, we consider some algebra of function\footnote{
    Following the tradition of noncommutative field theories, we consider the elements of this space of functions to be formal polynomials in element of $G_\ell$ (that are the $p$'s). The commutative counterpart of such a space is not precise, but one can think of continuous functions that vanish at infinity $C_0(G_\ell)$, smooth functions $C^\infty(G_\ell)$ or Schwarz functions $\mathcal{S}(G_\ell)$. The precise nature of $\Galg_\ell$ is actually fixed by the one of $\Malg_\ell$ since the former is defined by the latter via Fourier transform.
}
on $G_\ell$, that we call $\Galg_\ell$. The definition of the Fourier transform \eqref{eq:Md_Ft}, together with the combination of plane waves, imposes that $\Galg_\ell$ forms a group for the convolution product. Indeed,
\begin{align*}
    (fg)(x)
    &= f(x)g(x)
    = \int \Haar{p} \Haar{q} f(p) g(q) e^{ipx} e^{iqx} \\
    &= \int \Haar{p} \Haar{q} f(p) g(q) e^{i(p \dplus q)x} \\
    &\hspace{-.4cm}\overset{(p \to p \dminus q)}{=} \int \Haar{p} \left( \int \Haar{q} f(p \dminus q) g(q) \mathscr{I}(q) \right) e^{ipx}.
\end{align*}
Therefore, if we consider that $fg \in \Malg_\ell$ multiplied by the product of $\Malg_\ell$, should transform as a product of functions, we end-up with the convolution product of momentum space:
\begin{align}
    \label{eq:Md_conv_prod_def}
    (f \cp g)(p)
    &= \int \Haar{q} f(p \dplus q) g(\dminus q).
\end{align}
The unit of $(\Galg_\ell, \cp)$ is given by the $\delta$ function. Explicitly,
\begin{align*}
    (f \cp \delta)(p)
    &= \int \Haar{q} f(p \dplus q) \delta(\dminus q)
    = f(p), \\
    (\delta \cp f)(p)
    &= \int \Haar{q} \delta(p \dplus q) f(\dminus q)
    = f(p).
\end{align*}

\section{Noncommutative multi-particle space}
\label{sec:mps}
\paragraph{}

Since we are interesting in $n$-point correlation functions, we need to make sense of functions of more than one variable, such as $f(x_1, \ldots, x_n)$. More physically, we want to build the space of multiple particles. From the observation that\footnote{
    Note that, given two differentiable manifolds $\mathcal{M}$ and $\mathcal{N}$, the space $C^\infty(\mathcal{M \times N})$ is not isomorphic to the algebraic tensor product of the two space of functions $C^\infty(\mathcal{M}) \otimes C^\infty(\mathcal{N})$, since the latter only contains finite sums of simple tensor and cannot reproduce all functions of $C^\infty(\mathcal{M \times N})$. However, it is possible to complete the tensor product into a projective tensor product of nuclear Fréchet spaces and obtain that $C^\infty(\mathcal{M \times N}) \simeq C^\infty(\mathcal{M}) \,\widehat{\otimes}\, C^\infty(\mathcal{N})$, where $\widehat{\otimes}$ is the completed tensor product. See for example Theorem 51.6 of \cite{Treves_1967}. Considering $\widehat{\otimes}$ instead of $\otimes$ would normally means that we are trading finite sums for infinite power series for which we need to ensure convergence. Since our discussion is purely algebraic in nature, we consider in this paper formal power series without discussing their convergence.
}
$C^\infty(\mathcal{M} \times \cdots \times \mathcal{M}) \simeq C^\infty(\mathcal{M}) \otimes \cdots \otimes C^\infty(\mathcal{M})$, for $\mathcal{M}$ a differentiable manifold, it seems natural to construct the multi-particle space using tensor products of $\Malg_\ell$. In the following, we note $\Malg_\ell^{\tpog n} = \Malg_\ell \tpog \cdots \tpog \Malg_\ell$ ($n$ times), with the convention $\Malg_\ell^{\tpog 0} = \Cpx$. Because the tensor product only knows \textit{a priori} about the linear structure, this definition of $\Malg_\ell^{\otimes n}$ only makes it into a vector space. However, we would like  $\Malg_\ell^{\otimes n}$ to form a $\Poin_\ell$-module algebra, just like $C^\infty(\mathcal{M} \times \cdots \times \mathcal{M})$ does in the commutative theory. In other words, we would like that the multi-particle space inherits the properties of the single particle space: the fact that it is a space of functions (\ie it is an algebra) and that is has quantum Poincaré symmetries (\ie it is a $\Poin_\ell$-module).

One possible way to form an algebra structure on $\Malg_\ell^{\tpog n}$ is by equipping it with the canonical term-by-term product \eqref{eq:mps_gpdt}. But, as first noted in \cite{Fiore_2007}, this product consists in a ``commutative'' (or rather ``bosonic'') construction of the $n$-point space $\Malg_\ell^{\tpog n}$, in the sense that $[x^\mu_1, x^\nu_2] = 0$, for $x^\mu_1 = x^\mu \tpog 1$ and $x^\nu_2 = 1 \tpog x^\nu$. This ``bosonic'' algebra $\Malg_\ell^{\tpog n}$ cannot be compatible with the quantum Poincar\'{e} symmetries. More explicitly, if $\Malg_\ell$ is a $\Poin_\ell$-module, the ``bosonic'' algebra $\Malg_\ell^{\tpog n}$ is not a $\Poin_\ell$-module itself. The reason for this consists in the fact that the bosonic statistics is not compatible with the noncocommutativity of $\Poin_\ell$. We detail this fact in Appendix \ref{subapx:mps_cp}. 

Another possibility has been put forward by Majid: in \cite{Majid_1995} (see section 9.2), he explains that quantum group modules should inherit the braided statistics of the quantum group. A \emph{quasitriangular} quantum group possesses a braiding $\Psi$, that consists in a generalised statistics of particles, together with an associated braided tensor product $\underline{\otimes}$ such that $\Malg_\ell^{\underline{\otimes} n}$ naturally forms a $\Poin_\ell$-module. One then defines a product on $\Malg_\ell^{\underline{\otimes} n}$, that is compatible with the $\Poin_\ell$-module structure, by generalising the canonical product \eqref{eq:mps_gpdt} to a braided product 
\eqref{eq:mps_braid_pdt}.

\paragraph{}
We adopt a deliberately constructive standpoint on braidings, which we make explicit here. In the categorical approach of \cite{Majid_1995, Oeckl_2001} a braiding is, by definition, a covariant morphism; in the (co)quasitriangular case it is then automatically induced by the (co)$\qRm$-matrix through $\Psi_{V,W}(v \otimes w) = \tau\big(\qRm \actl (v\otimes w)\big)$, so covariance is built in from the outset. We do not start from that definition. Instead we consider general invertible linear maps $\Psi : \Malg_\ell^{\tpb 2} \to \Malg_\ell^{\tpb 2}$ subject to the hexagon relations \eqref{eq:mps_braid_defprop} and the unit conditions \eqref{eq:mps_braid_unit} (requirement \ref{it:mps_sp}), and we impose covariance \eqref{eq:mps_braid_cov} as a separate \emph{criterion} (requirement \ref{it:mps_act}) rather than as an axiom. This is what lets us keep candidate braidings in play even when $\Poin_\ell$ carries no (co)quasitriangular structure: the Hopf braiding \eqref{eq:mps_glb} is precisely such a map, defined for any Hopf algebra $\Malg_\ell$ but, as shown in Appendix~\ref{subsubapx:mps_glb}, failing the covariance criterion \eqref{eq:mps_braid_cov} in general. Where a (co)quasitriangular structure \emph{is} available the two standpoints agree: any covariant braiding then derives from the (co)$\qRm$-matrix as above and \ref{it:mps_act} holds automatically. We find the constructive presentation more transparent for readers meeting braidings for the first time, and it is the reason \ref{it:mps_act} appears among our requirements as a condition to be checked rather than as part of the definition of $\Psi$.

\paragraph{}
In this section, we build the braided tensor algebra $\Malg_\ell^{\underline{\otimes} n}$ and show that it satisfies the following properties:
\begin{enumerate}[label=(\roman*)]
	%\item \label{it:mps_tp_asso}
    %$\tpundef$ is associative: $(f_1 \tpundef f_2) \tpundef f_3 = f_1 \tpundef (f_2 \tpundef f_3)$, for any $f_1, f_2, f_3 \in \Malg_\ell$,
	\item \label{it:mps_sp}
    the product of $\Malg_\ell$ extends to an associative product $\star_\Psi : \Malg_\ell^{\underline{\otimes} n} \otimes \Malg_\ell^{\underline{\otimes} n} \to \Malg_\ell^{\underline{\otimes} n}$,
    \item \label{it:mps_act}
    the action of $\Poin_\ell$ on $\Malg_\ell$ extends to an action $\actl : \Poin_\ell \otimes \Malg_\ell^{\underline{\otimes} n} \to \Malg_\ell^{\underline{\otimes} n}$ which is compatible with the product $\star_\Psi$, \ie such that it satisfies the module algebra condition\footnote{
        Throughout the document, we use Sweedler's notations. For example here, the coproduct of $\Poin_\ell$ is $\Delta X = X_{(1)} \otimes X_{(2)}$. Note that the Sweedler's notations understands a summation, like $\Delta X = \sum_\alpha X_{(1)}^\alpha \otimes X_{(2)}^\alpha$, that we omit.
    }
    $X \actl (F_1 \star_\Psi F_2) = (X_{(1)} \actl F_1) \star_\Psi (X_{(2)} \actl F_2)$, for any $X \in \Poin_\ell$ and $F_1, F_2 \in \Malg_\ell^{\underline{\otimes} n}$.
\end{enumerate} 

This approach has already been realised in the literature for specific noncommutative deformations of Minkowski. The first construction of the multi-particle algebra has been carried out for Moyal in \cite{Fiore_2007}. It has also been constructed for the $\kappa$-Minkowski space, for which only the light-like case has been worked out \cite{Lizzi_2021, Fabiano_2024}. Indeed, the light-like case consists in a triangular deformation, for which there is a natural braiding. The multi-particle space has been built for all the $T$-Minkowski models \cite{Mercati_2024} along the same lines. The specific case of time-like $\kappa$-Minkowski, for which no $\qRm$-matrix exists in $3+1$-dimension (see \cite{Maslanka_1993}), is discussed in \cite{Arzano_2023}, where the authors used the braiding \eqref{eq:mps_glb} we dubbed ``Hopf braiding''.

Throughout the paper, we write $\otimes$ instead of $\underline{\otimes}$ for convenience, since we only deal with braided tensor products of $\Malg_\ell$.

\paragraph{}
A canonical way to construct the action $\actl: \Poin_\ell \otimes \Malg_\ell^{\otimes n} \to \Malg_\ell^{\otimes n}$ is through the following
\begin{align}
    \label{eq:mps_gact}
    X \actl (f_1 \otimes \cdots \otimes f_n)
    &= (X_{(1)} \actl f_1) \otimes \cdots \otimes (X_{(n)} \actl f_n),
\end{align}
with $X \in \Poin_\ell$, $f_1, \ldots, f_n \in \Malg_\ell$. Explicitly, we require that $X_{(j)}$ ``commutes'' with all the $f_k$ for $k < j$. In the context of the Poincar\'{e} group, this means that the Poincar\'{e} elements commute with the $x^\mu$'s.

In what follows, we construct the product of $\Malg_\ell^{\tpb 2}$, and study its covariance with respect to \eqref{eq:mps_gact}, in the sense of \ref{it:mps_act}. Its generalisation to any $n$ is done in section \ref{subsec:mps_An}. For an extended discussion on the construction of products on tensor product of algebras and their covariance, we refer to Appendix \ref{apx:mps}.

\subsection{Braided tensor product algebra}
\paragraph{}
As introduced in \cite{Majid_1995}, one can equip $\Malg_\ell^{\tpb 2}$ with a braiding, that is a linear map
\begin{align}
	\Psi : \Malg_\ell^{\tpb 2} \to \Malg_\ell^{\tpb 2},
	\label{eq:mps_braid_def}
\end{align}
which stands as a generalisation of the flip map $\tau(f \otimes g) = g \otimes f$. It is defined as being invertible and satisfying the two conditions, for any spaces $U$, $V$ and $W$
\begin{align}
    \label{eq:mps_braid_defprop}
    \Psi_{U,V\tpb W}
    &= (\id \otimes \Psi_{U,W}) \circ (\Psi_{U,V} \otimes \id), &
    \Psi_{U\tpb V, W}
    &= (\Psi_{U,W} \otimes \id) \circ (\id \otimes \Psi_{V,W}),
\end{align}
where $\Psi_{U,V} : U \tpb V \to V \tpb U$ is the braiding between $U$ and $V$. This axiom states how the braiding is handling tensor product of spaces. The requirement \eqref{eq:mps_braid_defprop} allows for higher order braidings (\ie braiding of $\Malg_\ell^{\tpb n}$ for $n > 2$) to be expressed in terms of $\Psi$ only. The braiding corresponds to a generalisation of the notion of particle's statistics since it expresses how two different particles are treated jointly via the (braided) tensor product. In that respect, the trivial braiding $\Psi = \tau$ corresponds to a bosonic statistic, and $\Psi = -\tau$ to a fermionic statistic. As detailed in \cite{Majid_1995}, any superalgebra can be encompassed within the formalism of a braided tensor product algebra. 

The product of $\Malg_\ell^{\tpb 2}$ associated to the braiding $\Psi$ is
\begin{align}
	(f_1 \tpb f_2) \star_\Psi (g_1 \tpb g_2)
	&= f_1 \Psi(f_2 \tpb g_1) g_2,
	\label{eq:mps_braid_pdt}
\end{align}
for any $f_1,f_2,g_1,g_2 \in \Malg_\ell$. Here $f_1$ and the first element of $\Psi(f_2 \tpb g_1)$ are multiplied thanks to the product of $\Malg_\ell$, and similarly for $g_2$ and the second element of $\Psi(f_2 \tpb g_1)$. The idea of introducing $\Psi$ is that $f_2$ and $g_1$ do not commute anymore (contrary to the canonical product \eqref{eq:mps_gpdt} of $\Malg_\ell^{\tpb 2}$) and their noncommutativity is captured by the braiding. We give below two explicit examples of braided product.

\paragraph{}
In the specific case where the quantum group of symmetries (here $\Poin_\ell$) has a quasitriangular structure, one can define the following braiding
\begin{align}
	\Psi(f \tpb g)
	&= (\qRm_{(2)} \actl g) \tpb (\qRm_{(1)} \actl f),
	\label{eq:mps_qt_braid}
\end{align}
where $\qRm = \qRm_{(1)} \otimes \qRm_{(2)} \in \Poin_\ell \otimes \Poin_\ell$ is the $\qRm$-matrix. We refer to appendix \ref{apx:Ha} for more details on (co)quasitriangular structures. The fact that the braiding depends on the action of $\Poin_\ell$ on $\Malg_\ell$ is precisely what makes it covariant, as shown in appendix \ref{apx:mps}.

The braided product \eqref{eq:mps_braid_pdt} associated to the braiding \eqref{eq:mps_qt_braid} then writes
\begin{align}
	(f_1 \tpb f_2) \star_\Psi (g_1 \tpb g_2)
	&= f_1 (\qRm_{(2)} \actl g_1) \tpb (\qRm_{(1)} \actl f_2) g_2,
	\label{eq:mps_qtpdt}
\end{align}
for any $f_1 \tpb f_2, g_1 \tpb g_2 \in \Malg_\ell^{\tpb 2}$.

\paragraph{}
Another braiding can be defined if $\Malg_\ell$ is a Hopf algebra. In that case, $\Malg_\ell$ acts on itself via the adjoint action. This action can be used to define the braiding
\begin{align}
	\Psi(f \tpb g)
	&= f_{(1)} g S(f_{(2)}) \tpb f_{(3)}
	\label{eq:mps_glb}
\end{align}
which is expressed through the coproduct $\Delta f = f_{(1)} \tpb f_{(2)}$ and antipode $S$ of $\Malg_\ell$. We call \eqref{eq:mps_glb} the ``Hopf'' braiding. Note that when considering plane waves $f = e^{i p x}$ and $g = e^{i q x}$, one obtains\footnote{
    We use here the fact that $\Delta x = x \otimes 1 + 1 \otimes x$ is cocommutative as the dual of the commutative algebra of the $P$'s.
}
\begin{align*}
	\Psi(e^{i p x} \tpb e^{i q x})
	&= e^{i (p \dplus q \dminus p) x} \tpb e^{i p x}
\end{align*}
and therefore $\Psi$ corresponds to the canonical braiding on the momentum group $G_\ell$. The product \eqref{eq:mps_braid_pdt} defined through the braiding \eqref{eq:mps_glb} is given by
\begin{align}
	(f_1 \tpb f_2) \star_\Psi (g_1 \tpb g_2)
	&= f_1 f_{2(1)} g_1 S(f_{2(2)}) \tpb f_{2(3)} g_2.
	\label{eq:mps_glbpdt}
\end{align}

\subsection{Braided diagrams}
\label{subsec:mps_bd}
\paragraph{}
Before going on, we make a short parenthesis on how to read or construct braided diagrams. They are to be read from top to bottom and display several lines that corresponds to elements of possibly different spaces. A crossing of lines correspond to the application of a braiding or its inverse, that is $\Psi_{V, W} =
\vcenter{\hbox{\scalebox{.3}{%
\begin{tikzpicture}[braid/.cd,]
    \pic[braid/number of strands = 2, braid/gap = .2,]
        (b) {
            braid={a_1}
    };
\end{tikzpicture}
}}}$ and $\Psi^{-1}_{V, W} =
\vcenter{\hbox{\scalebox{.3}{%
\begin{tikzpicture}[braid/.cd,]
    \pic[braid/number of strands = 2, braid/gap = .2,]
        (b) {
            braid={a_1^{-1}}
    };
\end{tikzpicture}
}}}$,
where the left line correspond to an element of $V$ and the right one to an element of $W$ at the start/top. In the case of Hopf algebras, one can add the representation for the product and coproduct as
\begin{align*}
    m_{\Malg} : \Malg \otimes \Malg \to \Malg \
    \Longleftrightarrow \
    \vcenter{\hbox{\scalebox{.7}{%
    \begin{tikzpicture}
        \draw (0,0) to (0,-.3)
            to[out=-90, in=-90] node (p1){} (1,-.3) 
            to (1,0);
        \draw (p1.center) -- +(0,-.3) node (p2){};
    \end{tikzpicture}%
    }}}, &&
    \Delta : \Malg \to \Malg \otimes \Malg \
    \Longleftrightarrow \
    \vcenter{\hbox{\scalebox{.7}{%
    \begin{tikzpicture}
        \draw (0,0) to (0,.2)
            to[out=90, in=90] node (p1){} (1,.2) 
            to (1,0);
        \draw (p1.center) -- +(0,.4) node (p2){};
    \end{tikzpicture}%
    }}}.
\end{align*}
Note that the symbol $\, \vcenter{\hbox{\scalebox{.4}{%
\begin{tikzpicture}
    \draw (0,0) to (0,-.3)
        to[out=-90, in=-90] node (p1){} (1,-.3) 
        to (1,0);
    \draw (p1.center) -- +(0,-.3) node (p2){};
\end{tikzpicture}%
}}} \,$
can also be used to denote the application of a linear function, say $L \in \mathrm{Lin}(V,W)$, to an element, say $v \in V$. Therefore the left line corresponds to $L$, the right line corresponds to $v$ and the bottom line to $L(v) \in W$. The different use of this symbol is clear from context throughout this paper.

\subsection{Braiding properties}
\paragraph{}
From now on, we consider a general braiding with associated product \eqref{eq:mps_braid_pdt}. We introduce the ``$n$-point coordinates'' as
\begin{align}
    \label{eq:mps_npt_subscr}
	x_1 = x \tpb 1 \tpb 1 \tpb \cdots, &&
	x_2 = 1 \tpb x \tpb 1 \tpb \cdots, 
\end{align}
and so on. Thus, a function $f \in \Malg_\ell$ can be evaluated at one point $x_j$ through the following
\begin{align}
	f(x_j) = 1 \tpb \cdots \tpb 1 \tpb f(x) \tpb 1 \tpb \cdots,
\end{align}
where $f(x)$ is now treated as an element of $\Malg_\ell$. It consists of a trivial lift of elements of $\Malg_\ell$ to elements of $\Malg_\ell^{\tpb n}$.

We impose on $\Psi$ the requirement that the braided product of two elements coming from the same copy of $\Malg_\ell$ simply gives the product of $\Malg_\ell$, that is
\begin{align}
	\label{eq:mps_bprod_comp}
	f(x_1) \star_\Psi g(x_1) 
	= (f(x) \tpb 1) \star_\Psi (g(x) \tpb 1) 
	= (f(x) g(x) \tpb 1)
	= (f g)(x_1),
\end{align}
where the last product $f g$ is actually the product of $\Malg_\ell$. Similarly for $x_2$, we have $f(x_2) \star_\Psi g(x_2) = (fg)(x_2)$. Going back to the definition \eqref{eq:mps_braid_pdt}, these two equalities hold if and only if
\begin{align}
	\Psi(1 \tpb f)
	&= f \tpb 1, &
	\Psi(f \tpb 1)
	&= 1 \tpb f, 
	\label{eq:mps_braid_unit}
\end{align}
for any $f \in \Malg_\ell$. We therefore require that \eqref{eq:mps_braid_unit} holds hereafter. Note that \eqref{eq:mps_braid_unit} is actually satisfied by the two braidings discussed above \eqref{eq:mps_qt_braid} and \eqref{eq:mps_glb}.

\paragraph{}
Moreover, we can define a map that relates $f(x_1) g(x_2)$ and $g(x_2) f(x_1)$, for any $f, g \in \Malg_\ell$. Explicitly, let us note $\Psi_\tau = \Psi \circ \tau$, where $\tau(f \tpb g) = g \tpb f$ is the flip map. Then, one can write
\begin{align*}
	\Psi_\tau \big(f(x_1) \star_\Psi g(x_2)\big)
	&= \Psi \circ \tau \big( f(x) \tpb g(x) \big)
	= \Psi \big( g(x) \tpb f(x) \big)
	= \big(1 \tpb g(x) \big) \star_\Psi \big(f(x) \tpb 1 \big)
	= g(x_2) \star_\Psi f(x_1).
\end{align*}
By noting $\Psi_\tau^{-1} = \tau \circ \Psi^{-1}$, we have therefore proved that
\begin{align}
	\Psi_\tau \big( f(x_1) \star_\Psi g(x_2) \big)
	&= g(x_2) \star_\Psi f(x_1), &
	\Psi_\tau^{-1} \big( g(x_2) \star_\Psi f(x_1) \big)
	&= f(x_1) \star_\Psi g(x_2),
	\label{eq:mps_braid_symm}
\end{align}
for any braiding $\Psi$. Note that in the case of bosons $\Psi = \tau$, then $\Psi_\tau = \id$ and \eqref{eq:mps_braid_symm} directly boils down to the commutativity of $\Malg_\ell^{\tpb 2}$.

\paragraph{}
Finally, we discuss the possibility that the action of the braiding on plane waves corresponds to plane waves, meaning that the braiding can be expressed purely as a momentum transformation:
\begin{align}
	e^{i q x_2} \star_\Psi e^{i p x_1}
	&= \Psi_\tau(e^{ipx_1} \star_\Psi e^{iqx_2}) 
	= e^{i \Psi_{\tau(1)}(p,q) x_1} \star_\Psi e^{i \Psi_{\tau(2)}(p,q) x_2}
	\label{eq:mps_braid_mom}
\end{align}
where $\Psi_{\tau(j)} : G_\ell \times G_\ell \to G_\ell$ is the momenta transformation. This implies that similar transformation exists for $\Psi_\tau^{-1}$. In the following, we \textit{do not} make the hypothesis that \eqref{eq:mps_braid_mom} hold in general. However, we may use it as proof of concept for specific computations.

Going back to the two examples of braiding, the relation \eqref{eq:mps_braid_mom} is satisfied straightforwardly by the Hopf braiding \eqref{eq:mps_glb} with $\Psi_{\tau(1)}(p, q) = p \dplus q \dminus p$ and $\Psi_{\tau(2)}(p, q) = q$. It is also satisfied by certain quasitriangular braiding \eqref{eq:mps_qt_braid}, as discussed in \cite{Fabiano_2025}. In this case, the momentum transformation writes in term of a momentum $\qRm$-matrix.

\subsection{Generalisation to \tops{$\Malg_\ell^{\tpb n}$}{A\^{}n}}
\label{subsec:mps_An}
\paragraph{}
The generalisation of the previous construction to any number of copies of $\Malg_\ell$ is done by a defining property of the braiding \eqref{eq:mps_braid_defprop}. By setting $U = V = W = \Malg_\ell$, the axiom \eqref{eq:mps_braid_defprop} states that there are two different braidings on $\Malg_\ell^{\tpb 3}$, that are fully defined thanks to $\Psi_{1}$ and $\Psi_{2}$, the latter being given by
\begin{align*}
    \Psi_{1} ( f_1 \tpb f_2 \tpb f_3)
    &= \Psi(f_1 \otimes f_2) \tpb f_3, &
    \Psi_{2}(f_1 \tpb f_2 \tpb f_3)
    &= f_1 \tpb \Psi(f_2 \otimes f_3),
\end{align*}
for $f_1, f_2, f_3 \in \Malg_\ell$. Indeed, we have $\Psi_{\Malg_\ell, \Malg_\ell^{\tpb 2}} = \Psi_{2} \circ \Psi_{1}$, and $\Psi_{\Malg_\ell^{\tpb 2}, \Malg_\ell} = \Psi_{1} \circ \Psi_{2}$.

By straightforward generalisation, the axiom \eqref{eq:mps_braid_defprop} suggests to define on $\Malg_\ell^{\tpb n}$ the braidings $\Psi_{j}$, for $j \in \{1, \ldots, n\}$, corresponding to the braiding of the $j^{\text{th}}$ and $(j+1)^{\text{th}}$ elements of the tensor product. These braiding allow to define higher order braidings involving $\Malg_\ell^{\tpb n}$.

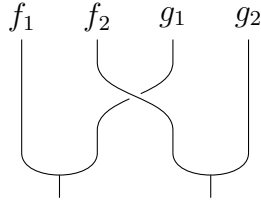
\begin{figure}
    \centering
    \begin{tikzpicture}[braid/.cd,]
        \pic[braid/number of strands = 4,]
            (b) {
                braid={a_2}
        };
        \node[at=(b-1-s), yshift=.3cm] {$f_1$};
        \node[at=(b-2-s), yshift=.3cm] {$f_2$};
        \node[at=(b-3-s), yshift=.3cm] {$g_1$};
        \node[at=(b-4-s), yshift=.3cm] {$g_2$};

        \draw (b-rev-1-e) to[out=-90, in=-90] node (p1){} (b-rev-2-e);
        \draw (b-rev-3-e) to[out=-90, in=-90] node (p2){} (b-rev-4-e);
        \foreach \p in {1,2}
            \draw (p\p.center) -- +(0,-.3);
    \end{tikzpicture}
    \caption{Diagrammatic version of the braided product of $2$-point functions \eqref{eq:mps_braid_pdt}.}
    \label{fig:braid_prod2}
\end{figure}

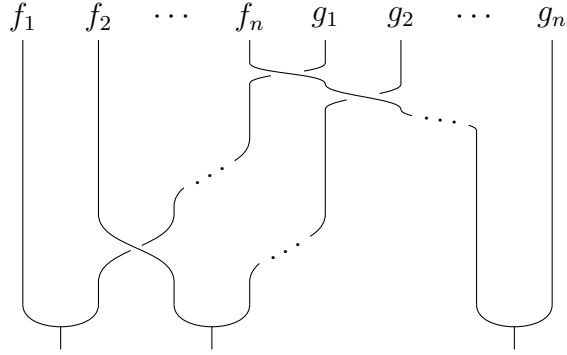
\begin{figure}
    \centering
    %\iffalse % Begin commenting out
    \begin{tikzpicture}[braid/.cd,]
        \pic[
            braid/crossing convention=under,
            braid/set symbols=under,
            braid/number of strands = 8,
            braid/strand 3/.style={white},
            braid/strand 7/.style={white}]
            (b) {
                braid={s_{4-7} s_3  |s_2-s_4| }
        };
        \node[fill=white, rotate=-8, xshift=-.5cm, yshift=.12cm] at (b-4-1) {$\ldots$};
        \node[fill=white, rotate=28, xshift=-.25cm, yshift=.65cm] at (b-3-2) {$\ldots$};
        \node[fill=white, rotate=28, xshift=-.25cm, yshift=.65cm] at (b-3-3) {$\ldots$};
        
        \node[at=(b-1-s), yshift=.3cm] {$f_1$};
        \node[at=(b-2-s), yshift=.3cm] {$f_2$};
        \node[at=(b-3-s), yshift=.3cm] {$\cdots$};
        \node[at=(b-4-s), yshift=.3cm] {$f_n$};
        \node[at=(b-5-s), yshift=.3cm] {$g_1$};
        \node[at=(b-6-s), yshift=.3cm] {$g_2$};
        \node[at=(b-7-s), yshift=.3cm] {$\cdots$};
        \node[at=(b-8-s), yshift=.3cm] {$g_n$};
    
        \draw (b-rev-1-e) to[out=-90, in=-90] node (p1){} (b-rev-2-e);
        \draw (b-rev-3-e) to[out=-90, in=-90] node (p2){} (b-rev-4-e);
        \draw (b-rev-7-e) to[out=-90, in=-90] node (p3){} (b-rev-8-e);
    
        \foreach \p in {1,2,3}
            \draw (p\p.center) -- +(0,-.3);
    \end{tikzpicture}
    %\fi % End commenting out
    \caption{Diagrammatic version of the braided product of $n$-point functions \eqref{eq:mps_braid_pdt_An}.}
    \label{fig:braid_prodn}
\end{figure}

\paragraph{}
In a similar way to how the product of $\Malg_\ell^{\tpb 2}$ is defined as
\begin{align} 
    (f_1 \tpb f_2) \star_\Psi (g_1 \tpb g_2)
    &= (m_{\Malg_\ell} \tpb m_{\Malg_\ell}) \circ \Psi_2 \big( f_1 \tpb f_2 \tpb g_1 \tpb g_2 \big),
    \tag{\ref{eq:mps_braid_pdt}}
\end{align}
see Figure \ref{fig:braid_prod2}, we define the product of $\Malg_\ell^{\tpb n}$ as
\begin{align}
    \label{eq:mps_braid_pdt_An}
\begin{aligned}
    (&f_1 \tpb \cdots \tpb f_n) \star_\Psi (g_1 \tpb \cdots \tpb g_n)
    = (m_{\Malg_\ell} \otimes \cdots \otimes m_{\Malg_\ell}) \ \circ \\
    & \Psi_2 \circ \cdots \circ (\Psi_{2n-4} \circ \cdots \circ \Psi_{n-1}) \circ (\Psi_{2n-2} \circ \cdots \circ \Psi_n) \big(f_1 \tpb \cdots \tpb f_n \tpb g_1 \tpb \cdots \tpb g_n \big),
\end{aligned}
\end{align}
as pictured in Figure \ref{fig:braid_prodn}. The term $\Psi_{2n-2} \circ \cdots \circ \Psi_{n}$ corresponds to sending $f_n$ from the $n^{\text{th}}$ position to the $(2n-1)^{\text{th}}$ position so that it can be multiplied with $g_n$. Correspondingly, the $\Psi_{2n-4} \circ \cdots \circ \Psi_{n-1}$ term send $f_{n-1}$ from the $(n-1)^{\text{th}}$ position to the $(2n-3)^\text{th}$ position in order to be multiplied with $g_{n-1}$. The iteration goes on to make $f_j$ next to $g_j$ until only $f_2$ needs to be flipped with $g_1$ thanks to a $\Psi_2$ as in \eqref{eq:mps_braid_pdt}.

One should note that \eqref{eq:mps_braid_pdt_An} corresponds to \eqref{eq:mps_braid_pdt} when $n = 2$. Moreover, as shown in appendix \ref{subapx:mps_An}, the product \eqref{eq:mps_braid_pdt_An} is well-defined and associative, thanks to the properties of the braiding. Finally, the product of $\Malg_\ell^{\tpb n}$ is compatible with the notation \eqref{eq:mps_npt_subscr}. Indeed, thanks to \eqref{eq:mps_braid_unit}, one can check that the following holds
\begin{align}
    \label{eq:mps_npt_npdt}
    f_1(x) \tpb \cdots \tpb f_n(x)
    &= f_1(x_1) \star_\Psi \cdots \star_\Psi f_n(x_n)
    := (f_1(x) \tpb 1 \tpb \cdots \tpb 1) \star_\Psi \cdots \star_\Psi (1 \tpb \cdots 1 \tpb f_n(x)).
\end{align}
In that way, any $n$-point function can be written as an ordered product of $n$ $1$-point functions.

\paragraph{}
In the remaining of this paper, except in appendix \ref{apx:mps}, the product $\star_\Psi$ is considered as the default product of $\Malg_\ell^{\tpb n}$ and therefore implied.

\subsection{Covariance of the tensor product algebra}
\label{subsec:mps_cov}
\paragraph{}
Having define the braided product \eqref{eq:mps_braid_pdt}, we can now turn to the requirement \ref{it:mps_act} from the beginning if the section: to what condition on the braiding does $\Malg_\ell^{\tpb n}$ form a $\Poin_\ell$-module algebra? Recall that it was the primary concern of why we had to consider the braided product in the first place.

One can show that the module algebra requirement \ref{it:mps_act} boils down to a compatibility relation between the action and the braiding that writes
\begin{align}
     \Psi \circ \actl
	&= \actl \circ (\id_{\Poin_\ell} \otimes \Psi).
	\label{eq:mps_braid_cov}
\end{align}
(See appendix \ref{apx:mps} for more details.) We say that a braiding is covariant if it satisfies this relation \eqref{eq:mps_braid_cov}. We make the hypothesis that the braiding we consider in the following is covariant.

\paragraph{}
At this point, it is essential to underline that there exist braiding satisfying \eqref{eq:mps_braid_cov}.

First, the quasitriangular braiding \eqref{eq:mps_qt_braid} is built to be covariant. As explicitly shown in appendix \ref{apx:mps}, the appearance of the $\qRm$-matrix is here to compensate for the noncommutativity of elements of $\Poin_\ell$ when acting on a product.

On the other hand, the Hopf braiding \eqref{eq:mps_glb} is not always covariant. Since it is defined thanks to the action of $\Malg_\ell$ on itself, it is covariant with respect to the action defined by the product of $\Malg_\ell$, and not the action of $\Poin_\ell$. Since we are interested in deformations of Minkowski seen as homogeneous space of a quantum Poincar\'{e} algebra, the product of $\Malg_\ell$ is actually inherited from the product of translations in $\Poin_\ell$. The part of the full $\Poin_\ell$ action that may break the invariance of the Hopf braiding therefore corresponds to Lorentz transformations, as shown in appendix \ref{subsubapx:mps_glb}. In general, Lorentz transformations are also deformed leading to a non-covariant Hopf braiding, however the $3$-dimensional $\kappa$-deformation of Poincar\'{e} has a covariant Hopf braiding, as shown in section \ref{subsec:ex_3qg}.

\paragraph{}
We now turn to the study of $\Poin_\ell$-invariant $n$-point functions. This necessitates the introduction of quantum Poincar\'{e} transformations in momentum space, that we do here.

Given a quantum Poincar\'{e} transformation $X \in \Poin_\ell$, we consider its action on the momentum convolution algebra $\Galg_\ell$, noted $\actlm : \Poin_\ell \otimes \Galg_\ell \to \Galg_\ell$, which is defined as follows
\begin{align}
    (X \actl f)(x)
    &= \int \Haar{p} (X \actlm f)(p) e^{ipx}.
\end{align}
In other words, if we make an explicit mention of the Fourier transform $\mathcal{F} :\Malg_\ell \to \Galg_\ell$ the definition of $\actlm$ reads
\begin{align}
    \mathcal{F}(X \actl f)
    = X \actlm (\mathcal{F}f).
\end{align}
Since $\actl$ is a left action, $\actlm$ satisfies the properties of a left action, that is $X \actlm (Y \actlm f) = (XY)\actlm f$ for any $X,Y \in \Poin_\ell$ and $f \in \Galg_\ell$. Given this we have
\begin{align*}
    (X \actl 1)(x)
    &= \varepsilon(X) 1(x)
    = \int \Haar{p} (X \actlm \delta)(p) e^{ipx} \\
    (X \actl fg)(x)
    &= \int \Haar{p} (X \actlm fg)(p) e^{ipx} \\
    = (X_{(1)} \actl f) (X_{(2)} \actl g)(x)
    &= \int \Haar{p} \big( (X_{(1)} \actlm f) \cp (X_{(2)} \actlm g)\big)(p) e^{i p x}.
\end{align*}
This shows that $\Galg_\ell$ is a $\Poin_\ell$-module algebra for the action $\actlm$.

\paragraph{}
The definition of the action $\actl$ on the multi-particle algebra \eqref{eq:mps_gact} to the action $\actlm$ transposes as 
\begin{align}
    (X \actl f)(x_1, \ldots, x_n)
    &= \int \Haar{p_1} \cdots \Haar{p_n} (\Delta^n X \actlm f)(p_1, \ldots,p_n) e^{ip_1x_1} \cdots e^{ip_n x_n},
\end{align}
which for a pure tensor product $f = f_1 \tpb \cdots \tpb f_n$ reads
\begin{align}
    \label{eq:mps_tPoin_npt_action}
    X \actl \big( f_1(x_1) \cdots f_n(x_n) \big)
    &= \int \Haar{p_1} \cdots \Haar{p_n} (X_{(1)} \actlm f_1)(p_1) \cdots (X_{(n)} \actlm f_n)(p_n) e^{ip_1x_1} \cdots e^{ip_n x_n}.
\end{align}
As an important notice, it should be emphasised that the order of $x$'s does not influence this computation, only the order of $p$'s does.  Explicitly,
\begin{align}
    X \actl \big( f_1(x_{j_1}) \cdots f_n(x_{j_n}) \big)
    &= \int \Haar{p_1} \cdots \Haar{p_n} (X_{(1)} \actlm f_1)(p_1) \cdots (X_{(n)} \actlm f_n)(p_n) e^{ip_1x_{j_1}} \cdots e^{ip_n x_{j_n}},
\end{align}
for any set of integers $j_1, \ldots, j_n \in \NInt$. Stated differently, the action does not see the braiding, as it was defined as the canonical action (see the comments below \eqref{eq:mps_gact}).

\paragraph{}
We make the hypothesis that the integral is invariant under $\Poin_\ell$. This is the case for example for all $T$-Minkowski models \cite{Mercati_2024} or $\kappa$-Minkowski \cite{Poulain_2018}. This invariance reads
\begin{align} 
    \label{eq:mps_int_cov}
    \int \tdl{x} (X \actl f)(x)
    = \varepsilon(X) \int \tdl{x} f(x).
\end{align}
for any $X \in \Poin_\ell$ and $f \in \Malg_\ell$. By writting the counit as $\varepsilon(X) 1 = S(X_{(1)}) X_{(2)}$ and $\varepsilon(X) 1 = S^{-1}(X_{(2)}) X_{(1)}$, we deduce from the invariance of the integral the change of variable property:
\begin{align}
    \label{eq:mps_chg_var_nc}
    \begin{aligned}
    \int \tdl{x} (X_{(1)} \actl f)(x) (X_{(2)} \actl g)(x)
    &= \int \tdl{x} \big( (S^{-1}(X_{(2)})X_{(1)}) \actl f \big)(x) g(x) \\
    &= \int \tdl{x} f(x) \big( (S(X_{(1)})X_{(2)}) \actl g \big)(x),
    \end{aligned}
\end{align}
which holds for any $f,g \in \Malg_\ell$. In the commutative case, this corresponds to a change of variable of the form $x^\mu \to \Lambda_\nu{}^\mu (x^\nu - a^\nu)$, given that $X$ is the Poincaré transform $x^\mu \to \Lambda^\mu{}_\nu x^\nu + a^\mu$. The equation \eqref{eq:mps_chg_var_nc} therefore express that the measure $\tdl{x}$ is invariant under that change of variable. In momentum space formalism the equation \eqref{eq:mps_chg_var_nc} becomes
\begin{align}
    \label{eq:mps_int_cov_mom}
    \begin{aligned}
    \varepsilon(X) \int \tdl{x} f(x)g(x)
    &= \int \tdl{x} (X_{(1)} \actl f)(x) (X_{(2)} \actl g)(x) \\
    = \varepsilon(X) (f \cp g)(0)
    &= \big( (X_{(1)} \actlm f) \cp (X_{(2)} \actlm g) \big) (0) \\
    = \varepsilon(X) \int \Haar{p} f(p) g(\dminus p)
    &= \int \Haar{p} (X_{(1)} \actlm f)(p) (X_{(2)} \actlm g)(\dminus p)
    \end{aligned}
\end{align}
The different rows of the previous equation correspond to different formulation of the equation \eqref{eq:mps_int_cov}. The different columns of \eqref{eq:mps_int_cov_mom} corresponds to different equality regarding the integral expressed first as a convolution product evaluated at $0$ and then by explicitly writing the convolution product thanks to \eqref{eq:Md_conv_prod_def}. 

By splitting $\varepsilon(X) 1 = S(X_{(1)}) X_{(2)} = S^{-1}(X_{(2)})X_{(1)}$, the last line of \eqref{eq:mps_int_cov_mom} implies
\begin{align}
    \label{eq:mps_chg_var_mom}
\begin{aligned}
    \int \Haar{p} (X_{(1)} \actlm f)(p) (X_{(2)} \actlm g)(\dminus p)
    &= \int \Haar{p} \big((S^{-1}(X_{(2)})X_{(1)}) \actlm f\big)(p) g(\dminus p) \\
    &= \int \Haar{p} f(p) \big( (S(X_{(1)})X_{(2)}) \actlm g\big)(\dminus p)
\end{aligned}
\end{align}
The latter equation corresponds to the momentum space counterpart of \eqref{eq:mps_chg_var_nc}. However, the relation \eqref{eq:mps_chg_var_mom} may not hold whenever $p$ and $\dminus p$ are reversed. Explicitly, in general
\begin{align}
    \label{eq:mps_chg_var_mom_wrong}
\begin{aligned}
    \int \Haar{p} (X_{(1)} \actlm f)(\dminus p) (X_{(2)} \actlm g)(p)
    &\neq \varepsilon(X) \int \Haar{p} f(\dminus p) g(p).
\end{aligned}
\end{align}
Contrary to direct space, momentum space is commutative, e.g.~$f(p) g(\dminus p) = g(\dminus p) f(p)$, so that \eqref{eq:mps_chg_var_mom_wrong} looks really like \eqref{eq:mps_chg_var_mom}. Yet, they differ in that one is acted upon by $\Delta X$ and the other by $\Delta^{\mathrm{op}}X = X_{(2)} \otimes X_{(1)}$. But the opposite coproduct $\Delta^{\mathrm{op}}$ may not satisfy the Hopf algebra relation $m_{\Poin_\ell} \circ (\id \otimes S) \circ \Delta = \varepsilon 1$. From a different point of view, \eqref{eq:mps_chg_var_mom} and \eqref{eq:mps_chg_var_mom_wrong} are related by a modular function, through the change of variable $p \to \dminus p$. But, in general, the modular function $\mathscr{I}$ is not covariant.\footnote{
    As an example, the modular function of $\kappa$-Minkowski reads, in $3+1$ dimensions, $\mathscr{I}(p) = \exp\left(\frac{-3p_0}{\kappa}\right)$, which is not invariant under $\kappa$-Lorentz transformation. To be fully explicit, the modular function can be written in direct space as an element of $\Poin_\kappa$, that is $\mathscr{I} = \exp\left(\frac{-3P_0}{\kappa}\right)$, then the condition \eqref{eq:mps_op_cov} is fulfilled if and only if $\mathscr{I}$ commutes with all elements of $\Poin_\kappa$. But since $P_0$ does not commute with the boosts, nor does $\mathscr{I}$: $[\mathscr{I}, K_j] = i \frac{3}{\kappa} P_j \mathscr{I} \neq 0$. In conclusion, $\mathscr{I}$ is not $\kappa$-Poincaré invariant.
}

\paragraph{}
Having defined the notion of $\Poin_\ell$-invariance of the integral, we now go to the one of linear operators. Given a generic linear operator $O : \Malg_\ell \to \Malg_\ell$, we say that $O$ is $\Poin_\ell$-covariant given that
\begin{align}
    \label{eq:mps_op_cov}
    O[X \actl \phi]
    &= X \actl (O[\phi]),
\end{align}
for any $\phi \in \Malg_\ell$ and any $X \in \Poin_\ell$. Should $O$ express as a kernel operator in momentum space, that is 
\begin{align*} 
    O[\phi](x) = \int \Haar{p} \left( \int \Haar{q} O(p,q) \phi(q) \right) e^{ipx},
\end{align*}
the definition of covariance \eqref{eq:mps_op_cov} would read
\begin{align*}
    X \actlm \left( \int \Haar{q} O(p,q) \phi(q) \right)
    = \int \Haar{q} O(p,q) X \actlm \phi(q).
\end{align*}
Moreover, since $\delta$ is the unit of the convolution algebra $\Galg_\ell$, the previous equation imposes for $\phi = \delta$
\begin{align}
    \label{eq:mps_op_nloc_cov_mom_2}
    X \actlm O(p,0)
    &= \varepsilon(X) O(p,0)
\end{align}
An ultra-local operator expresses as $O(p,q) = L(\dminus p) \delta(p \dplus q)$, and in that case, we have
\begin{subequations}
\begin{align}
    \label{eq:mps_op_cov_mom}
    X \actlm (L \cdot \phi)
    &= L \cdot (X \actlm \phi). \\
    \label{eq:mps_op_cov_mom_2}
    X \actlm L
    &= \varepsilon(X) L.
\end{align}
\end{subequations}
The last equation \eqref{eq:mps_op_cov_mom_2} can also be computed directly from \eqref{eq:mps_op_cov} as
\begin{align*}
    \int \Haar{p} (X \actlm L)(p) e^{ipx}
    &= X \actl L[1]
    = L[X \actl 1]
    = \varepsilon(X) L[1]
    = \varepsilon(X) \int \Haar{p} L(p) e^{ipx},
\end{align*}
then using the inverse Fourier transform, we obtain \eqref{eq:mps_op_cov_mom_2}. Note however, that \eqref{eq:mps_op_cov_mom_2} is only defined weakly but \eqref{eq:mps_op_nloc_cov_mom_2} is not.

\paragraph{}
Given a propagator $\Pi^{-1}$ fulfilling \eqref{eq:mps_op_cov}, a free action given by
\begin{align}
    S_0(\phi)
    &= \int \tdl{x} \phi(x) \Pi^{-1}[\phi](x)
\end{align}
is quantum Poincaré invariant. Indeed, given any $X \in \Poin_\ell$,
\begin{align*}
    S_0(X \actl \phi)
    &= \int \tdl{x} (X_{(1)} \actl \phi)(x) \Pi^{-1}[X_{(2)} \actl \phi](x) \\
    &= \int \tdl{x} (X_{(1)} \actl \phi)(x) (X_{(2)} \actl \Pi^{-1}[\phi])(x) \\
    &= \int \tdl{x} \big( (S^{-1}(X_{(2)})X_{(1)}) \actl \phi \big)(x) \Pi^{-1}[\phi](x) \\
    &= \varepsilon(X) S_0(\phi).
\end{align*}
Note that a $\phi^n$ interaction term (where the $\phi$'s are multiplied by the product of $\Malg_\ell$) is also $\Poin_\ell$-invariant from the invariance of the integral \eqref{eq:mps_int_cov}. In the commutative case, $\Pi^{-1}[\phi] = (\partial^\mu \partial_\mu + m^2)[\phi]$ and the definition \eqref{eq:mps_op_cov} is akin to
\begin{align*}
    (\partial^\mu \partial_\mu + m^2)\big[\phi(\Lambda^\mu{}_\nu x^\nu + a^\mu) \big]
    &= \big((\Lambda^\mu{}_\rho \partial^\rho \Lambda_\mu{}^\lambda \partial_\lambda + m^2)[\phi] \big) (\Lambda^\mu{}_\nu x^\nu + a^\mu) \\
    &= \big( (\partial^\mu \partial_\mu + m^2)[\phi]\big) (\Lambda^\mu{}_\nu x^\nu + a^\mu).
\end{align*}

\paragraph{}
In the specific case of pure translations, \eqref{eq:Md_Pev} gives us the explicit formula
\begin{align}
    (P_\mu \actlm f)(p)
    &= p_\mu f(p),
\end{align}
which is a left multiplication. Then, combining the definition \eqref{eq:Md_Pev} of $\dplus$ and \eqref{eq:mps_tPoin_npt_action}, we obtain, for any set of integers $j_1, \ldots, j_n \in \NInt$,
\begin{align}
    P_\mu \actl \big( e^{ip_1x_{j_1}} \cdots e^{ip_n x_{j_n}} \big)
    &= (p_1 \dplus \cdots \dplus p_n)_\mu e^{ip_1x_{j_1}} \cdots e^{ip_n x_{j_n}}
\end{align}
which vanishes for any spacetime index $\mu$ if and only if
\begin{align}
    \label{eq:mps_npt_cov_tran}
    p_1 \dplus \cdots \dplus p_n = 0.
\end{align}
The relation \eqref{eq:mps_npt_cov_tran} is the deformed energy-momentum conservation rule, and we simply recovered here the deformed analogue of the fact that translation invariance is associated to an energy-momentum conservation law.

\paragraph{}
Concerning pure Lorentz transformations, we do not specify an explicit formula for the action $M \actl f$ or $M \actlm f$. However, the action of a Lorentz transformation on the momentum space, at the commutative limit, should read (up to some $i$ factor)
\begin{align}
    (M_{\mu\nu} \actlm f)(p)
    \underset{\ell \to 0}{\longrightarrow} (p_\mu \partial_\nu - p_\nu \partial_\mu)f(p).
\end{align}

\paragraph{}
Consider a generic $n$-point function of the form, for any $j_1, \ldots, j_n \in \NInt$,
\begin{align*}
    W(x_{j_1}, \ldots, x_{j_n})
    &= \int \Haar{p_1} \cdots \Haar{p_n} \tilde{W}(p_1, \ldots, p_n) e^{i p_1 x_{j_1}} \cdots e^{i p_n x_{j_n}}.
\end{align*}
The translation invariance condition \eqref{eq:mps_npt_cov_tran} imposes that $p_1 \dplus \cdots \dplus p_n = 0$ in this integral. We enforce this by adding by hand some delta function, that is
\begin{align*}
    W&(x_{j_1}, \ldots, x_{j_n})
    = \int \Haar{p_1} \cdots \Haar{p_n} \delta(p_1 \dplus \cdots \dplus p_n) \tilde{W}(p_1, \ldots, p_n) e^{i p_1 x_{j_1}} \cdots e^{i p_n x_{j_n}} \\
    &= \int \Haar{p_2} \cdots \Haar{p_n} \mathscr{I}(p_2 \dplus \cdots \dplus p_n)  \tilde{W}(\dminus (p_2 \dplus \cdots \dplus p_n), p_2 , \ldots, p_n) e^{i (\dminus (p_2 \dplus \cdots \dplus p_n)) x_{j_1}} e^{i p_2 x_{j_2}} \cdots e^{i p_n x_{j_n}} \\
    &\hspace{-.5cm}\overset{(p_k \to \dminus p_k)}{=} \int \Haar{p_2} \cdots \Haar{p_n} \tilde{W}(p_n \dplus \cdots \dplus p_2, \dminus p_2, \ldots, \dminus p_n) e^{i(p_n \dplus \cdots \dplus p_2)x_{j_1}} e^{i (\dminus p_2) x_{j_2}} \cdots e^{i (\dminus p_n) x_{j_n}} \\
    &= \tilde{W}[(e^{i \cdot x_{j_1}} \cp \cdots \cp e^{i \cdot x_{j_n}})(0)].
\end{align*}
The second equality corresponds to the integration of the $\delta$ with respect to $p_1$. The third equality has been obtained by the change of variables $p_k \to \dminus p_k$ for $k = 2, \ldots, n$. Finally, the last equality is a rewriting using the convolution product\footnote{
    The convolution product \eqref{eq:Md_conv_prod_def} has been defined only for elements of $\Galg_\ell$ and the noncommutative exponentials $e^{i \cdot x_{j_k}} : p \mapsto e^{i p x_{j_k}}$ do not belong to $\Galg_\ell$ in general. Yet, we define the convolution product of these exponentials formally here through the same formula \eqref{eq:Md_conv_prod_def}.
}
\eqref{eq:Md_conv_prod_def} of $n$ exponentials $e^{i \cdot x_{j_k}} : p \mapsto e^{i p x_{j_k}}$. Moreover, $\tilde{W}$ is here regarded as an operator over these exponentials. In summary, the translationally-invariant quantities corresponds to convolution products of functions evaluated at zero.

Regarding the quantum Lorentz invariance, the quantity $W$ has two pieces. The convolution product and the $\tilde{W}$. Since the convolution product is evaluated at zero, the change of variable \eqref{eq:mps_int_cov_mom} makes this term Lorentz-invariant. All the remaining transformation is therefore stacked on the $\tilde{W}$ term. Explicitly, one has, for any Lorentz transformation $M \in \Poin_\ell$
\begin{align*}
    (M \actl W)(x_{j_1}, \ldots, x_{j_n})
    &= \int \Haar{p_2} \cdots \Haar{p_n} \big((\Delta^n M) \actlm \tilde{W} \big)(p_n \dplus \cdots \dplus p_2, \ldots, p_n) e^{i(p_n \dplus \cdots \dplus p_2)x_{j_1}} e^{ip_2 x_{j_2}} \cdots e^{i p_n x_{j_n}}
\end{align*}
If one considers that $\tilde{W}$ is now $\Poin_\ell$-invariant, according to \eqref{eq:mps_op_cov_mom_2}, $\Delta^n M \actlm \tilde{W} = \varepsilon(M) \tilde{W}$ and the $n$-point function $W$ is covariant. We therefore have a criterion for a $n$-point function $W$ to be $\Poin_\ell$-invariant: its Fourier transform should write
\begin{align}
    \label{eq:mps_cov_npf}
    \delta(p_1 \dplus \cdots \dplus p_n) \tilde{W}(p_1, \ldots, p_n)
\end{align}
with $\tilde{W}$ covariant in the sense of \eqref{eq:mps_op_cov}. One should note that the order of $x$'s (\ie the explicit values of $j_1, \ldots, j_n$) does not play any role in the covariance study. This is property is indeed a direct consequence of the covariance of the braiding.

From this analysis, it comes out that for any covariant linear operator $O$ the quantity 
\begin{align}
    \label{eq:mps_inv_2pf}
    \int \Haar{p} O(p) e^{i p x_a} e^{i (\dminus p) x_b}    
\end{align}
is $\Poin_\ell$-invariant, but the quantity
\begin{align}
    \label{eq:mps_not_inv_2pf}
    \int \Haar{p} O(p) e^{i (\dminus p) x_a} e^{i p x_b}
\end{align}
is not necessarily quantum Lorentz invariant. Both expression are indeed related by a modular function that is in general not covariant (as discussed above).

\subsection{Integration}
\paragraph{}
Having defined the multi-particle algebra, we now want to extend the notion of integration on $\Malg_\ell$, that corresponds to the weight $\int : \Malg_\ell \to \Cpx$ introduced in section \ref{sec:Md}, to any $\Malg_\ell^{\tpb n}$.

Let us first define $\int \tdl{x_1}: \Malg_\ell^{\tpb 2} \to \Malg_\ell$ as 
\begin{align}
	\int \tdl{x_1} f(x_1)
	&= \int \tdl{x_1} f(x) \tpb 1
	= \left(\int \tdl{x} f(x) \right) 1 \tpb 1.
\end{align}
Using the property \eqref{eq:mps_bprod_comp}, one computes that, for any $f,g \in \Malg_\ell$,
\begin{align*}
	\int \tdl{x_1} f(x_1) g(x_2)
	&= \int \tdl{x_1} (f(x) \tpb 1)(1 \tpb g(x)) \\
	&= \left( \int \tdl{x_1} (f(x) \tpb 1) \right) (1 \tpb g(x)) \\
	&= \left( \int \tdl{x} f(x) \right) (1 \tpb g(x))
	= \left( \int \tdl{x} f(x) \right) g(x_2).
\end{align*}

Generalising this to any number of particles, one has that the integral $\int \tdl{x_j}: \Malg_\ell^{\tpb n} \to \Malg_\ell^{\tpb n-1}$, for any $j=1, \ldots, n$, satisfy
\begin{align}
\begin{aligned}
	\int \tdl{x_{j}} f_1(x_1) \cdots f_{n}(x_{n})
	&= \int \tdl{x_{j}} f_1(x) \tpb \cdots \tpb f_{n}(x) \\
	&= \left( \int \tdl{x} f_{j}(x) \right) f_1(x) \tpb \cdots \tpb f_{j-1}(x) \tpb f_{j+1}(x) \cdots \tpb f_n(x) \\
	&= \left( \int \tdl{x} f_{j}(x) \right) f_1(x_1) \cdots f_{j-1}(x_{j-1}) f_{j+1}(x_{j+1}) \cdots f_n(x_n),
\end{aligned}
\end{align}
for any $f_1, \ldots, f_{n} \in \Malg_\ell$. Note that the order of integration does not matter, \ie~$\int \tdl{x_j} \tdl{x_k} = \int \tdl{x_k} \tdl{x_j}$.

The order of elements of the same component of the tensor product is controlled by the modular function $\mathscr{I}$ due to the generalisation of \eqref{eq:Md_int_tcyc} to
\begin{align}
\begin{aligned}
	\int & \tdl{x_1} f(x_1) g(x_1) h(x_2)
	= \left( \int \tdl{x} f(x) g(x) \right) h(x_2) \\
	&= \left( \int (\mathscr{I} \actl g)(x) f(x) \right) h(x_2)
	= \int \tdl{x_1} (\mathscr{I} \actl g)(x_1) f(x_1) h(x_2).
\end{aligned}
\end{align}
However, the order of elements of different components of the tensor product is controlled by the braiding $\Psi$ thanks to \eqref{eq:mps_braid_symm}. Explicitly,
\begin{align*}
	\int \tdl{x_1} f(x_1) g(x_2) h(x_1)
	= \int \tdl{x_1} f(x_1) \Psi_\tau \big( h(x_1) g(x_2) \big)
	= \int \tdl{x_1} f(x) \Psi \big( g(x) \tpb h(x) \big).
\end{align*}
However, in some specific cases the braiding may have a trivial action via \eqref{eq:mps_braid_unit}. For example, for $f = 1$ above, one computes
\begin{align*}
    \int \tdl{x_1} g(x_2) h(x_1)
    &= \int \tdl{x_1} \Psi_\tau \big(h(x_1) g(x_2) \big)
    = \Psi_\tau\left( \left(\int h\right) 1 \tpb g \right)
    = \left( \int h \right) g(x_2)
    = \int \tdl{x_1} h(x_1) g(x_2).
\end{align*}
As a consequence we have the following equalities
\begin{align}
    \label{eq:mps_int_choice}
    \begin{aligned}
    \left( \int \tdl{x} f(x) \right) g(x_2)
    &= \int \tdl{x_1} f(x_1) g(x_2)
    = \int \tdl{x_3} g(x_2) f(x_3), \\
    \left( \int \tdl{x} f(x) \right) g(x_2)
    &= \int \tdl{x_1} g(x_2) f(x_1) = \int \tdl{x_3} f(x_3) g(x_2).
    \end{aligned}
\end{align}
The scalar $\int f$ can be introduced on the right or on the left of $g$ equivalently. One deduces that
\begin{align}
	\left( \int \tdl{x} J(x) \phi(x) \right)^n
	= \int \tdl{x_1} \cdots \tdl{x_n} J(x_1) \phi(x_1) \cdots J(x_n) \phi(x_n),
\end{align}
for any $J,\phi \in \Malg_\ell$.

\paragraph{}
Another key ingredient in this construction is the delta function in $x$ space. The intuition toward its use can be found in the Fourier transform \eqref{eq:Md_Ft}. Indeed, by performing a Fourier transform and then its inverse\footnote{
    Note that some ambiguity could have appeared in the following equality concerning the position of the scalar $f(p)$. Before Fourier transforming back, the term $f(p)$ could be placed on the right or left of the $e^{ipx}$ term therefore generating an ambiguity since by inverse Fourier transform $f(p)$ contains an $x$-dependence: the two placement are \textit{a priori} inequivalent because of noncommutativity. However, the equality between the formulations \eqref{eq:mps_int_choice} specifically address that issue.
}
of a function $f \in \Malg_\ell$, one obtains
\begin{align*}
	f(x_1)
	&= \int \Haar{p} e^{ipx_1} f(p)
	= \frac{1}{(2\pi)^{d+1}} \int \tdl{x_2} \Haar{p} e^{ipx_1} e^{i(\dminus p)x_2} f(x_2).
\end{align*}
Since the final result should be $f(x_1)$ again, it is very tempting to define the noncommutative delta function as
\begin{align}
    \label{eq:mps_xdelta}
	\delta(x_1, x_2)
	= \frac{1}{(2\pi)^{d+1}} \int \Haar{p} e^{ip x_1} e^{i (\dminus p) x_2}.
\end{align}
By doing so, one can verify that the following properties are fulfilled
\begin{align}
	\int\tdl{x_2} \delta(x_1,x_2) f(x_2)
	&= f(x_1), &
	\int \tdl{x_1} f(x_1) \delta(x_1,x_2)
	&= f(x_2).
	\label{eq:mps_xdelta_prop}
\end{align}
Furthermore, this $\delta$ function is $\Poin_\ell$ invariant because it is of the form \eqref{eq:mps_inv_2pf}.

\begin{comp}[ams align*]
	\int \tdl{x_2} \delta(x_1,x_2) f(x_2)
	&= \frac{1}{(2\pi)^{d+1}} \int \tdl{x_2} \Haar{p} \Haar{q} e^{ipx_1} e^{i(\dminus p)x_2} e^{iq x_2} f(q) \\
	&= \int \Haar{p} \Haar{q} e^{ipx_1} \delta(\dminus p \dplus q) f(q)
	= \int \Haar{p} e^{ipx_1} f(p)
	= f(x_1), \\
	\int \tdl{x_1} f(x_1) \delta(x_1,x_2)
	&= \frac{1}{(2\pi)^{d+1}} \int \tdl{x_1} \Haar{p} \Haar{q} e^{iqx_1} f(q) e^{ipx_1} e^{i(\dminus p)x_2} \\
	&= \int \Haar{p} \Haar{q} f(q) \delta(q \dplus p) e^{i(\dminus p) x_2}
	= \int \Haar{p} f(p) e^{ipx_2}
	= f(x_2)
\end{comp}
Finally, note that $\delta$ is not symmetric in the general case, since $\delta(x_1,x_2) = \Psi^{-1}(\delta(x_2,x_1)) \neq \delta(x_2,x_1)$. 

\paragraph{}
Up to this point the construction holds for an arbitrary covariant braiding. The clean form of the $x$-Leibniz quantisation developed in Section~\ref{subsec:ncpi_uxl}, however, rests on one further structural assumption, which we state here once and for all and refer to as the \emph{ribbon assumption}. It requires the braided category of $\Malg_\ell$-modules to be \emph{ribbon} (equivalently, balanced): there exists a covariant, invertible element $\nu \in \Galg_\ell$, the \emph{ribbon element}, whose action diagonalises the self-braiding of a paired plane wave,
\begin{align}
    \label{eq:mps_ribbon_hypo}
    \int \Haar{p}\, L(p)\, e^{ipx_2} e^{i(\dminus p)x_1}
    = \int \Haar{p}\, L(p)\, \Psi\!\big(e^{ipx_1} e^{i(\dminus p)x_2}\big)
    = \int \Haar{p}\, L(p)\, \nu(p)\, e^{ipx_1} e^{i(\dminus p)x_2},
\end{align}
for every covariant weight $L \in \Galg_\ell$ satisfying the parity condition $L(p) = L(\dminus p)$. This is a genuine restriction: not every covariant braided Hopf algebra is ribbon, and the diagonal form \eqref{eq:mps_ribbon_hypo}, with a single scalar eigenvalue $\nu(p)$, presupposes that the paired wave $e^{ipx_1}e^{i(\dminus p)x_2}$ behaves as an eigenobject of the balancing -- automatic when ``momentum'' labels an irreducible, as on the quantum sphere, but not guaranteed for a generic continuous momentum group $G_\ell$. We treat \eqref{eq:mps_ribbon_hypo} as a standing assumption from Section~\ref{subsec:ncpi_uxl} onwards, and flag it wherever it is used. In our examples it is either verified or trivial: it holds genuinely for the quantum $2$-sphere $S^2_q$ (Section~\ref{subsec:ex_SUq2}), where $\nu|_{V_\ell} = q^{-2\ell(\ell+1)}$ is the bona fide ribbon element of $U_q(\mathfrak{su}(2))$, and trivially for the unimodular $T$-Minkowski models (Section~\ref{subsec:ex_TM}), where $\nu = 1$. The general non-ribbon case lies outside the scope of the present construction.

One last remark on the parity condition $L(p)=L(\dminus p)$: in \eqref{eq:mps_ribbon_hypo} the weight $L$ plays the role of the propagator $\Pi$ in the $2$-point function, and $L(p)=L(\dminus p)$ is the same evenness condition later imposed on the propagator in \eqref{eq:ncpi_prop_requirement}. It is what lets the two slots $p$ and $\dminus p$ of the paired wave enter symmetrically, so that the balancing acts on the pair as the \emph{single} scalar $\nu(p)$ rather than mixing $p$ with $\dminus p$; equivalently, it makes $L$ a function of the unordered pair $\{p,\dminus p\}$. Both $\Pi$ and $\nu$ satisfy it.

As a consequence of \eqref{eq:mps_ribbon_hypo}, the symmetry of $\delta$ now reads (with $L = 1$)
\begin{align}
    \label{eq:mps_delta_sym}
\begin{aligned}
    \delta(x_2,x_1)
    &= \frac{1}{(2\pi)^{d+1}} \int \Haar{p} e^{ipx_2} e^{i(\dminus p)x_1}
    = \frac{1}{(2\pi)^{d+1}} \int \Haar{p} \nu(p) e^{ipx_1} e^{i(\dminus p)x_2}
    = \nu \actl \delta(x_1,x_2).
\end{aligned}
\end{align}
Supposing that \eqref{eq:mps_ribbon_hypo} holds, the covariance of $\delta$ and of the braiding, expressed in \eqref{eq:mps_braid_cov}, impose that $\nu$ is itself covariant, in the sense of \eqref{eq:mps_op_cov}. Moreover, since $\Psi$ is invertible, $\nu$ also is. Finally, in the commutative limit, we should have $\nu = 1$.

\paragraph{}
The delta function could also be defined as $\tilde{\delta}(x_1,x_2) = \frac{1}{(2\pi)^{d+1}} \int \Haar{p} e^{i(\dminus p)x_1} e^{ipx_2}$. In this case, it can be shown to satisfy
\begin{align*}
    \int \tdl{x_2} \tilde{\delta}(x_1,x_2) f(x_2)
    &= (\mathscr{I} \actl f)(x_1), &
    \int \tdl{x_1} f(x_1) \tilde{\delta}(x_1,x_2)
    &= (\mathscr{I}^{-1} \actl f)(x_2).
\end{align*}
However, $\tilde{\delta}$ is of the form \eqref{eq:mps_not_inv_2pf} and is therefore not $\Poin_\ell$-invariant in general, as explained above. In the following, we only work with $\delta$ since it is $\Poin_\ell$-invariant.

\section{Noncommutative path integral quantisation}
\label{sec:ncpi}
\paragraph{}
We now turn to the core result of this paper that consists in implementing the braided statistics of fields in the path integral formalism. The latter uses the notion of functional derivative to define $n$-point functions. However, due to noncommutativity, the notion of functional derivative needs a proper treatment and we put forward here that several different definition of noncommutative functional derivatives are possible. We focus on two that we dub ``$x$-Leibniz'' and ``$p$-Leibniz'', referring to how they behave on products of fields. 

The $p$-Leibniz derivative is shown to bear the undeformed statistics and therefore consists in the ``usual'' way to define path integral quantisation in a noncommutative context. It is also proven that such a quantisation leads to non-invariant $n$-point functions and also to the famous UV/IR mixing of \cite{Minwalla_2000}.

On the other hand, the $x$-Leibniz is tailored to bear the braided statistics. Moreover, it is shown to give rise to fully invariant $n$-point functions and to be free of UV/IR mixing, at least in the sense of \cite{Minwalla_2000}. It is also recalled, with now stronger arguments, that only the behaviour of the propagator, or more precisely its integral over all momenta, is responsible for the divergence (that is usually taken care of by renormalisation). A noncommutative regularisation scheme, as the one pioneered by Snyder \cite{Snyder_1947}, should thus be more focused on the deformed propagator.

\paragraph{}
The results obtained in this section using path integral quantisation also echoes some results obtained recently for free theory thanks to canonical quantisation in \cite{Fabiano_2025}, and are summarised with our formalism in appendix \ref{apx:cq}.

\subsection{Noncommutative functional derivatives}
\label{subsec:ncpi_fd}
\paragraph{}
The notion of path integral quantisation is very dependent on the notion of functional derivative associated to it, as any correlation function can be derived thanks to functional derivatives of the generating functional \eqref{eq:ncpi_gen_func} in the perturbative approach. In the commutative case, the only properties needed for the functional derivative are the Leibniz rule and the delta function obtained from a field deriving itself. We have already seen that the noncommutative case has its own notion of delta function, but this section \ref{subsec:ncpi_fd} goes further: the notion of Leibniz rule itself may be deformed, leading to different possible cases that correspond to different quantisations.

\paragraph{}
Let us consider a functional $F: \Malg_\ell^{\tpb n} \to \Malg_\ell^{\tpb m}$. Its variation $\delta F$ is related to the functional derivative $\frac{\partial}{\partial \phi}$ by\footnote{
	Note that the functional derivative could be defined in two ways as a left or a right action via (here $\phi^{\tpb n} = \phi_1 \tpb \cdots \tpb \phi_n$ for short)
	\begin{align*}
        \delta F(\phi^{\tpb n})
        &= \int \tdl{x} \delta\phi(x) \left(\frac{\partial}{\partial \phi(x)} \actl F(\phi^{\tpb n})\right)
        = \int \tdl{x} \left(F(\phi^{\tpb n}) \actr \frac{\partial}{\partial \phi(x)} \right) \delta\phi(x),
    \end{align*}
	depending on where the $\delta\phi$ is. Their expression for elements of $\Malg_\ell^{\tpb n}$ differ as (derived as \eqref{eq:ncpi_funcder_def})
	\begin{align*}
        \frac{\partial}{\partial \phi(x_a)} \actl \phi(x_b)
        &= \delta(x_a, x_b), &
        \phi(x_b) \actr \frac{\partial}{\partial \phi(x_a)}
        &= \delta(x_b, x_a).
    \end{align*}
    Yet, both expressions conserve the ordering between $x_a$ and $x_b$, which makes the choice of left or right derivative irrelevant in our discussion. In the following, we choose to work with the left action only and we omit the action symbol ($\actl$).
}
\begin{align*}
	\delta F(\phi_1 \tpb \cdots \tpb \phi_n)(x)
	&= \int \tdl{x_1} \delta\phi(x_1) \frac{\partial F(\phi_1 \tpb \cdots \tpb \phi_n)(x_2)}{\partial \phi(x_1)},
\end{align*}
where $\phi_1 \tpb \cdots \tpb \phi_n \in \Malg_\ell^{\tpb n}$ and $\delta \phi \in \Malg_\ell$. Then, from this definition, with $n=m=1$ and $F(\phi) = \phi$,
\begin{align*}
	\delta\phi(x)
	&= \int \tdl{x_1} \delta\phi(x_1) \delta(x_1,x_2)
\end{align*}
so that
\begin{align}
	\label{eq:ncpi_funcder_def}
	\frac{\partial}{\partial \phi(x_1)} \phi(x_2)
	&= \delta(x_1,x_2).
\end{align}
We have therefore defined the functional derivative of one field, \ie of an element of $\Malg_\ell$. However, in order to define the functional derivative on $\Malg_\ell^{\otimes n}$, we need to specify a Leibniz rule.

\paragraph{}
Let us start with the case where the functional derivative fulfils the Leibniz rule in momentum space, that we call the ``$p$-Leibniz'' functional derivative. We denote it by $\pLd{\phi}$. Explicitly, we require that
\begin{align}
    \label{eq:ncpi_uplr}
    \pLd{\phi(p)} \big( \phi(q) \phi(r) \big)
    &= \pLd[\phi(q)]{\phi(p)} \phi(r) + \phi(q) \pLd[\phi(r)]{\phi(p)},
\end{align}
where we defined the functional derivative in momentum space as
\begin{align}
    \label{eq:ncpi_mom_funcder}
	\frac{\partial}{\partial \phi(p)}
	&= (2\pi)^{d+1} \int \tdl{x} e^{ipx} \frac{\partial}{\partial \phi(x)}, &
	\frac{\partial}{\partial \phi(x)}
	&= \int \Haar{p} e^{i (\dminus p) x} \frac{\partial}{\partial \phi(p)}.
\end{align}
Note the inversion of ``sign'' in the exponentials of \eqref{eq:ncpi_mom_funcder} compared to expressions \eqref{eq:Md_Ft}, that stands for the fact that the Fourier transformed functions are in the ``denominator''.

In position space, the $p$-Leibniz rule \eqref{eq:ncpi_uplr} leads to a braided Leibniz rule
\begin{align}
    \label{eq:ncpi_bxlr}
    \pLd{\phi(x_1)} \big(\phi(x_2)\phi(x_3) \big)
    &= \pLd[\phi(x_2)]{\phi(x_1)} \phi(x_3)
    + \Psi_\tau^{-1} \left(\phi(x_2) \pLd{\phi(x_1)} \right) \phi(x_3).
\end{align}
\begin{comp}[ams align*]
    \pLd{\phi(x_1)} & \big(\phi(x_2)\phi(x_3) \big)
    = \int \Haar{p} \Haar{q} \Haar{r} e^{i(\dminus p)x_1} e^{iqx_2} e^{irx_3} \pLd{\phi(p)} \big( \phi(q) \phi(r) \big) \\
    &= \int \Haar{p} \Haar{q} \Haar{r} e^{i(\dminus p)x_1} e^{iqx_2} e^{irx_3} \left( \pLd[\phi(q)]{\phi(p)} \phi(r) + \phi(q) \pLd[\phi(r)]{\phi(p)} \right) \\
    &= \int \Haar{p} \Haar{q} \Haar{r} \left( e^{i(\dminus p)x_1} e^{iqx_2} e^{irx_3} \pLd[\phi(q)]{\phi(p)} \phi(r) + \Psi^{-1}_\tau(e^{iqx_2} e^{i(\dminus p)x_1}) e^{irx_3} \phi(q) \pLd[\phi(r)]{\phi(p)} \right) \\
    &= \pLd[\phi(x_2)]{\phi(x_1)} \phi(x_3)
    + \Psi_\tau^{-1} \left(\phi(x_2) \pLd{\phi(x_1)} \right) \phi(x_3)
\end{comp}

\paragraph{}
The path integral based on the functional variation associated to a $p$-Leibniz rule, has been used in the original UV/IR mixing paper \cite{Minwalla_2000}, a result which has been extended to this more general formalism in \cite{Hersent_2024a}. However, we can already anticipate here the main result of this paper: this functional derivative does not preserve invariance. This is shown in detail in section \ref{subsec:ncpi_upl}, but a simple computation can be performed here to picture how it goes. The computation of a free $4$-point function of a scalar field theory can be obtained by an expression like:
\begin{align}
    \label{eq:ncpi_2fd2pt_uplr}
    \pLd{\phi(x_1)} \pLd{\phi(x_2)}\big(\phi(x_3) \phi(x_4) \big)
    &= \int \Haar{p} \Haar{q} \big( e^{i(\dminus p)x_1} e^{i(\dminus q)x_2} e^{iqx_3} e^{i p x_4} + e^{i(\dminus p)x_1} e^{i(\dminus q)x_2} e^{ipx_3} e^{iqx_4} \big).
\end{align}
The second term in the right hand side of \eqref{eq:ncpi_2fd2pt_uplr} does not fulfil the requirement of \eqref{eq:mps_npt_cov_tran} because $\dminus p \dminus q \dplus p \dplus q \neq 0$ and therefore is not (translation) invariant.
\begin{comp}
    In order to obtain \eqref{eq:ncpi_2fd2pt_uplr}, we need the result of \eqref{eq:ncpi_mom_funcder_delta}. It was postponed to the end of this section for simplicity, but can be proved beforehand. And so we have,
    \begin{align*}
        & \pLd{\phi(x_1)} \pLd{\phi(x_2)}\big(\phi(x_3) \phi(x_4) \big)
        = \int \Haar{p} \Haar{q} \Haar{r} \Haar{s} e^{i(\dminus p)x_1} e^{i(\dminus q)x_2} e^{irx_3} e^{isx_4} \pLd{\phi(p)} \pLd{\phi(q)}\big(\phi(r) \phi(s) \big) \\
        &= \int \Haar{p} \Haar{q} \Haar{r} \Haar{s} e^{i(\dminus p)x_1} e^{i(\dminus q)x_2} e^{irx_3} e^{isx_4} \pLd{\phi(p)} \big( \mathscr{I}(p)\delta(p\dminus r) \phi(s) + \mathscr{I}(p)\delta(p \dminus s) \phi(r) \big) \\
        &= \int \Haar{p} \Haar{q} \Haar{r} \Haar{s} e^{i(\dminus p)x_1} e^{i(\dminus q)x_2} e^{irx_3} e^{isx_4} \big( \mathscr{I}(p)\delta(p\dminus r) \mathscr{I}(q) \delta(q \dminus s) \\
        & \hspace{.5\textwidth} + \mathscr{I}(p)\delta(p \dminus s) \mathscr{I}(q) \delta(q \dminus r) \big) \\
        &= \int \Haar{p} \Haar{q} \big( e^{i(\dminus p)x_1} e^{i(\dminus q)x_2} e^{iqx_3} e^{i p x_4} + e^{i(\dminus p)x_1} e^{i(\dminus q)x_2} e^{ipx_3} e^{iqx_4} \big). 
    \end{align*}
\end{comp}

\paragraph{}
The question of how to define a functional derivative that transforms covariantly has actually been addressed at about the same time as the discovery of the UV/IR mixing, thanks to the work of Oeckl \cite{Oeckl_2001}. The idea is that the functional variation $\delta$ should respect the symmetries of the quantum Poincar\'{e} group and therefore be braided.  This formalism has been applied to the $3$-dimensional quantum gravity model (this model is detailed in section \ref{subsec:ex_3qg}) in \cite{Sasai_2007}. The authors of \cite{Sasai_2007} remarked that the translation invariance of the $2$-point function is related to a Leibniz rule in position space (see equation (106) of \cite{Sasai_2007}). 

Motivated by these ideas, we consider in the following a functional derivative satisfying the Leibniz rule in position space, that we call ``$x$-Leibniz'' and denote by $\xLd{\phi}$ hereafter, \ie
\begin{align}
    \label{eq:ncpi_uxlr}
    \xLd{\phi(x_1)} \big( \phi(x_2) \phi(x_3) \big)
    &= \xLd[\phi(x_2)]{\phi(x_1)} \phi(x_3) + \phi(x_2) \xLd[\phi(x_3)]{\phi(x_1)}.
\end{align}

It is first natural to ask the question of how $\pLd{\phi}$ compares to $\xLd{\phi}$. Their position space formulation differs by a braiding factor: compare \eqref{eq:ncpi_bxlr} and \eqref{eq:ncpi_uxlr}. Moreover, under the hypothesis \eqref{eq:mps_braid_mom}, it can be shown that the $x$-Leibniz rule corresponds to a braided $p$-Leibniz rule.
\begin{comp}
    We perform the Fourier transform of \eqref{eq:ncpi_bxlr}, using the definition \eqref{eq:ncpi_mom_funcder} and \eqref{eq:mps_braid_mom}:
    \begin{align*}
        \xLd{\phi(x_1)} & \big( \phi(x_2) \phi(x_3) \big)
        = \int \Haar{p} \Haar{q} \Haar{r} e^{i (\dminus p) x_1} e^{i q x_2} e^{i r x_3} \xLd{\phi(p)} \big( \phi(q) \phi(r) \big) \\
        &= \xLd[\phi(x_2)]{\phi(x_1)} \phi(x_3) + \phi(x_2) \xLd[\phi(x_3)]{\phi(x_1)} \\
        &= \int \Haar{p} \Haar{q} \Haar{r} e^{i (\dminus p) x_1} e^{i q x_2} e^{i r x_3} \xLd[\phi(q)]{\phi(p)} \phi(r) + e^{i q x_2} e^{i (\dminus p) x_1} e^{i r x_3} \phi(q) \xLd[\phi(r)]{\phi(p)}\\
        &= \int \Haar{p} \Haar{q} \Haar{r} e^{i (\dminus p) x_1} e^{i q x_2} e^{i r x_3} \xLd[\phi(q)]{\phi(p)} \phi(r) + \Psi_\tau\big(e^{i (\dminus p) x_1} e^{i q x_2}\big) e^{i r x_3} \phi(q) \xLd[\phi(r)]{\phi(p)} \\
	    &\hspace{-1.1cm}\overset{(\dminus p,q \to \Psi^{-1}_\tau(\dminus p,q))}{=} \int \Haar{p} \Haar{q} \Haar{r} e^{i (\dminus p) x_1} e^{i q x_2} e^{i r x_3} \Big( \xLd[\phi(q)]{\phi(p)} \phi(r) \\
        &\qquad + J^{-1}_{\Psi_\tau}(\dminus p,q) \phi\big(\Psi^{-1}_{\tau(2)}(\dminus p,q)\big) \xLd[\phi(r)]{\phi \big(\dminus \Psi^{-1}_{\tau(1)}(\dminus p,q) \big)} \Big)
    \end{align*}
    where $J_{\Psi_\tau}(p,q)$ is the Jacobian of the transformation $p,q \to \Psi_\tau(p,q)$.
\end{comp}
We notice the reader that the braided $p$-Leibniz rule derived above is not used throughout this manuscript since it is tedious and needs the hypothesis \eqref{eq:mps_braid_mom}. We still derived it for completeness.

As a final word on the difference between $x$-Leibniz and $p$-Leibniz, let us go back to our proof of concept for invariance \eqref{eq:ncpi_2fd2pt_uplr}. Straightforwardly using \eqref{eq:ncpi_funcder_def} and \eqref{eq:ncpi_uxlr}, we derive that
\begin{align}
    \label{eq:ncpi_2fd2pt_uxlr}
    \xLd{\phi(x_1)} \xLd{\phi(x_2)} \big( \phi(x_3) \phi(x_4) \big)
    &= \delta(x_2,x_3) \delta(x_1,x_4) + \delta(x_1,x_3) \delta(x_2,x_4),
\end{align}
which is $\Poin_\ell$-covariant because $\delta$ is. Recall that the expression \eqref{eq:mps_xdelta} of the $\delta$ is covariant because it is of the form \eqref{eq:mps_inv_2pf}: a Lorentz invariant prefactor (here $1$) together with a product of exponential with momenta $p$ and $\dminus p$ (so that $p \dminus p = 0$, \ie \eqref{eq:mps_npt_cov_tran} is satisfied).

\paragraph{}
We summarise the result of this section \ref{subsec:ncpi_fd} with the following diagram. 
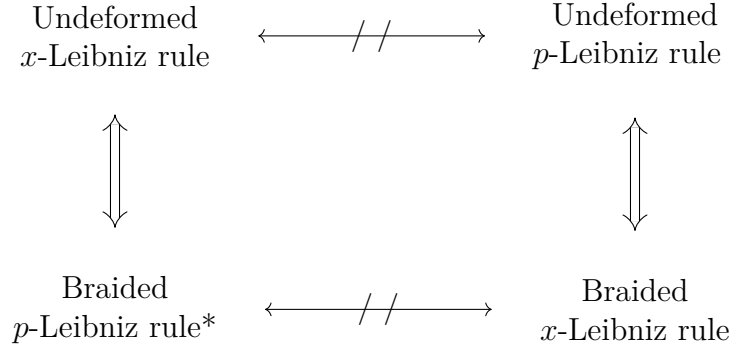
\begin{center}
	\begin{tikzpicture}[every node/.style={align=center, outer sep=.5cm}]
		\node (uxlr) {Undeformed\\ $x$-Leibniz rule};
		\node[right= 3cm of uxlr] (uplr) {Undeformed\\ $p$-Leibniz rule\phantom{*}};
		\node[below= 1.5cm of uxlr] (bplr) {Braided\\ $p$-Leibniz rule*};
		\node[right= 3cm of bplr] (bxlr) {Braided\\ $x$-Leibniz rule};	
		
		\draw[{To}-{To}] (uxlr.east) to node{/ /} (uplr.west);
		\draw[{To}-{To}] (bplr.east) to node{/ /} (bxlr.west);
		\draw[{Implies[]}-{Implies[]}, double distance=3pt] (uxlr.south) to (bplr.north);
		\draw[{Implies[]}-{Implies[]}, double distance=3pt] (uplr.south) to (bxlr.north);
	\end{tikzpicture}
\end{center}
Recall that the braided $p$-Leibniz rule only exists under the hypothesis \eqref{eq:mps_braid_mom}, as symbolised by the *. Moreover, it is important to note that these two cases ($p$ or $x$-Leibniz rule) may not be the only ones to form a coherent way to perform functional derivatives in a noncommutative context.

\paragraph{}
Both notions of functional derivatives lead to a coherent way to perform path integral quantisation. In the following subsections, we are investigating the possible difference of a quantum theory built thanks to functional derivatives that either follows the $p$-Leibniz rule (section \ref{subsec:ncpi_upl}), or the $x$-Leibniz rule (section \ref{subsec:ncpi_uxl}).

\paragraph{}
Finally, one can show that the different variants of Leibniz rule for functional derivatives give the same identities for momentum space derivatives
\begin{align}
    \label{eq:ncpi_mom_funcder_delta}
    \frac{\partial}{\partial \phi(p_1)} \phi(p_2)
    &= \mathscr{I}(p_1) \delta(p_1 \dminus p_2).
\end{align}
\begin{comp}
    In the $p$-Leibniz case, one computes
    \begin{align*}
    	\pLd{\phi(p_1)} \phi(p_2)
	   &= \int \tdl{x_1} \tdl{x_2} e^{i(\dminus p_2)x_2} e^{i p_1 x_1} \pLd{\phi(x_1)} \phi(x_2)
    	= \int \tdl{x_1} \tdl{x_2} e^{i(\dminus p_2)x_2} e^{ip_1 x_1} \delta(x_1,x_2) \\
    	&= \frac{1}{(2\pi)^{d+1}} \int \tdl{x} e^{i(\dminus p_2)x} e^{ip_1x}
    	= \delta(\dminus p_2 \dplus p_1)
    	= \mathscr{I}(p_1) \delta(p_1 \dminus p_2),
    \end{align*}
    while in the $x$-Leibniz case, one rather has
    \begin{align*}
        \xLd{\phi(p_1)} \phi(p_2)
    	&= \int \tdl{x_1} e^{ip_1x_1} \xLd{\phi(x_1)} \left( \int \tdl{x_2} e^{i(\dminus p_2) x_2} \phi(x_2) \right) \\
    	&= \int \tdl{x_1} e^{ip_1x_1} \xLd{\phi(x_1)} \left( \int \tdl{x_2} \phi(x_2) (\mathscr{I} \actl e^{i(\dminus p_2) x_2}) \right) \\
    	&= \int \tdl{x_1} \tdl{x_2} e^{ip_1x_1} \delta(x_1,x_2) \mathscr{I}(p_2)  e^{i(\dminus p_2) x_2}
    	= \frac{1}{(2\pi)^{d+1}} \int \tdl{x} \mathscr{I}(p_2) e^{ip_1 x} e^{i (\dminus p_2) x} \\
    	&= \mathscr{I}(p_2) \delta(p_1 \dminus p_2),
    \end{align*}
    the two are equal on the support of the delta functions.
\end{comp}

\subsection{Path integral and generating functional}
\label{subsec:ncpi_pi}
\paragraph{}
We consider the following free action
\begin{align}
	S_0(\phi)
	&= \frac{1}{2} \int \tdl{x} \phi(x) \Pi^{-1}[\phi(x)]
	= \frac{1}{2} \int \Haar{p} \phi(p) \Pi(p)^{-1} \phi(\dminus p),
	\label{eq:ncpi_free_action}
\end{align}
where $\Pi$ is the propagator term of the action, that is, in a commutative context, $\Pi(p) = p^2 + m^2$, where $m$ would be the mass of the scalar field.
The product between fields in $S_0$ is the product of $\Malg_\ell$.

We would like to introduce a path integral, which require to make sense of a Gaussian measure $\mathscr{D}\phi e^{-S_0}$. A natural possibility is to exploit the fact that, through the Fourier transform, we can map the space of noncommutative functions $\mathbb{A}_\ell$ (or at least a linear subspace of Fourier-transformable functions) to a space of commutative functions, $\phi(p)$. On this space, the generating functional can be explicitly calculated as
\begin{equation}
    \label{eq:ncpi_gen_func}
\begin{aligned}
	\mathcal{Z}(J)
	&= \frac{1}{\mathcal{Z}(0)} \int \mathscr{D}\phi\ \exp\left( - S_0(\phi) + \frac{1}{2} \int \tdl{x} \bigl(\phi (x) \, J (x)+ J(x) \,\phi(x) \bigr)  \right)\\
	&= \exp \left( - \frac{1}{2} \int \Haar{k} J(k) \Pi(k) J(\dminus k) \right).
	\end{aligned}
\end{equation}
The second equality of \eqref{eq:ncpi_gen_func} is obtained by performing the standard change of variable $\phi \to \phi + \Pi J$ (working with Fourier-transformed fields) and using the fact that
\begin{align}
    \label{eq:ncpi_prop_requirement}
    \Pi(\dminus k) = \Pi(k),
\end{align}
for any momenta $k$. This computation is detailed in \cite{Hersent_2024a}. In the following, we do not specify an explicit expression for $\Pi$ here, but simply require that \eqref{eq:ncpi_prop_requirement} holds.

Following the standard procedure, we want to compute the (free) $n$-point function defined as
\begin{equation}
	\label{eq:ncpi_npf_def}
\begin{aligned}
    	&\langle \phi(x_1) \cdots \phi(x_n) \rangle_0
	= \frac{1}{\mathcal{Z}(0)} \int \mathscr{D}\phi\ \phi(x_1) \cdots \phi(x_n) e^{-S_0(\phi)}
    \\
    &=\frac{1}{\mathcal{Z}(0)} \int \Haar{p_1} \dots \Haar{p_n} \, e^{i p_1 x_1}  e^{i p_2 x_2}  \dots   e^{i p_n x_n} \int \mathscr{D}\phi\ \phi(p_1) \cdots \phi(p_n) e^{-S_0(\phi)}
    \\
    &=\frac{1}{\mathcal{Z}(0)} \int \Haar{p_1} \dots \Haar{p_n} \, e^{i p_1 x} \tpb e^{i p_2 x} \tpb \dots \tpb  e^{i p_n x} \int \mathscr{D}\phi\ \phi(p_1) \cdots \phi(p_n) e^{-S_0(\phi)},
\end{aligned}
\end{equation}
where the subscript $0$ is here to specify that we take expectation values of the free theory. Moreover, we are also interested in performing calculations on some interacting theory. For simplicity, let us consider a noncommutative $\phi^4$ theory. Its action writes
\begin{align}
	S_4(\phi)
	&= S_0(\phi) + S_{\mathrm{int}}(\phi)
	= S_0(\phi) + \frac{\lambda}{4!} \int \tdl{x} \phi(x) \phi(x) \phi(x) \phi(x)
	\label{eq:ncpi_phi4_action}
\end{align}
where $S_0$ is the free action \eqref{eq:ncpi_free_action}, $\lambda$ is the coupling constant and the noncommutative product has been used in the $\phi^4$ term. In particular, we want to compute the $2$-point function at one-loop order. The full $2$-point function reads
\begin{align}
	\langle \phi(x_1) \phi(x_2) \rangle_4
	&= \frac{1}{\mathcal{Z}(0)} \int \mathscr{D}\phi\ \phi(x_1) \phi(x_2) e^{-S_{\mathrm{int}}(\phi)} e^{-S_0(\phi)},
\end{align}
where the subscript $4$ stands for the $\phi^4$ interaction term. The one-loop approximation consists in considering only one vertex, that is the first order in $\lambda$ of the preceding equation. This is done by expanding the exponential as an infinite sum and keeping only the first order term. Therefore,
\begin{align}
	\langle \phi(x_1) \phi(x_2) \rangle_4^{\text{1-loop}}
	&= - \frac{1}{\mathcal{Z}(0)} \int \mathscr{D}\phi\ \phi(x_1) \phi(x_2) S_{\mathrm{int}}(\phi) e^{-S_0(\phi)}
    = - \langle  \phi(x_1) \phi(x_2) S_{\mathrm{int}}(\phi) \rangle_0.
	\label{eq:ncpi_2pf_1l}
\end{align}

\paragraph{}
In Sec.~\ref{subsec:ncpi_upl}, we will show that the $n$-point function \eqref{eq:ncpi_npf_def}  can be calculated by taking $n$ $p$-Leibniz derivatives of $\mathcal{Z}$ from Eq.~\eqref{eq:ncpi_gen_func} with respect to $J$ and then setting $J = 0$. As shown in section \ref{subsec:ncpi_upl}, this quantisation leads to $n$-point functions that are not quantum Poincar\'{e} invariant and gives rise to UV/IR mixing effect.

To address this issue, inspired by Oeckl \cite{Oeckl_2000}, who argued that a braided notion of path integral is necessary to preserve covariance, we will introduce an invariant Gaussian path integral by postulating directly the form of the (free) generating functional \eqref{eq:ncpi_gen_func}:
\begin{equation}
    \label{eq:ncpi_gen_func_xLeib}
    \mathcal{Z}_0[J] = \exp \left( - \frac{1}{2} \int \Haar{k} J(k) \Pi(k) J(\dminus k) \right) \,.
\end{equation}
The free $n$-point functions will then be defined with the use of the x-Leibniz derivatives of \eqref{eq:ncpi_gen_func_xLeib}. The correlation functions of the ``$p$-Leibniz'' and ``$x$-Leibniz'' cases are different, so we will indicate the $n$-point functions of the former by $\langle \cdot \rangle$ as above, and the ones of the latter by $\dlangle \cdot \drangle$. The main focus of section \ref{subsec:ncpi_uxl} consists in the computation of the $n$-point functions with $x$-Leibniz derivative. It turns out that such correlation functions are quantum Poincar\'{e} invariant and do not produce UV/IR mixing.

\subsection{Undeformed \tops{$p$}{p}-Leibniz rule}
\label{subsec:ncpi_upl}

\subsubsection{Correlation functions}
\paragraph{}
We consider here the case of a functional derivative satisfying the $p$-Leibniz rule. First, we prove that the relation between the $n$-point function \eqref{eq:ncpi_npf_def} and the derivatives of the generating functional is
\begin{align}
	\langle \phi(x_1) \cdots \phi(x_n) \rangle_0
	&= \left. \left( \left( \frac{1 + \mathscr{I}}{2} \right)^{-1} \actl \pLd{J(x_1)} \right) 
	\cdots \left( \left( \frac{1 + \mathscr{I}}{2} \right)^{-1} \actl \pLd{J(x_n)} \right) 
	\mathcal{Z}(J) \right|_{J=0}.
	\label{eq:ncpi_upl_npf}
\end{align}
Note that $\left( \frac{1 + \mathscr{I}}{2} \right)^{-1}$ is well-defined since $\mathscr{I}$ is strictly positive. Moreover, in the unimodular case, $\frac{1 + \mathscr{I}}{2} = 1$ which simplifies the formula \eqref{eq:ncpi_upl_npf} to the usual one.

\begin{comp}
    We compute
\begin{align*}
	\pLd{J(x_1)} \mathcal{Z}(J)
	&= \int \Haar{p_1} e^{i (\dminus p_1) x_1} \pLd{J(p_1)} \frac{1}{\mathcal{Z}(0)} \int \mathscr{D}\phi\ \sum_{n=0}^{+\infty} \frac{1}{n! 2^n} \left(\int \phi J + J \phi \right)^n e^{-S(\phi)} \\
	&=  \int \Haar{p_1} e^{i (\dminus p_1) x_1} \frac{1}{\mathcal{Z}(0)} \int \mathscr{D}\phi\  \sum_{n=0}^{+\infty} \frac{1}{n! 2^n} \pLd{J(p_1)} \left(\int \phi J + J \phi \right)^n e^{-S(\phi)} \\
	&= \int \Haar{p_1} e^{i (\dminus p_1) x_1} \frac{1}{\mathcal{Z}(0)} \int \mathscr{D}\phi\  \sum_{n=0}^{+\infty} \frac{1}{n! 2^{n-1}} n \frac{1 + \mathscr{I}(p_1)}{2} \phi(\dminus p_1)  \left(\int \phi J + J \phi \right)^{n-1} e^{-S(\phi)} \\
	&= \frac{1}{\mathcal{Z}(0)} \int \mathscr{D}\phi\  \int \Haar{p_1} e^{i p_1 x_1} \frac{1 + \mathscr{I}(p_1)}{2} \phi(p_1) e^{-S(\phi) + \frac{1}{2} \int (\phi J + J \phi)}  \\
	&= \left\langle \left(\frac{1 + \mathscr{I}}{2} \actl \phi\right)(x_1) \right\rangle_0 \qquad \text{(for $J=0$),}
\end{align*}
where we used the fact that $\int (\phi J + J \phi)$ is a scalar and so commutes with itself. We also used the following computation
\begin{align*}
	\pLd{J(p_1)} \int \tdl{x} \big(\phi(x) J(x) + J(x) \phi(x)\big)
	&= \pLd{J(p_1)} \int \Haar{p}\Haar{q} \phi(p) J(q) \big( \delta(p \dplus q) + \delta(q \dplus p) \big) \\
	&= \pLd{J(p_1)} \int \Haar{p} \phi(p) J(\dminus p) \big( 1 + \mathscr{I}(p) \big) \\
	&= \int \Haar{p} \phi(p) \delta(p_1 \dplus p) \big( 1 + \mathscr{I}(\dminus p) \big) \\
	&= \phi(\dminus p_1) \big(1 + \mathscr{I}(p_1) \big).
\end{align*}
Then, for two derivatives, one has
\begin{align*}
	\pLd{J(x_2)} \pLd{J(x_1)} \mathcal{Z}(J)
	&= \! \int \! \Haar{p_2} e^{i(\dminus p_2) x_2} \frac{1}{\mathcal{Z}(0)} \! \int \! \mathscr{D}\phi \left(\frac{1 + \mathscr{I}}{2} \actl \phi\right)\!\!(x_1) e^{-S(\phi)}  \sum_{n=0}^{+\infty} \frac{1}{n! 2^n} \pLd{J(p_2)} \!\left(\int \! \phi J + J \phi \right)^{\!n} \\
	&= \frac{1}{\mathcal{Z}(0)} \int \mathscr{D}\phi\ \left(\frac{1 + \mathscr{I}}{2} \actl \phi\right)(x_2) \left(\frac{1 + \mathscr{I}}{2} \actl \phi\right)(x_1) e^{-S(\phi)}  \sum_{n=0}^{+\infty} \frac{1}{n! 2^n} \left(\int \phi J + J \phi \right)^n \\
	&= \left\langle \left(\frac{1 + \mathscr{I}}{2} \actl \phi\right)(x_2) \left(\frac{1 + \mathscr{I}}{2} \actl \phi\right)(x_1) \right\rangle_0 \qquad \text{(for $J=0$).}
\end{align*}
    The equation \eqref{eq:ncpi_upl_npf} is deduced recursively.
\end{comp}

\subsubsection{Free 4-point function}
\paragraph{}
Let us now consider the free $4$-point function, that is equation \eqref{eq:ncpi_upl_npf} with $n = 4$ and the action \eqref{eq:ncpi_free_action}. We compute
\begin{align}
	\begin{aligned}
	\langle & \phi(x_1) \cdots \phi(x_4)\rangle_0
	= \int \Haar{p} \Haar{q} \frac{2}{1 + \mathscr{I}(p)} \Pi(p) \frac{2}{1 + \mathscr{I}(q)} \Pi(q) \\
	&\qquad \times \Big( e^{i(\dminus p) x_1} e^{ip x_2} e^{i(\dminus q) x_3} e^{i q x_4}
		+ e^{i(\dminus p) x_1} e^{i(\dminus q) x_2} e^{i q x_3} e^{i p x_4}
		+ e^{i(\dminus p) x_1} e^{i(\dminus q) x_2} e^{i p x_3} e^{i q x_4} \Big).
	\end{aligned}
	\label{eq:ncpi_upl_4pf}
\end{align}
This expression is quite similar to the commutative one: the four points $x_1$, $x_2$, $x_3$ and $x_4$ are linked two by two by propagators with momenta $p$ and $q$ respectively. One of the main difference is the deformation of the propagator through the appearance of the modular function $\mathscr{I}$. Note that $\Pi$ can also have a complicated expression coming from the noncommutative action. The other main difference that is of primary importance here is that the exponentials do not commute anymore. 

\begin{comp}
Using expression \eqref{eq:ncpi_gen_func}, one computes
\begin{align*}
	\pLd{J(x_1)} & \mathcal{Z}(J)	
	= \int \Haar{p_1} e^{i(\dminus p_1)x_1} \pLd{J(p_1)} \mathcal{Z}(J) \\
	&= \int \Haar{p_1} e^{i(\dminus p_1)x_1} \pLd{J(p_1)} \exp\left( \frac{1}{2}\int \Haar{k} J(k) \Pi(k) J(\dminus k) \right) \\
	&= \frac{1}{2} \int \Haar{p_1} e^{i(\dminus p_1)x_1} \int \Haar{k} \big( \delta(k \dminus p_1) \mathscr{I}(k) \Pi(k) J(\dminus k) + J(k) \Pi(k) \mathscr{I}(\dminus k) \delta(p_1 \dplus k) \big) \\
	&\qquad \times \exp\left( \frac{1}{2}\int \Haar{k} J(k) \Pi(k) J(\dminus k) \right) \\
	&= \frac{1}{2} \int \Haar{p_1} e^{i(\dminus p_1)x_1} \big( \Pi(p_1) J(\dminus p_1) + J(\dminus p_1) \Pi(p_1) \mathscr{I}(p_1) \big) \exp\left( \frac{1}{2}\int \Haar{k} J(k) \Pi(k) J(\dminus k) \right) \\
	&= \int \Haar{p_1} e^{i(\dminus p_1)x_1} \frac{1 + \mathscr{I}(p_1)}{2} \Pi(p_1) J(\dminus p_1) \exp\left( \frac{1}{2}\int \Haar{k} J(k) \Pi(k) J(\dminus k) \right) \\
	&= - \left(\left(\frac{1 + \mathscr{I}}{2}\right) \actl \Pi[J(x_1)] \right) \mathcal{Z}(J)
\end{align*}
which implies that
\begin{align}
	\left( \left( \frac{1 + \mathscr{I}}{2} \right)^{-1} \actl \pLd{J(x_1)} \right) \mathcal{Z}(J)
	&= - \Pi[J(x_1)] \mathcal{Z}(J)
	\label{eq:ncpi_upl_gf_der}
\end{align}
Furthermore, one computes that
\begin{align*}
	\pLd{J(x_2)} \Pi[J(x_1)]
	&= \int \Haar{p_1}\Haar{p_2} e^{i (\dminus p_2) x_2} e^{i p_1 x_1} \Pi(p_1) \pLd{J(p_2)} J(p_1) \\
	&= \int \Haar{p_1}\Haar{p_2} e^{i (\dminus p_2) x_2} e^{i p_1 x_1} \mathscr{I}(p_1) \Pi(p_1) \delta(p_1 \dminus p_2) \\
	&= \int \Haar{p} e^{i (\dminus p) x_2} e^{i p x_1} \Pi(p) \\
	&= W(x_2, x_1),
\end{align*}
which implies that
\begin{align*}
	\left( \left( \frac{1 + \mathscr{I}}{2} \right)^{-1} \actl \pLd{J(x_2)} \right)  \Pi[J(x_1)]
	&= \int \Haar{p} e^{i (\dminus p) x_2} e^{i p x_1} \frac{2}{1 + \mathscr{I}(p)} \Pi(p)
	= W_{\mathscr{I}}(x_2, x_1)
\end{align*}
After performing the four derivatives of \eqref{eq:ncpi_upl_npf}, one obtains \eqref{eq:ncpi_upl_4pf}.
\end{comp}

\paragraph{}
Let us now look at the quantum Poincar\'{e} invariance of \eqref{eq:ncpi_upl_4pf}.
First, of all, the appearance of modular functions in \eqref{eq:ncpi_upl_4pf} could be detrimental for its quantum Lorentz invariance. An easy way out could be to consider only propagators $\Pi$ such that $\frac{2\Pi}{1 + \mathscr{I}}$ is $\Poin_\ell$-invariant. 
However, the translation invariant of \eqref{eq:ncpi_upl_4pf} is broken irreparably. Indeed, it contains three terms. The two first terms of \eqref{eq:ncpi_upl_4pf} satisfy the deformed conservation of energy/momenta \eqref{eq:mps_npt_cov_tran} and are therefore $\Tran_\ell$-invariant. However, the third term of \eqref{eq:ncpi_upl_4pf} has a deformed sum of momenta that reads
\begin{align*}
	\dminus p \dminus q \dplus p \dplus q
	\neq 0,
\end{align*}
which means that this term is not $\Tran_\ell$-invariant, since it does not fulfil \eqref{eq:mps_npt_cov_tran}. The breaking of the translation invariance is due to the fact that the momenta $p$ and $q$ are here intertwined, so that the previous expression cannot be simplified due to the noncommutativity of $\dplus$. 

Therefore, the $p$-Leibniz rule implies a breaking of the quantum Poincar\'{e} symmetry (notably through translations), even of the free theory.

\paragraph{}
As a direct example, let us consider the model of scalar field on Moyal spacetime studied in the original UV/IR mixing paper \cite{Minwalla_2000}, but only its free part. As already studied in~\cite{Hersent_2024a}, such a model is unimodular $\mathscr{I} = 1$. Moreover, the noncommutative product of plane waves produces a momentum-dependent phase, \ie $e^{ipx}e^{iqx} = e^{i(p+q)x} e^{-\frac{i}{2}p\Theta q}$. The phase originates from the fact that the algebra of $x$'s is a central extension of a Lie algebra. Following~\cite{Hersent_2024a}, we describe this model by enlarging momentum space to a $d+1$-dimensional space, and the momentum composition law generates nontrivial contributions to the additional $d+1$ component of momenta:
\begin{align*}
    (p \dplus q)_\mu
    &= p_\mu + q_\mu, &
    (\dminus p)_\mu
    &= -p_\mu, &
    (p \dplus q)_{d+1}
    &= p_{d+1} + q_{d+1} - \frac{i}{2} p_\mu \Theta^{\mu\nu} q_\nu, &
    (\dminus p)_{d+1}
    &= - p_{d+1},
\end{align*}
where $\mu,\nu = 1, \ldots ,d$. The additional $d+1$ component of momentum is zero when expanding fields in Fourier transform (it is not integrated over like the other components), and it appears only  when multiplying plane waves, in order to  take into account the phase term. This follows from writing the Moyal commutation relations as a central extension of the $d$-dimensional Abelian algebra, where the additional coordinate is fixed by the constraint $x^{d+1} = 1$~\cite{Hersent_2024a}.  The previous analysis therefore states that the free $4$-point function is not $\theta$-Poincar\'{e} invariant, especially not $\theta$-translation invariant. For example, a pure translation triggers a non-zero phase term
\begin{align}
    (\dminus p \dminus q \dplus p \dplus q)_{d+1}
    &= - i p_\mu \Theta^{\mu\nu} q_\nu.
\end{align}
This behaviour, already present in the quantisation of the free theory, has not been noticed by the authors of \cite{Minwalla_2000}. We claim here that this phenomenon has the same roots as the UV/IR divergence behaviour of the interacting theory that arose interest in the results of~\cite{Minwalla_2000} and is addressed by considering a fully covariant quantisation scheme. Other examples are treated in more details in the section \ref{sec:ex}. 

\paragraph{}
As a final note, let us observe that the (free) $2$-point function is $\Poin_\ell$-invariant, if we require that $\frac{2\Pi}{1 + \mathscr{I}}$ is covariant:
\begin{align}
	\langle \phi(x_1) \phi(x_2) \rangle_0
	&= \int \Haar{p} \frac{2}{1 + \mathscr{I}(p)} \Pi(p) e^{i(\dminus p) x_1} e^{i p x_2}
	= W_{\mathscr{I}}(x_1, x_2).
	\label{eq:ncpi_upl_2pf}
\end{align}

\subsubsection{2-point function of \tops{$\phi^4$}{phi\^{}4}-theory}
\paragraph{}
We now turn to the computation of the $2$-point function at $1$-loop order of the $\phi^4$ theory \eqref{eq:ncpi_phi4_action}. Using the definition \eqref{eq:ncpi_2pf_1l}, one has
\begin{align*}
	\langle \phi(x_1) \phi(x_2) \rangle_{4, \text{1-loop}}
	&= - \frac{\lambda}{4!} \int \tdl{x} \langle \phi(x_1) \phi(x_2) \phi(x)^4 \rangle_0.
\end{align*}
Then, one can use \eqref{eq:ncpi_upl_npf} and \eqref{eq:ncpi_upl_gf_der} to explicitly compute this $2$-point function and obtain
\begin{align}
\begin{aligned}
	\langle \phi(x_1) \phi(x_2) \rangle_4^{\text{1-loop}}
	&= \frac{\lambda}{4!} \int \Haar{p} e^{i  p x_1} e^{i (\dminus p) x_2} \frac{4}{1 + \mathscr{I}^{-1}(p)} \Pi(p)^2 \left( \int \Haar{q} \frac{2\big(3 + \mathscr{I}(q) \big)}{1 + \mathscr{I}(q)} \Pi(q) \right) \\
	&\phantom{=} + \frac{\lambda}{4!} \int \Haar{p_1} \Haar{p_2} e^{ip_1x_1} e^{i p_2 x_2} 4 \left( \frac{1}{1 + \mathscr{I}(p_1)} + \frac{1}{1 + \mathscr{I}(p_2)} \right) \Pi(p_1) \Pi(p_2) \\
	&\phantom{=} \times \left(  \int \Haar{q} \delta\big(\dminus q \dplus p_2 \dplus q \dplus p_1 \big) \frac{2}{1 + \mathscr{I}(q)} \Pi(q) \right) 
\end{aligned}
	\label{eq:ncpi_upl_2pf1l}
\end{align}
The previous expression splits into two contributions: the first one corresponds to the 8 planar diagrams and the second one to the 4 non-planar ones.\footnote{
	One can ``count'' the number of diagrams of each contribution by looking at the numerical prefactor of each term when $\mathscr{I} = 1$.
}
They correspond to the diagrams of Figure \ref{fig:planar} and Figure \ref{fig:nonplanar} respectively. Note that this result has already been derived in \cite{Hersent_2024a}.

\begin{figure}%[h]
    %\iffalse % Begin commenting out
    \begin{minipage}{.249\textwidth}
         \centering
    \begin{tikzpicture}[scale = 1.2]
        \draw[black] (-.7, -.7) node[anchor= east]{$p_2$} to (0, 0);
        \draw[black] (-.7,  .7) node[anchor= east]{$p_1$} to (0, 0);
        \draw[-{To}, black] (0, 0) to (.6, .6) to[out= 45, in= 90] (.9,0)
            node[anchor = east]{$q$};
        \draw[black] (0, 0) to (.6,-.6) to[out=-45, in=-90] (.9,0);
    \end{tikzpicture}
    \end{minipage}%
    \begin{minipage}{.249\textwidth}
         \centering
    \begin{tikzpicture}[scale = 1.2]
        \draw[black] (.7,  .7) node[anchor= west]{$p_2$} to (0, 0);
        \draw[black] (.7, -.7) node[anchor= west]{$p_1$} to (0, 0);
        \draw[black] (0, 0) to (-.6,-.6) to[out=-135, in=-90] (-.9,0);
        \draw[-{To}, black] (0, 0) to (-.6, .6) to[out= 135, in= 90] (-.9,0)
            node[anchor = west]{$q$};
    \end{tikzpicture}
    \end{minipage}%
    \begin{minipage}{.249\textwidth}
         \centering
    \begin{tikzpicture}[scale = 1.2]
        \draw[black] (-.7, -.7) node[anchor= east]{$p_1$} to (0, 0);
        \draw[black] ( .7, -.7) node[anchor= west]{$p_2$} to ( 0, 0);
        \draw[-{To}, black] (0, 0) to (-.6,  .6) to[out=135, in=180] (0, .9)
            node[anchor = north]{$q$};
        \draw[black] (0, 0) to (.6,.6) to[out=45, in=0] (0, .9);
    \end{tikzpicture}
    \end{minipage}%
    \begin{minipage}{.249\textwidth}
         \centering
    \begin{tikzpicture}[scale = 1.2]
        \draw[black] (-.7,.7) node[anchor= east]{$p_1$} to (0, 0);
        \draw[black] ( .7,.7) node[anchor= west]{$p_2$} to ( 0, 0);
        \draw[-{To}, black] (0, 0) to (-.6, -.6) to[out=-135, in=180] (0, -.9)
            node[anchor = south]{$q$};
        \draw[black] (0, 0) to (.6,-.6) to[out=-45, in=0] (0, -.9);
    \end{tikzpicture}
    \end{minipage}
    
    \begin{minipage}{.249\textwidth}
         \centering
        \begin{tikzpicture}[scale = 1.2]
        \draw[black] (-.7, -.7) node[anchor= east]{$p_1$} to (0, 0);
        \draw[black] (-.7,  .7) node[anchor= east]{$p_2$} to (0, 0);
        \draw[-{To}, black] (0, 0) to (.6, .6) to[out= 45, in= 90] (.9,0)
            node[anchor = east]{$q$};
        \draw[black] (0, 0) to (.6,-.6) to[out=-45, in=-90] (.9,0);
    \end{tikzpicture}
    \end{minipage}%
    \begin{minipage}{.249\textwidth}
         \centering
    \begin{tikzpicture}[scale = 1.2]
        \draw[black] (.7,  .7) node[anchor= west]{$p_1$} to (0, 0);
        \draw[black] (.7, -.7) node[anchor= west]{$p_2$} to (0, 0);
        \draw[black] (0, 0) to (-.6,-.6) to[out=-135, in=-90] (-.9,0);
        \draw[-{To}, black] (0, 0) to (-.6, .6) to[out= 135, in= 90] (-.9,0)
            node[anchor = west]{$q$};
    \end{tikzpicture}
    \end{minipage}%
    \begin{minipage}{.249\textwidth}
         \centering
    \begin{tikzpicture}[scale = 1.2]
        \draw[black] (-.7, -.7) node[anchor= east]{$p_2$} to (0, 0);
        \draw[black] ( .7, -.7) node[anchor= west]{$p_1$} to ( 0, 0);
        \draw[-{To}, black] (0, 0) to (-.6,  .6) to[out=135, in=180] (0, .9)
            node[anchor = north]{$q$};
        \draw[black] (0, 0) to (.6,.6) to[out=45, in=0] (0, .9);
    \end{tikzpicture}
    \end{minipage}%
    \begin{minipage}{.249\textwidth}
         \centering
    \begin{tikzpicture}[scale = 1.2]
        \draw[black] (-.7,.7) node[anchor= east]{$p_2$} to (0, 0);
        \draw[black] ( .7,.7) node[anchor= west]{$p_1$} to (0, 0);
        \draw[-{To}, black] (0, 0) to (-.6, -.6) to[out=-135, in=180] (0, -.9)
            node[anchor = south]{$q$};
        \draw[black] (0, 0) to (.6,-.6) to[out=-45, in=0] (0, -.9);
    \end{tikzpicture}
    \end{minipage}
    %\fi % End commenting out
    
    \caption{The four one-loop planar diagrams associated to the $2$-point function \eqref{eq:ncpi_upl_2pf1l}.}
    \label{fig:planar}
\end{figure}

\begin{figure}%[h]
    \begin{minipage}{.245\textwidth}
         \centering
    \begin{tikzpicture}[scale = 1.2]
        \draw[black] (-.7, -.7) node[anchor= east]{$p_1$} to (0, 0);
        \draw[black] ( .7,  .7) node[anchor= west]{$p_2$} to (0, 0);
        \draw[black,
            decoration={markings, mark=at position 0.75 with {\arrow{To}}},
            postaction={decorate}
        ] (0, 0) to (-.6,  .6) to[out=135, in=135] (.45, .55);
        \node at (-.1,1.1) {$q$};
        \draw[black] (0, 0) to ( .6, -.6) to[out=-45, in=-45] (.55, .45);
    \end{tikzpicture}
    \end{minipage}%
    \begin{minipage}{.245\textwidth}
         \centering
    \begin{tikzpicture}[scale = 1.2]
        \draw[black] (-.7, -.7) node[anchor= east]{$p_2$} to (0, 0);
        \draw[black] ( .7,  .7) node[anchor= west]{$p_1$} to (0, 0);
        \draw[black,
            decoration={markings, mark=at position 0.75 with {\arrow{To}}},
            postaction={decorate}
        ] (0, 0) to (-.6,  .6) to[out=135, in=135] (.45, .55);
        \node at (-.1,1.1) {$q$};
        \draw[black] (0, 0) to ( .6, -.6) to[out=-45, in=-45] (.55, .45);
    \end{tikzpicture}
    \end{minipage}%
    \begin{minipage}{.245\textwidth}
         \centering
    \begin{tikzpicture}[scale = 1.2]
        \draw[black] (-.7,  .7) node[anchor= east]{$p_1$} to (0, 0);
        \draw[black] ( .7, -.7) node[anchor= west]{$p_2$} to (0, 0);
        \draw[black,
            decoration={markings, mark=at position 0.75 with {\arrow{To}}},
            postaction={decorate}
        ] (0, 0) to ( .6,  .6) to[out=  45, in=  45] (.55, -.45);
        \node at (1,0) {$q$};
        \draw[black] (0, 0) to (-.6, -.6) to[out=-135, in=-135] (.45, -.55);
    \end{tikzpicture}
    \end{minipage}
    \begin{minipage}{.245\textwidth}
         \centering
    \begin{tikzpicture}[scale = 1.2]
        \draw[black] (-.7,  .7) node[anchor= east]{$p_2$} to (0, 0);
        \draw[black] ( .7, -.7) node[anchor= west]{$p_1$} to (0, 0);
        \draw[black,
            decoration={markings, mark=at position 0.75 with {\arrow{To}}},
            postaction={decorate}
        ] (0, 0) to ( .6,  .6) to[out=  45, in=  45] (.55, -.45);
        \node at (1,0) {$q$};
        \draw[black] (0, 0) to (-.6, -.6) to[out=-135, in=-135] (.45, -.55);
    \end{tikzpicture}
    \end{minipage}
    
    \caption{The four one-loop non-planar diagram associated to the $2$-point function \eqref{eq:ncpi_upl_2pf1l}.}
    \label{fig:nonplanar}
\end{figure}

\paragraph{}
One can analyse the formula \eqref{eq:ncpi_upl_2pf1l} as follows. If one removes the factors of $\mathscr{I}$, the two terms have the following form:
\begin{align*}
	\text{Planar:}&
	\quad \sim \delta(p_1 \dplus p_2) \int \Haar{q} \Pi(q), &
	\text{Non-planar:}&
	\quad \sim \int \Haar{q} \delta\big(\dminus q \dplus p_2 \dplus q \dplus p_1 \big) \Pi(q),
\end{align*}
where $p_1$ and $p_2$ are the ingoing and outgoing momenta and $q$ is the loop momenta. The planar diagrams correspond to the ``expected'' noncommutative generalisation of the $2$-point function, with a deformed conservation of momenta $p_1 \dplus p_2 = 0$, together with the integral of the propagator. However, the non-planar diagrams are a complete novelty compared to the commutative case and emerge specifically due to the noncommutativity of the momentum composition law. The loop momenta now intertwined with the external momenta in the conservation law and therefore affect the propagation, most of the time making the integral of the propagator finite. If one considers a commutative limit, or a limit where one of the external leg has a vanishing momenta (\ie $p_1 = 0$ or $p_2 = 0$), then it is immediate that $\dminus q$ and $\dplus q$ will cancel each other in the expression\footnote{
    If $p_1 = 0$, we use the cyclicity of the delta function \eqref{eq:Md_dcyc} to have $q$ and $\dminus q$ cancelling each other.
}
$\dminus q \dplus p_2 \dplus q \dplus p_1$. Therefore, we recover in this limit the (supposedly UV divergent) integral of the propagator. That is precisely what has been called ``UV/IR mixing'' in the noncommutative literature \cite{Minwalla_2000}: for a vanishing external momenta (the IR regime), one recovers a divergence of the propagator (in the UV regime). One can find an extensive discussion about this result and its consequences for the understanding of UV/IR mixing, as well as a more detailed list of references in \cite{Hersent_2024a}.

\paragraph{}
Returning to the example of Moyal, one computes that
\begin{align*}
    (\dminus q \dplus p_2 \dplus q \dplus p_1)_{d+1}
    &= (p_1)_{d+1} + (p_2)_{d+1} + i q \Theta p_1,
\end{align*}
with the notation $p \Theta q = p_\mu \Theta^{\mu\nu} q_\nu$ and where we used the anti-symmetry of $\Theta$, \ie $p \Theta q = - q \Theta p$. The $d+1$ components of $p_1$ and $p_2$ are here for computational reasons only and vanish in the final result. Yet, the Fourier transform of the $i p_1 \Theta q$ term yields a phase term: the non-planar contribution reads
\begin{align*}
    \frac{\lambda}{6} \int \Haar{p_1} \Haar{p_2} e^{ip_1x_1} e^{i p_2 x_2} \Pi(p_1) \Pi(p_2) \ \delta(p_1 + p_2) \int \Haar{q} \Pi(q) e^{i q \Theta p_1},
\end{align*}
which is exactly equation (3.3) of \cite{Minwalla_2000}, for $\Pi(q)^{-1} = q^2 + m^2$. More details can be found in section 4.1 of \cite{Hersent_2024a}.

\paragraph{}
Contrary to \cite{Hersent_2024a}, we now have the tools to study the quantum Poincar\'{e} invariance of this $2$-point function. First, the quantum Lorentz invariance of \eqref{eq:ncpi_upl_2pf1l}, should be treated with care because of the presence of the modular functions. However we would like to focus more on the translation invariance. The analysis of section \ref{subsec:mps_cov} showed that translation invariance is fulfilled iff the conservation of momenta is satisfied. Therefore, \eqref{eq:ncpi_upl_2pf1l} is $\Poin_\ell$-invariant iff $p_1 \dplus p_2 = 0$. For planar diagrams, this conservation of momenta is directly imposed by the delta function, so that those diagrams are quantum translation invariant. On the other hand, the specificity of the non-planar diagrams is that the conservation of energy is now mixed with the loop momenta $q$. As $q$ is integrated over, it can take arbitrary values, so that the non-planar diagrams are not $\Poin_\ell$-invariant in general. This analysis shows that quantum Poincar\'{e} invariance breaking and UV/IR mixing have a common origin.

The fact that momenta are mixed up, and that $\dplus$ is noncommutative, are the source of both the breaking of covariance, since the deformed energy conservation law \eqref{eq:mps_npt_cov_tran} does not hold anymore, and of the UV/IR mixing, as first analysed in \cite{Hersent_2024a}. Two important remarks are in order:
\begin{itemize}
    \item The notion of ``divergence'' is not linked to the intertwining of momenta (and so of the non-covariant quantisation procedure), but only to the behaviour of the propagator. Indeed, as developed in \cite{Hersent_2024a}, if one considers a theory where $\int \Haar{p} \Pi(p)$ is finite, then the IR limit $p_1, p_2 \to 0$ of non-planar diagrams will also be finite\footnote{
        Note that this property has been proved to be independent of the modular factors thanks to the requirement \eqref{eq:ncpi_prop_requirement}.
    }.
    The UV/IR mixing effect only generates divergences when the integral of the propagator diverges. The question of the regularisation of noncommutative QFTs, first put forward by Snyder \cite{Snyder_1947}, has therefore nothing to do with the UV/IR divergences of \cite{Minwalla_2000}: the first is related to the behaviour of the propagator, and the second comes from a non-covariant quantisation scheme.
    \item Even if a theory based on a covariant quantisation scheme was free of the UV/IR mixing of \cite{Minwalla_2000}, that does not mean such a theory has separated UV and IR regimes: there still could be UV/IR mixing phenomena coming from the fact that the classical theory is non-local.
\end{itemize}

\subsection{Undeformed \tops{$x$}{x}-Leibniz rule}
\label{subsec:ncpi_uxl}
\paragraph{}
Throughout this subsection we work under the ribbon assumption \eqref{eq:mps_ribbon_hypo} stated in Section~\ref{sec:mps}.

\subsubsection{Correlation functions}
\paragraph{}
We now define the free $n$-point functions from the generating functional \eqref{eq:ncpi_gen_func_xLeib} by means of $x$-Leibniz derivatives. The most direct prescription would be
\begin{align}
	\dlangle \phi(x_1) \cdots \phi(x_n) \drangle
	&= \left.  \xLd{J(x_n)} \dots \xLd{J(x_1)}  \mathcal{Z}_0(J) \right|_{J=0} \,,
	\label{eq:ncpi_uxl_npf_simplest}
\end{align}
which gives the (free) $2$-point function
\begin{align}
	\label{eq:ncpi_uxl_f2pf_alt}
	\dlangle \phi(x_1) \phi(x_2) \!\left.\drangle\right._{\!0}
	&= \int \Haar{q} \left( \frac{1 + \nu^{-1}(q)}{2}\right) \Pi(q) e^{i q x_1} e^{i (\dminus q) x_2}.
\end{align}
As it stands, \eqref{eq:ncpi_uxl_npf_simplest} is not yet the correct prescription, and the reason is structural rather than a matter of taste, as we now explain.

\paragraph{}
The source term entering $\mathcal{Z}_0$ is the \emph{symmetric} coupling $\tfrac{1}{2}\int \tdl{x}\big(\phi(x) J(x) + J(x) \phi(x)\big)$. The symmetrisation is forced: $\phi$ and $J$ do not commute, and a single ordering would privilege one of the two inequivalent pairings. As a consequence, a single derivative with respect to $J$ acts on \emph{both} orderings and therefore returns the field not bare, but dressed by a symmetrisation factor. The two orderings of the coupling are exchanged by the reordering identity that is active in the formalism at hand: the cyclicity of the integral \eqref{eq:Md_int_tcyc}, governed by the modular function $\mathscr{I}$, in the $p$-Leibniz case; and the braided reordering of paired plane waves \eqref{eq:mps_ribbon_hypo}, governed by the ribbon element $\nu^{-1}$ (equivalently the symmetry of the delta \eqref{eq:mps_delta_sym}), in the $x$-Leibniz case. Explicitly, one derivative of the coupling yields the field dressed by $\tfrac{1+\mathscr{I}}{2}$ in the former case (see the computation leading to \eqref{eq:ncpi_upl_npf}) and by $\tfrac{1+\nu^{-1}}{2}$ in the latter (see the computation below). This doubling is a property of the source \emph{pairing}, not of the field itself: for ``$n$ derivatives'' to mean ``$n$ field insertions'', it must be divided out.

\paragraph{}
We therefore adopt, in both quantisation schemes, one and the same definition: each insertion is the \emph{normalised} source derivative $\big(\tfrac{1+\,\cdot\,}{2}\big)^{-1} \actl \tfrac{\partial}{\partial J}$, with the dressing factor of the active reordering divided out. In the $p$-Leibniz case this rule \emph{is} \eqref{eq:ncpi_upl_npf}, where the factor $\tfrac{1+\mathscr{I}}{2}$ was \emph{derived} from the Gaussian integral. Transporting the identical rule to the $x$-Leibniz case, with the ribbon element $\nu^{-1}$ in place of $\mathscr{I}$, gives
\begin{align}
	\dlangle \phi(x_1) \cdots \phi(x_n) \drangle
	&= \left. \left( \left( \frac{1 + \nu^{-1}}{2} \right)^{-1} \actl \xLd{J(x_n)} \right) \cdots \left( \left( \frac{1 + \nu^{-1}}{2} \right)^{-1} \actl \xLd{J(x_1)} \right) \mathcal{Z}_0(J) \right|_{J=0} \,.
	\label{eq:ncpi_uxl_npf}
\end{align}
Thus \eqref{eq:ncpi_uxl_npf} is not an independent postulate but the image of \eqref{eq:ncpi_upl_npf} under a single prescription, fixed by the symmetric source coupling.

\paragraph{}
It is worth stressing that covariance does \emph{not} by itself select this normalisation. Since the ribbon element $\nu$ is automatically covariant (in the sense of \eqref{eq:mps_op_cov}), every power of $\tfrac{1+\nu^{-1}}{2}$ produces a covariant correlator; in particular both \eqref{eq:ncpi_uxl_npf_simplest} and \eqref{eq:ncpi_uxl_npf} are covariant. What singles out \eqref{eq:ncpi_uxl_npf} is precisely the source-pairing rule above. The resulting free $2$-point function,
\begin{align}
	\label{eq:ncpi_uxl_f2pf}
	\dlangle \phi(x_1) \phi(x_2) \!\left.\drangle\right._{\!0}
	&= \int \Haar{q} \frac{2}{1 + \nu^{-1}(q)} \Pi(q) e^{i q x_1} e^{i (\dminus q) x_2}
	= W_{\nu^{-1}}(x_1, x_2),
\end{align}
is the exact image, under the dictionary $\mathscr{I} \leftrightarrow \nu^{-1}$, of the $p$-Leibniz result $W_{\mathscr{I}}$ of \eqref{eq:ncpi_upl_2pf}; the naive choice \eqref{eq:ncpi_uxl_npf_simplest}, with $\tfrac{1+\nu^{-1}}{2}$ in the numerator \eqref{eq:ncpi_uxl_f2pf_alt}, has no $p$-Leibniz counterpart. The two prescriptions differ only by the field redefinition $\phi \mapsto \big(\tfrac{1+\nu^{-1}}{2}\big)^{-1} \actl \phi$, which can just be seen as a nonlinear redefinition of the kinetic operator in the free action.

\paragraph{}
A word on the status of $\mathcal{Z}_0$. In the $x$-Leibniz scheme we do not construct a Gaussian measure: $\mathcal{Z}_0$ is taken as the defining object \eqref{eq:ncpi_gen_func_xLeib}, the algebraic counterpart of the formal Gaussian integral. The construction is nonetheless self-contained, since $\mathcal{Z}_0$, together with the source-pairing normalisation \eqref{eq:ncpi_uxl_npf}, determines every free correlator. The symmetrisation factor $\tfrac{1+\nu^{-1}}{2}$ divided out in \eqref{eq:ncpi_uxl_npf} is precisely the one produced when a single source derivative acts on the symmetric coupling of $\mathcal{Z}_0$, so the prescription is fixed by the same data that defines $\mathcal{Z}_0$.
\begin{comp}
We derive here Eqs.~\eqref{eq:ncpi_uxl_f2pf_alt} and \eqref{eq:ncpi_uxl_f2pf}. The expression \eqref{eq:ncpi_gen_func_xLeib} of $\mathcal{Z}_0$ in momentum variable can be shown to reduce in position variable to
\begin{align*}
	\mathcal{Z}_0(J)
	= \exp\left( - \frac{1}{2} \int \tdl{x_1} \tdl{x_2} J(x_1) W(x_1,x_2) J(x_2) \right),
\end{align*}
since,
\begin{align*}
    \int \tdl{x_1} \tdl{x_2} J(x_1) W(x_1,x_2) J(x_2)
    &= \int \Haar{p} \tdl{x_1} \tdl{x_2} J(x_1) \Pi(p) e^{ipx_1} e^{i(\dminus p) x_2} J(x_2) \\
    &= \int \Haar{p} \left( \int \tdl{x_1} J(x_1) e^{ipx_1} \right) \Pi(p) \left( \int \tdl{x_2} e^{i(\dminus p)x_2} J(x_2) \right) \\
    &= \int \Haar{p} \left( \int \tdl{x_1} \mathscr{I}(p) e^{ipx_1} J(x_1) \right) \Pi(p) J(p) \\
    &\hspace{-.5cm}\overset{(p \to \dminus p)}{=} \int \Haar{p} \left( \int \tdl{x_1} e^{i(\dminus p)x_1} J(x_1) \right) \Pi(p) J(\dminus p) \\
    &= \int \Haar{p} J(p) \Pi(p) J(\dminus p).
\end{align*}
The functional derivative of $\mathcal{Z}_0$ thus reads
\begin{align*}
	\xLd{J(x_2)} \xLd{J(x_1)} \mathcal{Z}_0(J) \big|_{J=0}
	&= \frac{1}{2} \xLd{J(x_2)} \left( \int \tdl{x_3} \tdl{x_4} \delta(x_1,x_3) W(x_3,x_4) J(x_4) + J(x_3) W(x_3,x_4) \delta(x_1,x_4) \right) \\
	&= \frac{1}{2} \int \tdl{x_3} \tdl{x_4} \delta(x_1,x_3) W(x_3,x_4) \delta(x_2,x_4) + \delta(x_2,x_3) W(x_3,x_4) \delta(x_1,x_4) \\
    &= \frac{1}{2} \int \tdl{x_4} W(x_1,x_4) (\nu^{-1} \actl \delta)(x_4,x_2) + W(x_2,x_4) (\nu^{-1} \actl \delta)(x_4,x_1) \\
    &= \frac{1}{2} \big( \nu^{-1} \actl W(x_1,x_2) + \nu^{-1} \actl W(x_2,x_1) \big) \\
	&= \left( \frac{1 + \nu^{-1}}{2} \right) \actl W(x_1,x_2),
\end{align*}
Since $\nu$ acts diagonally, it follows
\begin{align*}
	\dlangle \phi(x_1)\phi(x_2) \!\left.\drangle\right._{\!0}
    &= \left( \frac{2}{1 + \nu^{-1}} \actl \xLd{J(x_2)} \right) \left( \frac{2}{1 + \nu^{-1}} \actl \xLd{J(x_1)} \right) \mathcal{Z}_0(J) \Big|_{J=0} \\
    &= \left( \left(\frac{2}{1 + \nu^{-1}} \right)^2 \frac{1 + \nu^{-1}}{2} \right) \actl W(x_1,x_2)
    = \frac{2}{1 + \nu^{-1}} \actl W(x_1,x_2)
    = W_{\nu^{-1}}(x_1,x_2).
\end{align*}
\end{comp}

\subsubsection{Free 4-point function}
\paragraph{}
We turn now to the computation of the (free) $4$-point function and its covariance. First, let us state that the (free) $2$-point function \eqref{eq:ncpi_uxl_f2pf} is $\Poin_\ell$-invariant, thanks to the criterion \eqref{eq:mps_cov_npf}: it is of the form \eqref{eq:mps_inv_2pf} since both $\Pi$ and $\nu$ are covariant.

The (free) $4$-point function is computed similarly as the $2$-point function:
\begin{align}
	\begin{aligned}
	\dlangle & \phi(x_1) \cdots \phi(x_4) \!\left.\drangle\right._{\!0}
	= W_{\nu^{-1}}(x_1, x_2) W_{\nu^{-1}}(x_3, x_4)
	+ W_{\nu^{-1}}(x_1, x_3) W_{\nu^{-1}}(x_2, x_4) \\
	&\hspace{3.2cm} + W_{\nu^{-1}}(x_1, x_4) W_{\nu^{-1}}(x_2, x_3) \\
	&= \int \Haar{p} \Haar{q} \frac{2}{1 + \nu^{-1}(p)} \Pi(p) \frac{2}{1 + \nu^{-1}(q)} \Pi(q) \\
	&\qquad \times \Big( e^{i p x_1} e^{i (\dminus p) x_2} e^{i q x_3} e^{i (\dminus q) x_4}
	+ e^{i p x_1} e^{i (\dminus p) x_3} e^{i q x_2} e^{i (\dminus q) x_4}
	+ e^{i p x_1} e^{i (\dminus p) x_4} e^{i q x_2} e^{i (\dminus q) x_3} \Big)
	\end{aligned}
	\label{eq:ncpi_uxl_4pf}
\end{align}
This expression is straightforwardly covariant since it consists in a (sum of) product of $2$-point functions which are themselves $\Poin_\ell$-covariant. Moreover, comparing \eqref{eq:ncpi_upl_4pf} and \eqref{eq:ncpi_uxl_4pf}, one observes that modular factors have been replaced by factors of $\nu^{-1}$. The momenta ($p$ and $q$) are ordered in \eqref{eq:ncpi_uxl_4pf} while they were mixed in \eqref{eq:ncpi_upl_4pf}: this is the main feature that ensure the invariance of $\Poin_\ell$, and will also ensure the non-appearance of UV/IR mixing of the $2$-point function at one-loop of the interacting theory, as discussed below. Remarkably, the positions (the $x_j$'s) are ordered in \eqref{eq:ncpi_upl_4pf} but scrambled     in \eqref{eq:ncpi_uxl_4pf}.

\paragraph{}
Therefore, already at the level of the free theory, the choice in the functional derivative influences dramatically the covariance properties of the $n$-point functions. One can even prove a ``deformed Wick theorem'' by recursion: any free $n$-point function of the $x$-Leibniz free theory is either $0$ (if $n$ is odd), or a product of $2$-point functions \eqref{eq:ncpi_uxl_f2pf} (if $n$ is even). It implies that any (free) $n$-point function is therefore quantum Poincar\'{e} invariant.

\subsubsection{2-point function of \tops{$\phi^4$}{phi\^{}4}-theory}
\paragraph{}
The formula \eqref{eq:ncpi_2pf_1l} for the $2$-point function at one-loop order with the $\phi^4$-action, writes
\begin{align}
    \dlangle \phi(x_1) \phi(x_2) \drangle_4^{\text{1-loop}}
    &= - \frac{\lambda}{4!} \dlangle[\Big] \left( \int \tdl{x_3} \phi(x_3)^4 \right) \phi(x_1) \phi(x_2) \!\left.\drangle[\Big]\right._{\!0},
\end{align}
which consists in a free $6$-point function computation. Therefore, using \eqref{eq:ncpi_uxl_f2pf}, one obtains that
\begin{align}
    \label{eq:ncpi_uxl_2pf1l}
	\dlangle \phi(x_1) \phi(x_2) \drangle_4^{\text{1-loop}}
	&= \frac{\lambda}{4!} \int \tdl{x_3} W_{\nu^{-1}}(x_3, x_3) \big( 6 W_{\nu^{-1}}(x_1 ,x_3) W_{\nu^{-1}}(x_2, x_3) + 6 W_{\nu^{-1}}(x_2, x_3) W_{\nu^{-1}}(x_1, x_3) \big).
\end{align}
First, let us note that the loop contribution $W_{\nu^{-1}}(x,x)$ is a number that commutes with all other quantities since
\begin{align}
    \label{eq:ncpi_loop_prop}
	W_{\nu^{-1}}(x,x)
	&= \int \Haar{k} \frac{2}{1 + \nu^{-1}(k)} \Pi(k) e^{i k x} e^{i (\dminus k) x}
	= \int \Haar{k} \frac{2\Pi(k)}{1 + \nu^{-1}(k)}.
\end{align}
Then, we have two sets of 6 diagrams each, which are not distinguished by planarity. The $6$ first diagrams $W_{\nu^{-1}}(x_1,x) W_{\nu^{-1}}(x_2,x)$ corresponds to $\phi(x_1)$ being the ``ingoing'' particle and $\phi(x_2)$ the ``outgoing'' one. Conversely, the 6 diagrams $W_{\nu^{-1}}(x_2,x) W_{\nu^{-1}}(x_1,x)$ corresponds to $\phi(x_2)$ being the ``ingoing'' particle and $\phi(x_1)$ the ``outgoing'' one. Note that this $6+6$ splitting was already found by Oeckl \cite{Oeckl_2001}. 

\begin{comp}
Using the hypothesis \eqref{eq:mps_ribbon_hypo}, the expression of \eqref{eq:ncpi_uxl_2pf1l} can be further simplified. Indeed, for any $a,b <3$, one computes
\begin{align*}
    \int \tdl{x_3} W_{\nu^{-1}}(x_a,x_3) W_{\nu^{-1}}(x_b,x_3)
    &= \int \tdl{x_3} W_{\nu^{-1}}(x_a,x_3) (\nu^{-1} \actl W_{\nu^{-1}})(x_3,x_b) \\
    &= \int \Haar{p} \Haar{q} \tdl{x_3} \frac{2\Pi(p)}{1 + \nu^{-1}(p)} \frac{2 \nu^{-1}(q) \Pi(q)}{1 + \nu^{-1}(q)} e^{i p x_a} e^{i(\dminus p) x_3} e^{i q x_3} e^{i (\dminus q) x_b} \\
    &= \int \Haar{p} \Haar{q} \frac{2\Pi(p)}{1 + \nu^{-1}(p)} \frac{2 \nu^{-1}(q) \Pi(q)}{1 + \nu^{-1}(q)} e^{i p x_a} \delta(\dminus p \dplus q) e^{i (\dminus q) x_b} \\
    &= \int \Haar{p} \nu^{-1}(p) \left( \frac{2 \Pi(p)}{1 + \nu^{-1}(p)} \right)^2 e^{i p x_a} e^{i (\dminus p) x_b}.
\end{align*}
Two terms are of this kind: one corresponds to $a = 1$, $b = 2$ and the other to $a = 2$, $b = 1$. They differ by a factor of $\nu^{-1}$ by \eqref{eq:mps_ribbon_hypo}. Therefore, we can write
\begin{align*}
    \int \tdl{x_3} \big( W_{\nu^{-1}}(x_1,x_3) W_{\nu^{-1}}(x_2,x_3) &+ W_{\nu^{-1}}(x_2,x_3) W_{\nu^{-1}}(x_1,x_3) \big) \\
    &= \int \Haar{p} (1 + \nu^{-1}(p)) \left( \frac{2 \Pi(p)}{1 + \nu^{-1}(p)} \right)^2 e^{i p x_1} e^{i (\dminus p) x_2} \\
    &= \int \Haar{p} \frac{4 \Pi(p)^2}{1 + \nu^{-1}(p)} e^{i p x_1} e^{i (\dminus p) x_2}
\end{align*}
\end{comp}

Finally, let us write the 1-loop correction to the $2$-point function in full length
\begin{align}
    \tag{\ref{eq:ncpi_uxl_2pf1l}}
    \begin{aligned}
	\dlangle \phi(x_1) \phi(x_2) \drangle_4^{\text{1-loop}}
	&= \lambda \int \Haar{p} \frac{2 \Pi(p)^2}{1 + \nu^{-1}(p)} e^{i p x_1} e^{i (\dminus p)x_2} \left( \int \Haar{k} \frac{2\Pi(k)}{1 + \nu^{-1}(k)} \right) \\
    &= \lambda \int \Haar{p} \Haar{q} \frac{2 \Pi(p)^2}{1 + \nu^{-1}(p)} \delta(p \dplus q) e^{i p x_1} e^{i q x_2} \left( \int \Haar{k} \frac{2\Pi(k)}{1 + \nu^{-1}(k)} \right),
	\end{aligned}
\end{align}
where the second equality is only here to emphasise the deformed energy/momentum conservation $p \dplus q = 0$.

\paragraph{}
As a first observation, the $2$-point function at one-loop order \eqref{eq:ncpi_uxl_2pf1l} is expressed as a sum of product of $2$-point functions which are quantum Poincar\'{e} invariant, and so is itself invariant. This can also be read from the final expression which is of the form \eqref{eq:mps_inv_2pf}.

Moreover, the loop integral, that is the term \eqref{eq:ncpi_loop_prop}, is isolated. Therefore, no intertwining of external momenta with the loop momenta leading to a UV/IR divergence, as discussed for \eqref{eq:ncpi_upl_2pf1l}, can occur. The expression \eqref{eq:ncpi_uxl_2pf1l} is free of UV/IR mixing. Regarding divergences, one should proceed as in the commutative case: one needs to study the convergence of \eqref{eq:ncpi_loop_prop} and regularise it if necessary. In that respect, note that \eqref{eq:ncpi_uxl_2pf1l} is not necessarily free of divergence (the latter depends on the behaviour of the propagator $\Pi$) but it is free of UV/IR mixing.

\subsection{Disentangling the divergences}
\label{subsec:ncpi_div}
\paragraph{}
As a last remark, we wish to underline here the role played by the propagator $\Pi$. The hope of Snyder \cite{Snyder_1947} for noncommutative field theory was that it could regularise the UV divergences of relativistic quantum mechanics.\footnote{
    We recall that Snyder's work precedes the renormalisation program.
}
In the quantum group approach, noncommutative regularization has been proposed as a covariant alternative to lattice regulators, which inherently break Lorentz and translation invariance~\cite{Majid_1990, Oeckl_2000}. However, subsequent work by Minwalla et al.~\cite{Minwalla_2000} revealed that certain noncommutative field theories still suffer from divergences, which are actually exacerbated by the UV/IR mixing effect. The goal of the present paper was to show that the UV/IR mixing effect, as introduced in \cite{Minwalla_2000}, is not intrinsic to noncommutativity itself: in the models considered here it arises from a non-covariant quantisation scheme. Yet, the question of divergences, as put forward by Snyder, has not been addressed yet and we wish to make a step forward in that direction too. In order to do so, we analyse the role of the propagator $\Pi$ in our computations.

\paragraph{}
On the one hand, since the plane waves are eigenfunctions of the propagator, \ie $\Pi[e^{ipx}] = \Pi(p) e^{ipx}$, the choice of propagator does not influence the quantum Poincaré invariance of the theory (as long as it is Lorentz-invariant). Therefore, by requiring only Poincar\'e covariance, one could be left, at the level of the scalar theory, with an infinite set of admissible propagators.\footnote{
    By ``admissible'', we here mean that the propagator needs to tend to the usual Klein-Gordon propagator in the commutative limit, \ie $\Pi(p) \underset{\ell \to 0}{\longrightarrow} (p^2 + m^2)^{-1}$.
}. However, it is straightforward to see that the propagator is determining the singularity structure of the theory, at the level of the $n$-point functions. Explicitly in \eqref{eq:ncpi_uxl_2pf1l}, the divergence of the $2$-point function is determined only\footnote{
    Indeed, it can be shown \cite{Hersent_2024a} that the modular function does not affect the divergence.
}
by the divergence of $\int \Pi$, the loop contribution. Therefore, depending on $\Pi$, there could be divergent $\int \Pi$, like in unimodular $T$-Minkowski models (see section \ref{subsec:ex_TM}), or there could be convergent $\int \Pi$ that are regularised thanks to the noncommutativity. For the latter case, we refer to the $3$-dimensional quantum gravity model (see section \ref{subsec:ex_3qg}), or the $\kappa$-Minkowski model discussed in section 4.2 of \cite{Hersent_2024a}.

As a consequence, it is important to underline two different phenomena, that we think need clarification but also to be treated separately:
\begin{itemize}
    \item Concerning the \textbf{regularisation} of the quantum noncommutative field theories, the essential piece is the propagator and its behaviour.
    \item Concerning the \textbf{covariance} of these theories, the crucial aspect is the quantization scheme and the braiding.
\end{itemize}
A related point concerns the renormalisation-group (RG) interpretation of this distinction. If noncommutativity is to play the role of a regulator, the relevant question is not only whether a given loop integral is finite at fixed deformation scale $\ell$, but also whether the induced effective theory remains inside the same quantum-group covariant theory space under coarse graining. In other words, covariance should not be imposed only on the bare action: the RG flow itself should preserve the braided/quantum-group structure of the theory, or else generate only counterterms compatible with it. From this perspective, a noncommutative model with finite loop integrals but a non-covariant quantisation prescription is not a satisfactory covariant regularisation, because the regularised theory has already left the symmetry class it was meant to define. Conversely, a covariant braided theory whose loop integrals still diverge may require renormalisation, but its counterterms are at least constrained by the deformed Poincaré symmetry. Thus noncommutativity is not, by itself, a regulator: it becomes one only when the covariant propagator, measure and braided perturbative expansion combine to produce a well-defined RG flow within the quantum-group invariant space of theories.

There are several examples of models that are finite, because $\int \Pi$ is finite and all loop divergences are regularised by the UV behaviour of the propagator, but the underlying quantization scheme breaks the quantum space(time) isometries~\cite{Grosse_2005a, Grosse_2003, Grosse_2005b, Gurau_2009} (\textit{e.g.} a scheme based on the $p$-Leibniz derivative), and as a consequence it exhibit non-covariant emerging effect such as the UV/IR mixing of \cite{Minwalla_2000}. This is problematic from several points of view: empirically, such models need to be reconciled with the stringent limits on Lorentz and Poincar\'e symmetry breaking~\cite{Addazi_2022}. From the perspective of model-building, a Lorentz/Poincaré breaking theory represents a significant headache, even assuming it is tuned to evade all experimental constraints on the relativity principle: Poincar\'e covariance represents a strong limitation to the space of theories that a model lives in, and from an effective field theory viewpoint, abandoning such space makes the theory immensely less predictive. Ultimately, one should ask what is the point of basing the whole construction on a deformed quantum group of spacetime isometries, just to break that symmetry at the level of the quantum theory?

Other models can preserve the relativity principle (although in the sense of a deformed Lorentz/Poincar\'e symmetry based on a quantum group), but fail to regularize the loop divergences due to $\int \Pi$. This leads to a theory that is not better behaved than commutative QFT. An example of such situation is provided by the $T$-Minkowski models considered in the next section, which, at least at the level of $n$-point functions, are indistinguishable from their commutative counterpart.

Finally, one can build theories that are both covariant and more regular than commutative QFT. Two examples of this is the effective $3$D quantum gravity model and the quantum 2-sphere (with the $x$-Leibniz prescription), described in the next section.

\section{Examples}
\label{sec:ex}
\paragraph{}
In this section, we discuss three applications of our formalism in order to illustrate the results of this paper. The first example consists in a class of Poincaré deformations that have been dubbed $T$-deformations. It has the advantage to cover various famous models, such as Moyal, $\rho$-Minkowski and lightlike $\kappa$-Minkowski, in one single formalism. The second example is the Ponzano-Regge model of $3$-dimensional quantum gravity coupled with a scalar field, that was showed to correspond to a $\kappa$-deformed field theory (with $\kappa = 1 / \ell$ the Planck energy) once the gravitational degrees of freedom were integrated over. This represents a very direct physical application, for which a quantisation scheme preserving the deformed Poincaré invariance has never been carried out to the best of the authors knowledge. Finally, we treat the example of the quantum $2$-sphere $S^2_q$, which consists in a non-triangular and non-unimodular deformation, unlike the two other models. This example was considered by Oeckl as a toy model for his braided QFT~\cite{Oeckl_2001}. We show that our formalism reproduces the \emph{structure} of his results -- the $SU_q(2)$-covariance of the correlators, the $6{+}6$ splitting of the one-loop $2$-point function, and the $q$-regularisation of the loop -- and that the only difference, a global $\ell$-dependent prefactor in the $n$-point functions, is fully accounted for by our symmetrised source coupling, as we explain at the end of Section~\ref{subsec:ex_SUq2}.

\subsection{\tops{$T$}{T}-Minkowski models}
\label{subsec:ex_TM}
\paragraph{}
The $T$-Minkowski models form a group of triangular deformations of Minkowski, covering $\theta$-Minkowski (also known as Moyal) \cite{Szabo_2003}, $\rho$-Minkowski \cite{Dimitrijevic_2018} or $\zeta$-Minkowski \cite{Lukierski_2006}, as introduced by the second author \cite{Mercati_2023, Mercati_2024}. This transversal formalism allows to treat different types of deformations. Through this example, we wish to show that our general formalism is also suited for all the $T$-Minkowski models. The triangularity requirement imposes that the braiding is involutive, that is $\Psi^2 = \id$, as one can deduce from expression \eqref{eq:mps_qt_braid} and the triangularity requirement \eqref{eq:Ha_Rmat_tri}. One should note that not all $T$-Minkowski models are unimodular.

\paragraph{}
The $T$-Poincaré algebra is constructed thanks to the FRT formalism \cite{Fadeev_1988}. It requires an $\qcRm$-matrix, or rather its matrix representation when expressed on a basis of the Poincaré group: the $T$'s. The deformation of the Poincaré group into the $T$-Poincaré quantum group is obtained thanks to the R$TT$-relations, the matrix representation form of the relation \eqref{eq:Ha_cRmat_dpdt}.

Then, the algebra of $T$-Minkowski\footnote{
    We stick to \cite{Mercati_2023, Mercati_2024} and consider the $3+1$-dimensional case, but other dimensions could have been considered in our discussion.
}
$C_\ell(\Real^{1,3})$ is defined as the algebra of functions over the noncommutative ring of coordinates $x^\mu$, which, thanks to the Poincaré-Birkhoff-Witt theorem, corresponds to the noncommutative ring of polynomials of $x$'s. The latter coordinates are obtained thanks to the so-called R$xx$-relations expressing the braided commutativity relation in the matrix representation (it corresponds to the $\qcRm$-matrix formulation of \eqref{eq:bc_def_nbra}).

Finally, the R$xy$-relations allow to construct the multi-particle algebra, since $y$ is here to be understood as the coordinate of a second particle, \ie $y = 1 \tpb x$ (corresponding to $x_2$ in our notations). Therefore, the R$xy$-relations correspond to the equation \eqref{eq:mps_qtpdt} for $f_1 \tpb f_2 = 1 \tpb x = y$, $g_1 \tpb g_2 = x \tpb 1 = x$ and expressed in the matrix representation.

The set of non-isomorphic triangular $\qRm$-matrices (and so $\qcRm$-matrices) of Poincaré has been classified by Zakrzewski \cite{Zakrzewski_1994, Zakrzewski_1997}, part of which were chosen to be the $T$-Minkowski models \cite{Mercati_2023}. The $T$-deformations were required to have a Lie algebra type\footnote{
    This Lie algebra might be centrally extended as $[x^\mu, x^\nu] = i \theta^{\mu\nu}(\ell) +  i c^{\mu\nu}{}_\rho(\ell) x^\rho$, but as noted in \cite{Hersent_2024a}, such a relation can still be made of Lie algebra type by introducing $1$ as a generator. Explicitly, let $x^{4} = 1$, then the Lie algebra writes
    \begin{align*}
        [x^\mu, x^\nu]
        &= i \theta^{\mu\nu}(\ell) x^{4} +  i c^{\mu\nu}{}_\rho(\ell) x^\rho, &
        [x^\mu, x^{4}] = 0.%
    \end{align*}%
}
coordinate algebra, that is  $[x^\mu, x^\nu] = i c^{\mu\nu}{}_\rho(\ell) x^\rho$, with $c \in \Cpx$, and to have commuting Lorentz transformations, \ie $[\Lambda, \Lambda] = 0$.

\paragraph{}
Being of Lie algebra type, the momentum space $G_\ell$ of $T$-Minkowski can be defined as the Lie group associated to this Lie algebra, a case already mentioned in section \ref{sec:Md} and more formally discussed in \cite{Mercati_2024, Hersent_2024a}.

The integral over $C_\ell(\Real^{1,3})$ is taken to be defined as the linear functional satisfying $\int \tdl{{}^4 x} e^{ipx} = \delta(p)$, for any $p \in G_\ell$. This relation is actually \eqref{eq:Md_mom_delta_def} so that this integral corresponds to the one considered in this discussion. Since $G_\ell$ is not unimodular, one can chose to work with the left or the right invariant Haar measure. In that respect, the author of \cite{Mercati_2024} defines left and right Fourier transforms. In this example, we keep on working with only the left invariant Haar measure, as mentioned in section \ref{sec:Md}, and so consider only the left Fourier transform \eqref{eq:Md_Ft}.

The $T$-Poincaré group transformations are considered explicitly in the discussion of the $T$-Minkowski models. Starting from the transformation of a single coordinate $x$, it is shown that such transformations are compatible with the algebra structure of $C_\ell(\Real^{1,3})$, with the multi-particle algebra $C_\ell(\Real^{1,3})^{\tpb n}$ and the explicit transformation of plane waves are computed. We do not focus on those here but retain only two essential consequences. First, thanks to the explicit transformation of plane waves, it is shown in \cite{Mercati_2024} that the integral is covariant. This has been given as an hypothesis of our discussion and can be shown to be equivalent to $\int \tdl{x} (\coactl e^{ipx}) = 1 \otimes \delta(p)$, a computation that can be carried out for $T$-Minkowski models (see equations (144) and (145) of \cite{Mercati_2024}). Secondly, the requirement \eqref{eq:mps_npt_cov_tran} that ensures the translation covariance of a $n$-point function is shown to hold in the case of $T$-Minkowski (see the equation (84) of \cite{Fabiano_2025}). Finally, the quantum Lorentz covariance is made explicit thanks to the introduction of the $\xi$ function below.

By introducing a differential calculus, two major ingredients of the $T$-Minkowski models appear. They relate the left and right module structure of the $1$-forms. Explicitly, one has
\begin{align}
    \td f 
    &= i \td x^\mu (\xi_\mu \actl f), &
    f \td x^\mu 
    &= \td x^\nu (\chi^\mu{}_\nu \actl f),
\end{align}
for any $f \in C_\ell(\Real^{1,3})$. The two functions $\xi$ and $\chi$ are momentum space functions, where $\xi$ has a translation-like behaviour and $\chi$ has a rotation-like (or group-like) behaviour:
\begin{align}
    \label{eq:ex_TM_xi_chi_coalg}
\begin{aligned}
    \Delta \xi_\mu 
    &= \xi_\mu \otimes 1 + \chi^\nu{}_\mu \otimes \xi_\nu, &
    S(\xi_\mu)
    &= - \xi_\nu (\chi^{-1})^\nu{}_\mu, &
    \varepsilon(\xi_\mu)
    &= 0, \\
    \Delta \chi^\mu{}_\nu 
    &= \chi^\mu{}_\rho \otimes \chi^\rho{}_\nu, &
    S(\chi^\mu{}_\nu)
    &= (\chi^{-1})^\mu{}_\nu, &
    \varepsilon(\chi^\mu{}_\nu)
    &= \delta^\mu_\nu.
\end{aligned}
\end{align}
These elements can be computed explicitly (see for example appendix A of \cite{Fabiano_2025}). Yet, it can be shown \cite{Mercati_2024} that the $\xi_\mu$ are the coordinates of momentum space for which the left Haar measure is the Lebesgue measure, \ie $\Haar{p} = \td^4\xi$. This result is essential because it shows that there exists a change of coordinate $p \mapsto \xi(p)$ for which the momentum space becomes undeformed. This is accompanied by the fact that $\xi$ transforms linearly under a (quantum) Lorentz transform so that $\td^4\xi$ is straightforwardly $T$-Poincaré covariant. However, it should be underlined that this change of coordinate is \textit{local} around the identity of $G_\ell$ and therefore might not cover all $G_\ell$. It was shown to be a global map for all unimodular $T$-Minkowski models \cite{Fabiano_2025}, but to cover only half of $G_\ell$ for the non-unimodular $\kappa$-light-like deformation.\footnote{
    Yet, it has also been shown that the $\xi$ map could be made global if the plane waves (and so the momentum space) were extended \cite{Fabiano_2024}.
}
In the following, we focus on the models that have globally defined $\xi$ functions which corresponds to the unimodlar models. Therefore, we set $\mathscr{I} = 1$.

\paragraph{}
By using the \textit{global} $\xi$ coordinates, the free action \eqref{eq:ncpi_free_action} writes
\begin{align}
    \label{eq:ex_TM_free_action}
    S_0(\phi)
    &= \frac{1}{2} \int \tdl{{}^4x} \phi(x) \Pi^{-1}[\phi](x)
    = \frac{1}{2} \int \tdl{{}^4\xi} \Pi(\xi) \phi(\xi) \phi(\dminus \xi).
\end{align}
One should note that $\dminus \xi = S(\xi)$ is not $-\xi$ in general--see~\eqref{eq:ex_TM_xi_chi_coalg}. Moreover, considering the reality condition $\phi(\dminus \xi) = \overline{\phi(\xi)}$ (the complex conjugate of $\phi(\xi)$) and the propagator
\begin{align} 
    \label{eq:ex_TM_comm_prop}
    \Pi^{-1}(\xi)
    = \xi^\mu \xi_\mu + m^2
    = (\dminus \xi)^\mu (\dminus \xi)_\mu + m^2,
\end{align}
the equation \eqref{eq:ex_TM_free_action} matches exactly the commutative free action of a complex scalar field.

The canonical quantisation of the free theory \eqref{eq:ex_TM_free_action} has already been developed in \cite{Fabiano_2025}. It was shown to have undeformed $n$-point functions, a result that we reproduce here thanks to our path integral formalism.

The first important property we need relates to the $2$-point function $W$. Indeed, since $\mathscr{I} = 1$, we have that $W_{\mathscr{I}^{-1}} = W$, defined in equations \eqref{eq:ncpi_uxl_f2pf}. The $2$-point function can be simplified by the following (equation (79) of \cite{Fabiano_2025})
\begin{align}
    \label{eq:ex_TM_exp_grouping}
    e^{ipx_a} e^{i(\dminus p)x_b}
    &= e^{i \xi(p) (x_a - x_b)}, &
    e^{i(\dminus p) x_a} e^{i p x_b}
    &= e^{i (\dminus \xi(p))(x_a-x_b)}.
\end{align}
These relations are obtained thanks to the introduction of the centre-of-mass coordinates $x_{\mathrm{cm}} = \frac{1}{n} \sum\limits_{a=1}^n x_a$, with our notation $x_a = 1 \tpb \cdots \tpb 1 \tpb x \tpb 1 \tpb \cdots$. Any coordinate can thus be split as $x_a = x_{\mathrm{cm}} + y_a$, where the $x_{\mathrm{cm}}$'s follow a similar Lie-algebra-type relation since the (single particle) $x$'s and the $y_a$'s commute. By reordering the exponentials one finds out that $e^{ipx_a} = e^{i \xi(p) y_a} e^{i p x_{\mathrm{cm}}}$ therefore leading to \eqref{eq:ex_TM_exp_grouping}. Note that the equation \eqref{eq:ex_TM_exp_grouping} implies that the requirement \eqref{eq:mps_ribbon_hypo} holds with $\nu = 1$.

As a result of \eqref{eq:ex_TM_exp_grouping}, the free $2$-point function \eqref{eq:ncpi_uxl_f2pf} now expresses as 
\begin{align}
    \label{eq:ex_TM_f2pf}
\begin{aligned}
    W(x_a,x_b)
    &= \int \Haar{p} \Pi(p) e^{ipx_a} e^{i(\dminus p)x_b}
    = \int \Haar{p} \Pi(p) e^{i \xi(p) (x_a - x_b)} \\
    &= \int \tdl{{}^4\xi} \Pi(\xi) e^{i \xi (x_a - x_b)}
\end{aligned}
\end{align}
for any $a < b$. This expression is fully undeformed (upon the requirement that $\Pi(\xi)$ actually fits the commutative propagator \eqref{eq:ex_TM_comm_prop}). Coupling this result with the expression of the free $4$-point function \eqref{eq:ncpi_uxl_4pf}, or the Wick theorem straightforwardly generalising it to any $n$-point function, we directly obtain that the free theory \eqref{eq:ex_TM_free_action} of the unimodular $T$-Minkowski models expresses as a sum of products of $W$ functions, and, upon requiering \eqref{eq:ex_TM_comm_prop}, is completely undeformed. One could also check that it is $T$-Poincaré invariant, whatever covariant $\Pi(\xi)$ we choose. We just established that the $n$-point functions are undeformed using the path integral formalism, but it has been put forward first for canonical quantisation in \cite{Fabiano_2025}.

Another important property of $T$-Minkowski deformations relies in the fact that $W$ commutes with any plane wave, that is $[e^{ipx_a}, W(x_b-x_c)] = 0$, for any $a,b,c$. The commutation of plane waves and $W$ is obtained via the property that
\begin{align}
    \label{eq:ex_TM_exp_swap}
    e^{ipx_a} e^{-i \xi (x_b - x_c)}
    &= e^{-i (\xi \cdot \chi(p)) (x_b - x_c)} e^{ipx_a},
\end{align}
where $(\xi \cdot \chi(p))_\mu = \xi_\nu \chi^\nu{}_\mu(p)$. Since \eqref{eq:ex_TM_exp_swap} is integrated over $\td^4\xi$ in $W$ and that $\chi$ consists in a rotation in $SO(1,3)$, one can change variable $\xi \to \xi \cdot \chi^{-1}(p)$. The measure $\td^4\xi$ being Poincaré invariant, if we make the hypothesis that $\Pi(\xi \cdot \chi) = \Pi(\xi)$ (invariance of the propagator under rotations), as satisfied by \eqref{eq:ex_TM_comm_prop}, then we just showed that $W$ commutes with plane waves.

\paragraph{}
One novelty of the present paper is the introduction of an interaction term of the form $\phi^4$. Indeed, by considering an interaction term of the form \eqref{eq:ncpi_phi4_action}, we end-up with a $2$-point function at one loop, computed with the $x$-Leibniz derivative, as \eqref{eq:ncpi_uxl_2pf1l}. As we consider unimodular $T$-Minkowski models having the property \eqref{eq:ex_TM_f2pf}, we obtain that
\begin{align}
    \label{eq:ex_TM_uxl_2pf1l}
    \langle \phi(x_1) \phi(x_2)\rangle_4^{\text{1-loop}}
    &= \frac{\lambda}{2} \int \tdl{{}^4x} \big( W(x_1,x) W(x_2,x) + W(x_2,x) W(x_1,x) \big) \left( \int \Pi \right),
\end{align}
where we noted $\int \Pi = \int \tdl{{}^4\xi} \Pi(\xi)$. Moreover, we can write
\begin{align*}
    \int \tdl{{}^4x} W(x_a,x) W(x_b,x)
    &\overset{\eqref{eq:ex_TM_f2pf}}{=} \int \tdl{{}^4x} \Haar{p} \Pi(p) e^{i(\dminus p)x_a} e^{ipx} W(x_b, x) \\
    &\overset{\eqref{eq:ex_TM_exp_swap}}{=}
    \int \tdl{{}^4x} \tdl{{}^4\xi} \Haar{p} \Pi(p) \Pi(\xi) e^{i(\dminus p)x_a} e^{-i\xi(x_b - x)} e^{ipx} \\
    &\overset{\eqref{eq:ex_TM_f2pf}}{=} \int \tdl{{}^4x} \Haar{p} \Haar{q} \Pi(p) \Pi(q) e^{i(\dminus p)x_a} e^{i(\dminus q)x_b} e^{iqx} e^{ipx} \\
    &\hspace{.2cm} =  \int \Haar{p} \Haar{q} \Pi(p) \Pi(q) e^{i(\dminus p)x_a} e^{i(\dminus q)x_b} \delta(q \dplus p) \\
    &\hspace{.2cm} = \int \Haar{p} \Pi(p)^2 e^{i(\dminus p)x_a} e^{i p x_b}
    = \int \tdl{{}^4\xi} \Pi(\xi)^2 e^{-i\xi(x_a - x_b)}
\end{align*}
This leads to
\begin{align}
    \tag{\ref{eq:ex_TM_uxl_2pf1l}}
    \dlangle \phi(x_1) \phi(x_2)\drangle_4^{\text{1-loop}}
    &= \lambda \int \tdl{{}^4\xi} \Pi(\xi)^2 e^{-i\xi(x_1 - x_2)} \left( \int \Pi \right),
\end{align}
Again, for the undeformed propagator \eqref{eq:ex_TM_comm_prop}, the $2$-point function at one-loop \eqref{eq:ex_TM_uxl_2pf1l} is exactly the undeformed one.

\paragraph{}
As a proof of concept, we also wish to write down the $2$-point function at one loop for the $p$-Leibniz rule here. It corresponds to \eqref{eq:ncpi_upl_2pf1l} which, in the unimodular $T$-Minkowski case, reads
\begin{align}
    \label{eq:ex_TM_upl_2pf1l}
    \begin{aligned}
    \dlangle \phi(x_1) \phi(x_2)\drangle_4^{\text{1-loop}}
    &= \frac{\lambda}{3} \int \tdl{{}^4\xi} \Pi(\xi)^2 e^{-i \xi (x_1-x_2)} \left( \int \Pi \right) \\
    &\ + \frac{\lambda}{6} \int \Haar{p_1} \Haar{p_2} e^{ip_1 x_1} e^{ip_2x_2} \Pi(p_1) \Pi(p_2) \int \Haar{q} \delta(\dminus q \dplus p_2 \dplus q \dplus p_1) \Pi(q).
    \end{aligned}
\end{align}
The first term is invariant, and even undeformed, as already shown by the ``planar equivalence theorem'' \cite{Meier_2023}. The second term however is not invariant under $T$-Poincaré.

\subsection{Effective 3D quantum gravity}
\label{subsec:ex_3qg}
\paragraph{}
We consider the Ponzano--Regge model, that is \emph{Euclidean} $3$-dimensional quantum gravity coupled to a scalar field. Once the gravitational degrees of freedom are integrated out, the matter sector becomes a noncommutative field theory whose residual quantum symmetry is the Euclidean Drinfel'd double $\mathfrak{dsu}(2) = D(SU(2))$, the $3$-dimensional analogue of $\kappa$-Poincar\'{e}, with deformation parameter the Planck length $\ell$ (equivalently the Planck energy $\kappa = 1/\ell$). Accordingly $\Poin_\ell$ is here $\mathfrak{dsu}(2)$, the ``translations'' curve into the group $SU(2)$, and all actions below are Euclidean. We refer to \cite{Freidel_2006} for the derivation and to \cite{Sasai_2007, Sasai_2009} for the field-theoretic setup.

\paragraph{}
We collect the few facts we need about $SU(2)$ and its Lie algebra $\mathfrak{su}(2)$. Both are spanned, over the reals, by the identity $\idu$ and the Pauli matrices $\sigma_j$ (with $\sigma_i \sigma_j = \delta_{ij}\idu + i\epsilon_{ijk}\sigma_k$, $\tr \sigma_j = 0$ and $\tr(\sigma_i \sigma_j) = 2\delta_{ij}$). We use this in two ways. The noncommutative coordinates are assembled into the \emph{Lie-algebra} element $x = x_j \sigma_j \in \mathfrak{su}(2)$, the $x_j$ being the spacetime coordinate functions; their algebra is of $\mathfrak{su}(2)$ type, $[x_i, x_j] = 2i\ell\,\epsilon_{ijk} x_k$. The momenta are assembled into the \emph{group} element
\begin{align}
    g = p_0\, \idu - i\ell\, p_j \sigma_j \in SU(2),
    && p_0 = \sqrt{1 - \ell^2 p_j p^j},
    \label{eq:ex_3dqg_group_param}
\end{align}
where unitarity ($\det g = p_0^2 + \ell^2 p_j p^j = 1$) fixes $p_0$ and confines $p$ to the ball $B(0,\ell^{-1})$ of \eqref{eq:ex_3dqg_int_norm}; the map $p \mapsto g$ thus charts (half of) $SU(2)$. Group inversion is $g^{-1} = p_0\,\idu + i\ell\, p_j \sigma_j$, \ie $p_j \mapsto -p_j$. The noncommutativity of spacetime is the image of the non-Abelian nature of $SU(2)$.

\paragraph{}
The plane waves are the pairing of a coordinate $x$ with a momentum $g$ through the trace,
\begin{align}
    e^{\frac{i}{2\ell} \tr(x g)} = e^{i p_j x^j},
    \label{eq:ex_3dqg_planewave}
\end{align}
the second equality following from $\tr(\sigma_i \sigma_j) = 2\delta_{ij}$ and $\tr \sigma_j = 0$. Products of plane waves compose the momenta by group multiplication, $e^{\frac{i}{2\ell} \tr(x g_1)} e^{\frac{i}{2\ell} \tr(x g_2)} = e^{\frac{i}{2\ell} \tr(x\, g_1 g_2)}$, so that the deformed momentum addition $p_1 \dplus p_2$ is read off from $g_1 g_2$, and momentum space carries the (left) Haar measure \eqref{eq:ex_3dqg_int_norm}; for a scalar field the momentum integration runs over $SO(3) = SU(2)/\mathbb{Z}_2$. One then define the momentum space as the Lie group $SO(3)$, and the Fourier transform as
\begin{align}
    \label{eq:ex_3dgq_Ft}
    \phi(x)
    &= \int_{SO(3)} \Haar{g} e^{\frac{i}{2 \ell} \tr(xg)} \phi(g), &
    \phi(g) &= \int_{\mathfrak{su}(2)} \tdl{x} e^{\frac{i}{2 \ell} \tr(x g^{-1})} \phi(x)
\end{align}
where $\Haar{g}$ is the Haar measure on the (unimodular) group $SO(3)$, $\ell = 4\pi G$ is the Einstein gravitational constant, with length dimension in 3d, and $\tr(xg)$ is the trace over Pauli matrices. 

Note that we have omitted the explicit mention of the (noncommutative) star-product in accordance with our notations. Since the product of two plane waves is given by
\begin{align}
    e^{\frac{i}{2 \ell} \tr(xg_1)} e^{\frac{i}{2 \ell} \tr(xg_2)}
    &= e^{\frac{i}{2 \ell} \tr(xg_1g_2)},
\end{align}
for any $g_1, g_2 \in SO(3)$, one can verify that the transforms \eqref{eq:ex_3dgq_Ft} are indeed inverse of one another. This check needs the definition of the delta function on $SO(3)$ and $\mathfrak{su}(2)$ respectively as
\begin{align*}
    \delta(g)
    &= \int_{\mathfrak{su}(2)} \tdl{x} e^{\frac{i}{2 \ell} \tr(xg)}, &
    \delta(x_1,x_2)
    &= \int_{SO(3)} \Haar{g} e^{\frac{i}{2 \ell} \tr(x_1g)} e^{\frac{i}{2 \ell} \tr(x_2g^{-1})}.
\end{align*}

\begin{comp}[ams align*]
    \int_{\mathfrak{su}(2)} \tdl{x} e^{\frac{i}{2 \ell} \tr(x g^{-1})} \phi(x)
    &= \int_{\mathfrak{su}(2)} \tdl{x} e^{\frac{i}{2 \ell} \tr(x g^{-1})} \left( \int_{SO(3)} \Haar{g'} e^{\frac{i}{2 \ell} \tr(xg')} \phi(g') \right) \\
    &= \int_{SO(3)} \Haar{g'} \left( \int_{\mathfrak{su}(2)} \tdl{x}  e^{\frac{i}{2 \ell} \tr(x g^{-1}g')} \right) \phi(g') \\
    &= \int_{SO(3)} \Haar{g'} \delta(g^{-1}g') \phi(g')
    = \phi(g), \\
    \int_{SO(3)} \Haar{g} e^{\frac{i}{2 \ell} \tr(xg)} \phi(g)
    &= \int_{SO(3)} \Haar{g} e^{\frac{i}{2 \ell} \tr(xg)} \left( \int_{\mathfrak{su}(2)} \tdl{y} e^{\frac{i}{2 \ell} \tr(y g^{-1})} \phi(y) \right) \\
    &= \int_{\mathfrak{su}(2)} \tdl{y} \left( \int_{SO(3)} \Haar{g} e^{\frac{i}{2 \ell} \tr(xg)} e^{\frac{i}{2 \ell} \tr(y g^{-1})} \right) \phi(y) \\
    &= \int_{\mathfrak{su}(2)} \tdl{y} \delta(x,y) \phi(y)
    = \phi(x). 
\end{comp}

\paragraph{}
The braiding used to define the multi-particle algebra, corresponds here to the Hopf braiding \eqref{eq:mps_glb} (see (98) of \cite{Freidel_2006}). Explicitly, we have for plane waves
\begin{align}
    \label{eq:ex_3dqg_glb}
    \Psi\big( e^{\frac{i}{2 \ell} \tr(xg_1)} \otimes e^{\frac{i}{2 \ell} \tr(xg_2)} \big)
    &= e^{\frac{i}{2 \ell} \tr(xg_2)} \tpb e^{\frac{i}{2 \ell} \tr(xg_2^{-1}g_1g_2)}.
\end{align}
It can be shown that this braiding \eqref{eq:ex_3dqg_glb} has actually a $\qcRm$-matrix form and so is also a coquasitriangular braiding \eqref{eq:mps_cqt_braid}. The product of exponentials can be deduced from \eqref{eq:ex_3dqg_glb}, via \eqref{eq:mps_braid_symm}, to be
\begin{align}
    e^{\frac{i}{2\ell}\tr(x_1g_1)} e^{\frac{i}{2\ell} \tr(x_2g_2)}
    &= e^{\frac{i}{2\ell}\tr(x_2(g_1g_2g_1^{-1}))} e^{\frac{i}{2\ell} \tr(x_1g_1)}, &
    e^{\frac{i}{2\ell}\tr(x_2g_2)} e^{\frac{i}{2\ell} \tr(x_1g_1)}
    &= e^{\frac{i}{2\ell}\tr(x_1g_1)} e^{\frac{i}{2\ell} \tr(x_2(g_1^{-1}g_2g_1))}.
\end{align}
This, together with unimodularity of $SO(3)$, implies that the hypothesis \eqref{eq:mps_ribbon_hypo} holds with $\nu = 1$.

\paragraph{}
The group of symmetry and its action on $\Malg_\ell$ has been studied in \cite{Sasai_2007}. It consists in the Drinfeld double of $SU(2)$ and we denote its quantum algebra as $\mathfrak{dsu}(2)$. It corresponds to the $\ell$-deformation of Poincar\'{e} in $2+1$-dimensions \cite{Freidel_2005}. One starts by decomposing $g \in SO(3)$ as $g = p_0 - i \ell p_j \sigma_j$, where the $\sigma_j$'s are the Pauli matrices, and $p_0 = \pm \sqrt{1 - \ell^2p^jp_j}$ by definition. Note that in these notations $g^{-1} = p_0 + i \ell p_j \sigma_j$. One then defines the action of the translations $P \in \mathfrak{dsu}(2)$ and the rotations $M \in \mathfrak{dsu}(2)$ as (see (74) and (83) of \cite{Sasai_2007})
\begin{align}
    \label{eq:ex_3dqg_PM_act}
    P_\mu \actl e^{\frac{i}{2\ell} \tr(xg)}
    &= P_\mu(g) \, e^{\frac{i}{2\ell} \tr(xg)}
    = p_\mu \, e^{\frac{i}{2\ell} \tr(xg)}, &
    M_h \actl e^{\frac{i}{2\ell} \tr(xg)}
    &= e^{\frac{i}{2\ell} \tr(x h^{-1} g h)},
\end{align}
for $h \in SO(3)$. The action of $M$ is the adjoint action of the Lie group. For simplicity, these actions can be expressed in momentum space as
\begin{align}
    \label{eq:ex_3dqg_MP_act_mom}
    P_\mu \actl \phi(g)
    &= P_\mu(g) \phi(g), &
    M_h \actl \phi(g)
    &= \phi(h g h^{-1}).
\end{align}
The Hopf algebra relations of the $P$'s and $M$'s can be deduced from the action \eqref{eq:ex_3dqg_PM_act} to be
\begin{align}
    \label{eq:ex_dsu2_algebra}
\begin{aligned}[]
    [P_\mu, P_\nu]
    &= 0, &
    \Delta(P_0)
    &= P_0 \otimes P_0 - \ell^2 P^j \otimes P_j, &
    S(P_0)
    &= P_0, \\
    \varepsilon(P_\mu)
    &= 0, &
    \Delta(P_j)
    &= P_0 \otimes P_j + P_j \otimes P_0 + i \ell \epsilon^{jkl} P_k \otimes P_l, &
    S(P_j)
    &= - P_j, \\
    M_{h_1} M_{h_2}
    &=  M_{h_2 h_1} , &
    \Delta(M_h)
    &= M_h \otimes M_h, \\
    \varepsilon(M_h) &= 1, &
    S(M_h)
    &= M_{h^{-1}}, \\
    [P_0, M_h]
    &= 0, &
    [P_j, M_h]
    &= 2 i \ell \epsilon^{jkl}P_0(h) P_l(h) M_h P_k,
\end{aligned}
\end{align}
where $\epsilon$ is the fully antisymmetric tensor.

\paragraph{}
We want to show that the Hopf braiding \eqref{eq:ex_3dqg_glb} is covariant with respect to the action \eqref{eq:ex_3dqg_MP_act_mom}, which would imply that the multi-particle algebra based on this braiding is a module algebra. 

The analysis of section \ref{subsubapx:mps_glb} showed that the Hopf braiding does not commute with the action in the general case. However, this $3$-dimensional case can be shown to be covariant. Indeed, the covariance of the braiding with respect to the action has already been shown in \cite{Sasai_2007}, and we repeat the computations here. First, let us note that the braiding can be expressed in the momentum space also through
\begin{align}
    \Psi\big( \phi(g_1) \otimes \tilde{\phi}(g_2) \big)
    &= \tilde{\phi}(g_2) \tpb \phi(g_2^{-1}g_1g_2).
\end{align}
There remains to show that this braiding commutes with the action, as in \eqref{eq:mps_braid_cov}. It reads
\begin{align*}
    P_\mu \actl \big( \Psi\big( \phi(g_1) \otimes \tilde{\phi}(g_2) \big) \big)
    &= P_\mu \actl \big( \tilde{\phi}(g_2) \tpb \phi(g_2^{-1}g_1g_2) \big)
    = \Delta P_\mu(g_2 \otimes g_2^{-1} g_1 g_2) \big( \tilde{\phi}(g_2) \tpb \phi(g_2^{-1}g_1g_2) \big) \\
    &= P_\mu(g_2 g_2^{-1} g_1 g_2) \big( \tilde{\phi}(g_2) \tpb \phi(g_2^{-1}g_1g_2) \big)
    = P_\mu (g_1g_2) \big( \tilde{\phi}(g_2) \tpb \phi(g_2^{-1}g_1g_2) \big) \\
    &= \Psi \big( P_\mu \actl \big( \phi(g_1) \otimes \tilde{\phi}(g_2) \big) \big) \\
    M_h \actl \big( \Psi\big( \phi(g_1) \tpb \tilde{\phi}(g_2) \big) \big)
    &= M_h \actl \big( \tilde{\phi}(g_2) \tpb \phi(g_2^{-1}g_1g_2) \big)
    = \Delta M_h \actl \big( \tilde{\phi}(g_2) \tpb \phi(g_2^{-1}g_1g_2) \big) \\
    &=  \tilde{\phi}(h g_2 h^{-1}) \tpb \phi(h g_2^{-1} g_1 g_2 h^{-1})
    = \Psi \big( \phi(h g_1 h^{-1}) \otimes \tilde{\phi}(hg_2h^{-1}) \big) \\
    &= \Psi \big( M_h \actl \big( \phi(g_1) \otimes \tilde{\phi(g_2)} \big) \big).
\end{align*}
where we used \eqref{eq:mps_gact}.

\paragraph{}
We now turn to the quantisation of its field theory, in the $x$-Leibniz formalism. In the  Ponzano-Regge model, the action \eqref{eq:ncpi_phi4_action} reads
\begin{align}
    \label{eq:ex_3dqg_action}
\begin{aligned}[]
    S(\phi)
    &= S_0(\phi) + S_{\text{int}}(\phi)
    = \frac{1}{8\pi \ell^3} \int_{\mathfrak{su}(2)} \tdl{x} \left( \frac{1}{2} (\partial_j \phi \partial^j \phi)(x) + \frac{1}{2} m_\ell^2 \phi^2(x) + \frac{\lambda}{4!} \phi^4(x)\right) \\
    &= \frac{1}{8\pi\ell^3} \int_{SO(3)} \Haar{g} \Pi(g)^{-1} \phi(g) \phi(g^{-1})
    + \frac{\lambda}{4!} \int \Haar{g_1} \Haar{g_2} \Haar{g_3} \Haar{g_4} \delta(g_1g_2g_3g_4) \phi(g_1) \phi(g_2) \phi(g_3) \phi(g_4),
\end{aligned}
\end{align}
where $\Pi(g)^{-1} = P^2(g) + m_\ell^2 = p_j p^j + m_\ell^2$, with $m_\ell^2 = \frac{\sin^2(m \ell)}{\ell^2}$. We use the normalisation corresponding to 
\begin{align} 
    \label{eq:ex_3dqg_int_norm}
    \int \Haar{g}
    &= \frac{\ell^3}{\pi^2} \int_{B(0,\ell^{-1})} \frac{\td^3 p}{\sqrt{1 - \ell^2p^jp_j}},
\end{align}
where $B(0,\ell^{-1})$ is the ball of radius $\ell^{-1}$. Before going into the path integral quantisation of this model, one needs to ensure that the condition \eqref{eq:ncpi_prop_requirement} is fulfilled. In this example, it writes $\Pi(g^{-1}) = \Pi(g)$ for any $g$. Since $m_\ell$ is independent of the momenta, one simply needs to show that $P^2(g^{-1}) = P^2(g)$. Indeed, since for $g = p_0 - i\ell p_j\sigma_j$ one has $g^{-1} = p_0 + i\ell p_j\sigma_j$, the group inversion $g \mapsto g^{-1}$ amounts to $p_j \mapsto -p_j$. Therefore, $P^2(g) = p_j p^j \mapsto P^2(g^{-1}) = p_jp^j$.

\paragraph{}
The free $2$-point function is directly computed from
\begin{align*}
    \dlangle \phi(x_1) \phi(x_2) \drangle_0
    &= \left. \xLd{J(x_2)}\Pi[J(x_1)]  \right|_{J=0}\\
    &=\xLd{J(x_2)} \left( \int_{SO(3)} \Haar{g} (P^2(g) - m_\ell^2)^{-1} e^{\frac{i}{2\ell} \tr(x_1g)} \int_{\mathfrak{su}(2)} \tdl{x_3} e^{\frac{i}{2\ell}\tr(x_3g^{-1})} J(x_3) \right)\\
    &= \int_{SO(3)} \Haar{g} \Pi(g) e^{\frac{i}{2\ell} \tr(x_1g)} \int_{\mathfrak{su}(2)} \tdl{x_3} e^{\frac{i}{2\ell}\tr(x_3g^{-1})} \delta(x_3, x_2) \\
    &= \int_{SO(3)} \Haar{g} \Pi(g) e^{\frac{i}{2\ell} \tr(x_1g)} e^{\frac{i}{2\ell} \tr(x_2g^{-1})} \\
    &= W(x_1, x_2).
\end{align*}
Note that thanks to the group-like form of the braiding one has
\begin{align} 
    \label{eq:ex_3dqg_f2pf_symm}
    W(x_1,x_2)
    = W(x_2,x_1),
\end{align}
which is not a general feature.

We deduce that the free $4$-point function writes
\begin{align*}
    \dlangle & \phi(x_1) \cdots \phi(x_4) \drangle_0
    = W(x_1,x_2) W(x_3,x_4) + W(x_1,x_3)W(x_2,x_4) + W(x_1,x_4)W(x_2,x_3) \\
    &= \int \Haar{g_1} \Haar{g_2} \Pi(g_1) \Pi(g_2) \left( e^{\frac{i}{2\ell}\tr(x_1g_1)} e^{\frac{i}{2\ell}\tr(x_2g_1^{-1})} e^{\frac{i}{2\ell}\tr(x_3g_2)} e^{\frac{i}{2\ell}\tr(x_4g_2^{-1})} \right.\\
    &\phantom{=} \left. + e^{\frac{i}{2\ell}\tr(x_1g_1)} e^{\frac{i}{2\ell}\tr(x_3g_1^{-1})} e^{\frac{i}{2\ell}\tr(x_2g_2)} e^{\frac{i}{2\ell}\tr(x_4g_2^{-1})}
    + e^{\frac{i}{2\ell}\tr(x_1g_1)} e^{\frac{i}{2\ell}\tr(x_4g_1^{-1})} e^{\frac{i}{2\ell}\tr(x_2g_2)} e^{\frac{i}{2\ell}\tr(x_3g_2^{-1})}\right) \\
    &= \int \Haar{g_1} \Haar{g_2} \Pi(g_1) \Pi(g_2) \left( e^{\frac{i}{2\ell}\tr(x_1g_1)} e^{\frac{i}{2\ell}\tr(x_2g_1^{-1})} e^{\frac{i}{2\ell}\tr(x_3g_2)} e^{\frac{i}{2\ell}\tr(x_4g_2^{-1})} \right.\\
    &\phantom{=} \left. + e^{\frac{i}{2\ell}\tr(x_1g_1)} e^{\frac{i}{2\ell}\tr(x_2g_2)} e^{\frac{i}{2\ell}\tr(x_3g_2^{-1}g_1^{-1}g_2)} e^{\frac{i}{2\ell}\tr(x_4g_2^{-1})}
    + e^{\frac{i}{2\ell}\tr(x_1g_1)} e^{\frac{i}{2\ell}\tr(x_2g_2)} e^{\frac{i}{2\ell}\tr(x_3g_2^{-1})} e^{\frac{i}{2\ell}\tr(x_4g_1^{-1})}\right).
\end{align*}
This expression is $\mathfrak{dsu}(2)$-invariant since it can be written under the form \eqref{eq:mps_cov_npf}. Explicitly, here \eqref{eq:mps_npt_cov_tran} translates as follows: any product of exponentials with momenta $g_1, \ldots, g_n$ is invariant if $g_1\cdots g_n = 1$. One can check that all three terms of the free $4$-point function satisfy this requirement:
\begin{align*}
    \text{Term 1: }
    g_1 g_1^{-1} g_2 g_2^{-1} = 1, &&
    \text{Term 2: }
    g_1 g_2 g_2^{-1} g_1^{-1} g_2 g_2^{-1} = 1, &&
    \text{Term 3: }
    g_1 g_2 g_2^{-1} g_1^{-1} = 1.
\end{align*}

\paragraph{}
The $2$-point function of the interacting theory at one-loop is then
\begin{align*}
    \dlangle & \phi(x_1) \phi(x_2)\drangle_4^{\text{1-loop}}
    = \frac{\lambda}{4!} \dlangle[\left] \left(\int \tdl{x} \phi^4(x) \right) \phi(x_1) \phi(x_2) \drangle[\right]_0 \\
    &= \frac{\lambda}{4} \int \tdl{x_3} W(x_3,x_3) \bigl( W(x_1,x_3) W(x_2,x_3) + W(x_2,x_3) W(x_1,x_3) \bigr)
\end{align*}

One notices that, like in \eqref{eq:ncpi_loop_prop}, a simplification occurs:
\begin{align}
    \label{eq:ex_3dqg_Wxx}
\begin{aligned}
    W(x_a, x_a)
    &= \int \Haar{g} \Pi(g) e^{\frac{i}{2\ell}\tr(x_ag)} e^{\frac{i}{2\ell}\tr(x_ag^{-1})}
    = \int \Haar{g} \Pi(g) \\
    &= \frac{4 \ell^3}{\pi} \int_0^{\ell^{-1}} \frac{\td p}{\sqrt{1 - \ell^2 p^2}} \frac{p^2}{p^2 + m_{\ell}^2} < \infty
\end{aligned}
\end{align}
and $W(x_a,x_a)$ is the integral of the propagator for any $a$. Note that $W(x_a,x_a)$ therefore commutes with everything, as it is a scalar. Moreover, the integral of the propagator is here finite.\footnote{
    Indeed, the factor $\frac{p^2}{p^2 - m_{\ell}^2}$ is finite for $p$ close to $\ell^{-1}$ and $\frac{1}{\sqrt{1 - \ell^2p^2}} \sim \frac{1}{\sqrt{2}} \frac{1}{\sqrt{1 - \ell p}}$ which has a finite integral near $\ell^{-1}$ because the square root function is integrable near $0$.
}
Furthermore, we compute
\begin{align*}
    \int \tdl{x_b} W(x_a, x_b) W(x_b, x_c)
    &= \int \tdl{x_b} \Haar{g_1} \Haar{g_2} \Pi(g_1) \Pi(g_2) e^{\frac{i}{2\ell}\tr(x_ag_1)} e^{\frac{i}{2\ell}\tr(x_bg_1^{-1})} e^{\frac{i}{2\ell}\tr(x_bg_2)} e^{\frac{i}{2\ell}\tr(x_cg_2^{-1})} \\
    &= \int \Haar{g_1} \Haar{g_2} \Pi(g_1) \Pi(g_2) \delta(g_1^{-1} g_2) e^{\frac{i}{2\ell}\tr(x_ag_1)} e^{\frac{i}{2\ell}\tr(x_cg_2^{-1})} \\
    &= \int \Haar{g} \Pi(g)^2 e^{\frac{i}{2\ell}\tr(x_ag)} e^{\frac{i}{2\ell}\tr(x_cg^{-1})}.
\end{align*}
One can then use the symmetry property \eqref{eq:ex_3dqg_f2pf_symm} and show that both terms of the $2$-point function at one-loop are actually equal. To summarise,
\begin{align}
    \label{eq:ex_3dqg_2pf1l}
    \dlangle & \phi(x_1) \phi(x_2)\drangle_4^{\text{1-loop}}
    = \frac{\lambda}{2} \left( \int \Pi \right)  \int \Haar{g} \Pi(g)^2 e^{\frac{i}{2\ell}\tr(x_1g)} e^{\frac{i}{2\ell}\tr(x_2 g^{-1})},
\end{align}
where $\int \Pi$ denotes the loop integral \eqref{eq:ex_3dqg_Wxx}. It is important to note that the $2$-point function at $1$-loop is invariant since it takes the form \eqref{eq:mps_inv_2pf}. 

\paragraph{}
The equation \eqref{eq:ex_3dqg_2pf1l} should be compared with its $p$-Leibniz counterpart (obtained from \eqref{eq:ncpi_upl_2pf1l}):
\begin{align}
    \label{eq:ex_3dqg_2pf1l_upl}
\begin{aligned}[]
    \langle \phi(x_1) \phi(x_2) \rangle_4^{\text{1-loop}}
    &= \frac{\lambda}{3} \left( \left( \int \Pi \right) \int \Haar{g} \Pi(g)^2 e^{\frac{i}{2\ell}\tr(x_1g)} e^{\frac{i}{2\ell}\tr(x_2 g^{-1})} \right) \\
    &\phantom{=} + \frac{\lambda}{6} \int \Haar{g_1} \Haar{g_2} \Pi(g_1) \Pi(g_2) e^{\frac{i}{2\ell}\tr(x_1g_1)} e^{\frac{i}{2\ell}\tr(x_2 g_2)} \int \Haar{h} \Pi(h) \delta(h^{-1} g_2 h g_1),
\end{aligned}
\end{align}
where the first term corresponds to planar diagrams and the second to the non-planar ones. The planar diagrams are $\mathfrak{dsu}(2)$-invariant because they have the form \eqref{eq:mps_inv_2pf}. However, non-planar diagrams are explicitly breaking the $\mathfrak{dsu}(2)$ symmetry since already the translation invariance, which requires $g_1g_2=1$, is broken by the delta function imposing $h^{-1} g_2 h g_1 = 1$ for all $h$.

\paragraph{}
Note that the $\phi^3$-theory has already been studied in \cite{Sasai_2009}, in which they compute both the $p$-Leibniz and the braided QFT results for the $2$-point function at one-loop order. As expected from section \ref{subsec:ncpi_upl}, the $p$-Leibniz $2$-point function in \cite{Sasai_2009} is found to split into two contributions: the planar and the non-planar diagrams. The planar diagrams have an overhaul $\delta$ function that ensures covariance, but the non-planar ones have a mixed $\delta$ of the form $\delta(g_1 h g_2 h^{-1})$, where $g_1$, $g_2$ are the external momenta and $h$ is the loop momenta (see equation (3.18) of \cite{Sasai_2009}). This is exactly the structure we found for the general case in section \ref{subsec:ncpi_upl}, and it can be shown similarly that such non-planar diagrams are not $\mathfrak{dsu}(2)$-invariant.

On the other hand, the braided QFT give diagrams that are all identical, similarly as in \eqref{eq:ex_3dqg_2pf1l}, and that correspond to the planar diagrams. As one can see by comparing \eqref{eq:ex_3dqg_2pf1l} and \eqref{eq:ex_3dqg_2pf1l_upl}, this is also the case for the $\phi^4$-theory. Since the planar diagrams are covariants, it means that the braided QFT result is also covariant \cite{Sasai_2009}.

\subsection{Scalar field theory on the quantum two-sphere}
\label{subsec:ex_SUq2}
\paragraph{}
In this section, we study scalar field theory on the quantum $2$-sphere $S^2_q$. In the same way as the $2$-sphere can be defined as the quotient of $SU(2)$ by $U(1)$, the quantum $2$-sphere $S^2_q$ is built as the quantum homogeneous space of $SU_q(2)$ which is invariant under $U(1)$. $S^2_q$ has a vast literature and we refer the reader to \cite{Klimyk_1997, Majid_1995, Oeckl_2001} for more details on this space. Note that, in accordance with the former references, we consider $q \in \Real$ and $q \neq 0$ in this section.

The Hopf algebra of $SU_q(2)$ expressed with Wigner matrix coefficients $t$ reads\begin{subequations}
    \label{eq:ex_SUq2_Ha}
\begin{align}
    \Delta \big(t^{(\ell)}_{mn}\big) = \sum_{k=-\ell}^{\ell} t^{(\ell)}_{mk} \otimes t^{(\ell)}_{kn}, \qquad
    \epsilon\!\big(t^{(\ell)}_{mn}\big) = \delta_{mn}, &&
    S\big(t^{(\ell)}_{mn}\big)
    &= (-q)^{m-n}\,t^{(\ell)}_{-n\,-m}, \\
    \sum_{k=-\ell}^{\ell} \big(t^{(\ell)}_{km}\big)^{*} t^{(\ell)}_{k\,n}
    = \delta_{mn}
    = \sum_{k=-\ell}^{\ell} t^{(\ell)}_{mk} \big(t^{(\ell)}_{nk}\big)^{*} &&
    \big(t^{(\ell)}_{mn}\big)^{*}
    &= (-q)^{n-m} t^{(\ell)}_{-m-n},
\end{align}
\end{subequations}
where $\ell \in \mathbb{N}$ corresponds to the orbital quantum number and $m,n = -\ell, \ldots, \ell$ to the magnetic quantum numbers.

\paragraph{}
When dealing with the quantum $2$-sphere $S^2_q$, the basis of functions used to express ``momentum space'' are no more plane waves, but corresponds to deformed spherical harmonics. Indeed, the so-called $q$-spherical harmonics $Y^m_\ell$ are defined thanks to the matrix coefficients $t^{(\ell)}_{mn} \in SU_q(2)$ through\footnote{
    $q$-numbers are defined as $[n]_q = \frac{q^n - q^{-n}}{q - q^{-1}}$, for $n \in \Int$.
} 
\begin{align}
    Y^m_\ell
    &= \sqrt{[2\ell+1]_q} (-q)^{-m} t_{-m0}^{(\ell)}
    = \sqrt{[2\ell+1]_q} \big( t_{m0}^{(\ell)} \big)^*.
\end{align}
The right coaction of $SU_q(2)$ on the $q$–spherical harmonics is obtained from the coproduct of the matrix coefficients \eqref{eq:ex_SUq2_Ha} and reads
\begin{equation}
    \label{eq:ex_SUq2_coact}
    \coactl Y^m_\ell 
    = \sum_{n=-\ell}^{\ell} S(t^{(\ell)}_{nm}) \otimes Y^n_\ell.
\end{equation}
Explicitly,
\begin{align*}
    \coactl Y^m_\ell
    &= \sqrt{[2\ell+1]_q} (-q)^{-m} \Delta t_{-m0}^{(\ell)}
    = \sqrt{[2\ell+1]_q} (-q)^{-m} \sum_{n=-\ell}^\ell t_{-mn}^{(\ell)} \otimes t^{(\ell)}_{n0} \\
    &= \sqrt{[2\ell+1]_q} \sum_{n=-\ell}^\ell S(t_{-nm}^{(\ell)}) \otimes (-q)^n t^{(\ell)}_{n0}
    = \sum_{n=-\ell}^\ell S(t^{(\ell)}_{nm}) \otimes Y^n_\ell.
\end{align*}
The analogue of the Fourier transform \eqref{eq:Md_Ft} writes in this context
\begin{align}
    \label{eq:ex_SUq2_Ft}
    J
    &= \sum_{\ell=0}^\infty \sum_{m=-\ell}^\ell J_{\ell m} Y^m_\ell, &
    J_{\ell m}
    &= \int_{S^2_q} (Y^m_\ell)^* J,
\end{align}
where $J \in S^2_q$ and $J_{\ell m} \in \Cpx$. Note that the ``integral'' of momentum space is replaced by a sum, since the $q$-harmonics are discrete. Moreover, we have introduced the integral over $S^2_q$, which actually corresponds to the invariant integral of $SU_q(2)$ (see for example section 4.3.2 of \cite{Klimyk_1997}). In order to prove that the Fourier transforms \eqref{eq:ex_SUq2_Ft} are inverse of one another, one needs the orthogonality relations of the $q$-spherical harmonics given by
\begin{subequations}
    \label{eq:ex_SUq2_ortho_sh}
\begin{align}
    \int_{S^2_q} Y^{m'}_{\ell'} (Y^m_\ell)^*
    &= q^{-2m} \delta_{\ell, \ell'} \delta_{m, m'}, &
    \int_{S^2_q} (Y^m_\ell)^* Y^{m'}_{\ell'}
    &= \delta_{\ell, \ell'} \delta_{m, m'},
\end{align}
or equivalently, thanks to $(Y^m_\ell)^* = (-q)^{-m} Y^{-m}_\ell$ ($q \in \Real$),
\begin{align}
    \int_{S^2_q} Y^m_\ell Y^{m'}_{\ell'}
    &= (-q)^{-m} \delta_{\ell, \ell'} \delta_{m, -m'}
\end{align}
\end{subequations}
From the previous relations, it is straightforward that $S^2_q$ is not unimodular. One can compute the modular function, using \eqref{eq:ex_SUq2_ortho_sh}, to be\footnote{
    This expression for the modular function is in agreement with the general theory of Haar integral on ribbon Hopf algebras. Indeed, in appendix \ref{subsubapx:SUq2_ribbon}, it is computed that the pivotal element $g$ of $U_q(\mathfrak{su}(2))$ is given by $g = q^{H}$. Recall that the pivotal element is the group-like element such that $S^2 = \mathrm{ad}_g$. Then, the ribbon element satisfies $\nu = g^{-1} \mathfrak{u}$ (that is our case here) if and only if the modular function matches the pivotal element $\mathscr{I}^2 = g^{-2} = q^{-2H}$ (see Lemma 2.18 of \cite{Faes_2025}), and so $\mathscr{I} \actl Y^m_\ell = q^{-H} \actl Y^m_\ell = q^{-2m} Y^m_\ell$.
}
$\mathscr{I} \actl Y^{m}_\ell = q^{-2m} Y^m_\ell$. Note that $\mathscr{I}$ is well-defined and positive since we consider here $q \in \Real$ (see Prop 3.2.6 of \cite{Majid_1995}) and $q \neq 0$. For a general $q\in \mathbb{C}^*$, one can actually show that $\mathscr{I}(q) = |q|^{2m}$, which is still well-defined. 

Given the orthogonality relation \eqref{eq:ex_SUq2_ortho_sh}, one can verify that the Fourier transform \eqref{eq:ex_SUq2_Ft} are inverse of one another. Explicitly,
\begin{align*}
    \int_{S^2_q} (Y^m_\ell)^* J
    &= \int_{S^2_q}  (Y^m_\ell)^* \left( \sum_{\ell',m'} J_{\ell'm'} Y^{m'}_{\ell'} \right)
    = \sum_{\ell',m'} J_{\ell'm'} \int_{S^2_q} (Y^m_\ell)^* Y^{m'}_{\ell'} \\
    &= \sum_{\ell',m'} J_{\ell'm'} \delta_{\ell, \ell'} \delta_{m, m'}
    = J_{\ell m}.
\end{align*}
Performing the transforms in the other order gives
\begin{align*}
    J
    &= \sum_{\ell,m} Y^m_\ell  J_{\ell m}
    = \sum_{\ell,m} (Y^{(1)})^m_\ell \int_{S^2_q}^{(2)} [(Y^{(2)})^m_\ell]^* J^{(2)}
    = \int_{S^2_q}^{(2)} \delta^{(1, 2)} J^{(2)}
    = J,
\end{align*}
where we have used the notation $\sum_{\ell,m} = \sum_{\ell=0}^\infty \sum_{m=-\ell}^\ell$. This notation is used throughout all the section \ref{subsec:ex_SUq2}. Most importantly, we have just defined the $q$-deformed delta function $\delta^{(1,2)} = \sum_{\ell m} (Y^{(1)})^m_\ell [(Y^{(2)})^m_\ell]^*$, in a similar fashion then \eqref{eq:mps_xdelta}. We also used its properties, namely
\begin{align}
    \label{eq:q_delta_prop}
    \int^{(1)}_{S^2_q} f^{(1)} \delta^{(1,2)}
    &= f^{(2)}, &
    \int^{(2)}_{S^2_q} \delta^{(1,2)} f^{(2)}
    &= f^{(1)},
\end{align}
with the notation $\int^{(a)} = \mathrm{id} \otimes \cdots \otimes \mathrm{id} \otimes \int{} \otimes \mathrm{id} \otimes \cdots$, which are analogues of \eqref{eq:mps_xdelta_prop}.

\paragraph{}
We need to compute the braiding that is
\begin{align}
    \Psi(f \otimes g)
    &= \qcRm(g^{(0)} \otimes f^{(0)}) g^{(1)} \tpb f^{(1)},
    \label{eq:mps_cqt_braid}
\end{align}
where $\qcRm$ is the $\qcRm$-matrix of $SU_q(2)$ and we have used Sweedler's notations for the coaction $\coactl f = f^{(0)} \otimes f^{(1)}$. We refer to appendix \ref{subapx:Ha_cqts} for the definition of coquasitriangular structure. It should be underlined that coquasitriangular structure work similarly as the quasitriangular case, detailed in appendix \ref{subsubapx:mps_qtb}. We refer to the section 9.2 of \cite{Majid_1995} for more details on this point.

Since the expression of the full $\qcRm$-matrix seems to be too complex as such, we rather compute useful formulas that we may recover in computations. We first start with the $\qRm$-matrix of $U_q(\mathfrak{su}(2))$ (see Proposition 2.1.8 of \cite{Majid_1995} for an explicit expression). By defining, $\mathfrak{u} = S(\qRm_{2}) \qRm_{1}$ and $\mathfrak{v} = \qRm_{1} S(\qRm_{2})$, one can compute that their action on the $q$-spherical harmonics are diagonal and writes (see proof in appendix \ref{subsubapx:SUq2_ribbon}):
\begin{align}
\begin{aligned}[]
    \mathfrak{u} \actl Y^m_\ell
    &= q^{-2\ell(\ell+1)} q^{2m} Y^m_\ell, & &&
    \mathfrak{v} \actl Y^m_\ell
    &= q^{-2\ell(\ell+1)} q^{-2m} Y^m_\ell, \\
    \mathfrak{u}^{-1} \actl Y^m_\ell
    &= q^{2\ell(\ell+1)} q^{-2m} Y^m_\ell, & &&
    \mathfrak{v}^{-1} \actl Y^m_\ell
    &= q^{2\ell(\ell+1)} q^{2m} Y^m_\ell,
\end{aligned}
\end{align}
Define now their dual counterparts, $u(a) = \qcRm(a_{(2)} \otimes S(a_{(1)}))$ and $v(a) = \qcRm(a_{(1)} \otimes S(a_{(2)}))$, for any $a \in U_q(\mathfrak{su}(2))$ (see Proposition 2.2.4 of \cite{Majid_1995}). By duality of the action $\actl$ and the coaction $\coactl$, one has the relation
\begin{align}
    \mathfrak{w} \actl Y^m_\ell
    &= \sum_{n=-\ell}^\ell \langle \mathfrak{w}, t_{nm}^{(\ell)} \rangle Y^n_\ell
    = \sum_{n = -\ell}^\ell w(t_{nm}^{(\ell)}) Y^n_\ell, &
    \text{for any } \mathfrak{w} \in U_q(\mathfrak{su}(2)),
\end{align}
where $\langle \cdot, \cdot \rangle : U_q(\mathfrak{su}(2)) \otimes SU_q(2) \to \Cpx$ denotes the dual coupling and $w = \langle \mathfrak{w}, \cdot \rangle \in SU_q(2)$ is the dual element of $\mathfrak{w}$. By duality between $\mathfrak{u}$, $\mathfrak{v}$ and $u$, $v$ respectively, one deduces that
\begin{align}
\begin{aligned}[]
    u(t_{nm}^{(\ell)})
    &= \delta_{nm} q^{-2\ell(\ell+1)} q^{2m}, & &&
    v(t_{nm}^{(\ell)})
    &= \delta_{nm} q^{-2\ell(\ell+1)} q^{-2m}, \\
    u^{-1}(t_{nm}^{(\ell)})
    &= \delta_{nm} q^{2\ell(\ell+1)} q^{-2m}, & &&
    v^{-1}(t_{nm}^{(\ell)})
    &= \delta_{nm} q^{2\ell(\ell+1)} q^{2m}. 
\end{aligned}
\end{align}
By inserting the defining expressions of $u$ and $v$ with the $\qcRm$-matrix, one obtains that
\begin{subequations}
    \label{eq:ex_SUq2_braid_tmat}
\begin{align}
    \sum_{n=-\ell}^\ell (-q)^{n} \qcRm\big( t_{mn}^{(\ell)} \otimes t_{m'-n}^{(\ell)} \big)
    = \sum_{n=-\ell}^\ell (-q)^{-n} \qcRm\big( t_{m-n}^{(\ell)} \otimes t_{m'n}^{(\ell)} \big)
    &= \delta_{m,-m'} q^{-2\ell(\ell+1)} (-q)^{-m}, 
    \label{eq:ex_SUq2_braid_tmat_1} \\
    \sum_{n=-\ell}^\ell (-q)^{-n} \qcRm\big( t_{nm}^{(\ell)} \otimes t_{-nm'}^{(\ell)} \big)
    = \sum_{n=-\ell}^\ell (-q)^{n} \qcRm\big( t_{-nm}^{(\ell)} \otimes t_{nm'}^{(\ell)} \big)
    &= \delta_{m,-m'} q^{-2\ell(\ell+1)} (-q)^{m},
    \label{eq:ex_SUq2_braid_tmat_2} \\
    \sum_{n=-\ell}^\ell (-q)^{-2n} \qcRm(t_{mn}^{(\ell)} \otimes t_{nm'}^{(\ell)})
    = \sum_{n=-\ell}^\ell (-q)^{2n} \qcRm(t_{nm'}^{(\ell)} \otimes t_{mn}^{(\ell)})
    &= \delta_{mm'} q^{2\ell(\ell+1)},
    \label{eq:ex_SUq2_braid_tmat_3}
\end{align}
\end{subequations}
where we changed variable $n \to -n$ to get the first equalities\footnote{
    As a remark, we could also have used $\qcRm(S(a)\otimes S(b)) = \qcRm(a \otimes b)$ to obtain those.
}.

\paragraph{}
Let us now go to the path integral. Here, we go again through the main computation of section \ref{subsec:ncpi_pi} for completeness. The free action \eqref{eq:ncpi_free_action} is given by
\begin{align}
    \label{eq:ex_SUq2_free_act}
    S_0(\phi)
    &= \int_{S^2_q} \phi \Pi^{-1}[\phi]
    = \sum_{\ell,m} (-q)^{-m} \Pi_\ell^{-1} \phi^\ell_m \phi^\ell_{-m},
\end{align}
where $\Pi_\ell^{-1} = [\ell+1]_q[\ell]_q + \mu^2$ is the propagator. One computes the generating functional \eqref{eq:ncpi_gen_func} to be equal to
\begin{align}
    \label{eq:ex_SUq2_free_gf}
    \mathcal{Z}_0(J)
    &= \exp\left( \frac{1}{2} \int_{S^2_q} J \Pi[J] \right)
    = \exp\left( \frac{1}{2} \sum_{\ell,m} (-q)^{-m} \Pi_\ell J_{\ell,m} J_{\ell,-m} \right).
\end{align}
Having define the braided tensor product algebra (through the braiding), one can also define the functional derivative like in section \ref{subsec:ncpi_fd} and obtain that
\begin{align}
    \label{eq:ex_SUq2_fd_def}
    \frac{\partial}{\partial J^{(1)}} J^{(2)}
    &= \delta^{(1,2)},
\end{align}
Since we work in direct space formalism it is more convenient to write $\mathcal{Z}_0$ as a Gaussian functional. In order to do so, we introduce the Green's function for the differential operator $\Pi^{-1}$, that we define by
\begin{align}
    W^{(1,2)}
    &= \sum_{\ell,m} \Pi_\ell (Y^{(1)})^m_\ell [(Y^{(2)})^m_\ell]^*.
\end{align}
Then, the generating functional of the free action reads
\begin{align}
    \mathcal{Z}_0
    &= \exp \left( \frac{1}{2} \int_{S^2_q}^{(1,2)} J^{(1)} W^{(1,2)} J^{(2)} \right),
\end{align}
where we noted $\int_{S^2_q}^{(1,2)} = \int_{S^2_q}^{(1)} \int_{S^2_q}^{(2)}$. Indeed,
\begin{align*}
    \int_{S^2_q}^{(1,2)} J^{(1)} W^{(1,2)} J^{(2)}
    &= \sum_{\ell,m,\ell',m',\ell'',m''} J_{\ell,m} \Pi_{\ell'} J_{\ell'',m''} \left( \int^{(1)} (Y^{(1)})^m_\ell (Y^{(1)})^{m'}_{\ell'} \right) \left( \int^{(2)} [(Y^{(2)})^{m'}_{\ell'}]^* (Y^{(2)})^{m''}_{\ell''} \right) \\
    &= \sum_{\ell,m,\ell',m',\ell'',m''} \Pi_{\ell'} J_{\ell,m} J_{\ell'',m''} \big( (-q)^{-m} \delta_{\ell,\ell'} \delta_{m,-m'} \big) \big( \delta_{\ell',\ell''} \delta_{m',m''} \big) \\
    &= \sum_{\ell,m} (-q)^{-m} \Pi_\ell J_{\ell,m} J_{\ell,-m}.
\end{align*}
The Green's function has nice properties. The first and most important being its $SU_q(2)$-coinvariance. Indeed,
\begin{align}
    \label{eq:ex_SUq2_W_cov}
\begin{aligned}
    \coactl W^{(1,2)}
    &= \sum_{\ell,m} \Pi_\ell (-q)^{-m} \coactl\big( (Y^{(1)})^{m}_\ell (Y^{(2)})^{-m}_\ell \big)
    = \sum_{\ell,m,n,n'} \Pi_\ell (-q)^{-m} S(t^{(\ell)}_{nm}) S(t^{(\ell)}_{n'\,-m}) \otimes Y^n_\ell \otimes Y^{n'}_\ell \\
    &= \sum_{\ell,m,n,n'} \Pi_\ell (-q)^{n'} S(t^{(\ell)}_{nm}) t^{(\ell)}_{m\,-n'} \otimes Y^n_\ell \otimes Y^{n'}_\ell
    = \sum_{\ell,n,n'} \Pi_\ell \delta_{-n,n'} 1 \otimes Y^n_\ell \otimes (-q)^{n'} Y^{n'}_\ell \\
    &= 1 \otimes W^{(1,2)},
\end{aligned}
\end{align}
where we have used the Hopf algebra property $S(a_{(1)})a_{(2)} = \epsilon(a) 1$. It is very important to note that, the fact that $\Pi$ does not depend on $m$ is central for $W$ to be invariant. Actually, one can trace back the importance of $\Pi$ being independent of $m$ to the covariance of the free action \eqref{eq:ex_SUq2_free_act}. 

The second property is its braided symmetry. Indeed, thanks to the braiding, one computes (see the detailed computation below)
\begin{align}
    \label{eq:ex_SUq2_Y_braid_relation}
\begin{aligned}[]
    \sum_m (Y^{(1)})^m_\ell [(Y^{(2)})^m_\ell]^*
    &= q^{2\ell(\ell+1)} \sum_m (Y^{(2)})^m_\ell [(Y^{(1)})^m_\ell]^*,
\end{aligned}
\end{align}
This implies that $W^{(1,2)}$ and $\delta^{(1,2)}$ are related to $W^{(2,1)}$ and $\delta^{(2,1)}$ by a factor of $q^{2\ell(\ell+1)}$ corresponding to the application of a ribbon element. In that respect, \eqref{eq:ex_SUq2_Y_braid_relation} corresponds to the hypothesis \eqref{eq:mps_ribbon_hypo}. We write
\begin{align}
    W^{(1,2)}
    &= \nu^{-1} \actl W^{(2,1)}, &
    \delta^{(1,2)}
    &= \nu^{-1} \actl \delta^{(2,1)}.
\end{align}
\begin{comp}
    Here is the detailed computation of \eqref{eq:ex_SUq2_Y_braid_relation}
    \begin{align*}
    \sum_m (Y^{(2)})^m_\ell [(Y^{(1)})^m_\ell]^*
    &= \sum_m (-q)^{-m} (Y^{(2)})^{m}_\ell (Y^{(1)})^{-m}_\ell \\
    &= \sum_m (-q)^{-m} \sum_{n,n'} \mathcal{R}\big(S(t^{(\ell)}_{n\, -m}) \otimes S(t^{(\ell)}_{n'm}) \big) (Y^{(1)})^n_\ell (Y^{(2)})^{n'}_\ell \\
    &= \sum_{n,n'} \left( \sum_m (-q)^{-m} \mathcal{R}(t^{(\ell)}_{n\,-m} \otimes t^{(\ell)}_{n'm})\right) (Y^{(1)})^n_\ell (Y^{(2)})^{n'}_\ell \\
    &\overset{\eqref{eq:ex_SUq2_braid_tmat_2}}{=} \sum_{n,n'} q^{-2\ell(\ell+1)} (-q)^{-n} \delta_{n,-n'} (Y^{(1)})^n_\ell (Y^{(2)})^{n'}_\ell \\
    &= q^{-2\ell(\ell+1)} \sum_m (Y^{(1)})^m_\ell [(Y^{(2)})^m_\ell]^*,
    \qedhere
    \end{align*}
\end{comp}

Finally, the third property of $W$ is that it indeed is the Green's function for $\Pi^{-1}$ and is invertible. Since $\Pi^{-1}(Y^m_\ell) = \Pi_\ell^{-1} Y^m_\ell$ by definition, one computes that 
\begin{align} 
    (1 \otimes \Pi^{-1}) [W^{(1,2)}]
    &= (\Pi^{-1} \otimes 1) [W^{(1,2)}]
    = \delta^{(1,2)}.
\end{align}
Moreover, let us define 
\begin{align}
    G^{(1,2)}
    &= \sum_{\ell,m} \Pi^{-1}_\ell (Y^{(1)})^m_\ell [(Y^{(2)})^m_\ell]^*,
\end{align}
then, one has
\begin{align}
    \int_{S^2_q}^{(2)} G^{(1,2)} W^{(2,3)}
    &= \delta^{(1,3)}
    = \int_{S^2_q}^{(2)} W^{(1,2)} G^{(2,3)}.
\end{align}
This means that $G$ is the ``inverse'' of $W$, and one can actually show that $G$ is the Green's function for $\Pi$.
\begin{comp}
    \begin{align*}
        \int_{S^2_q}^{(2)} G^{(1,2)} W^{(2,3)}
        &= \sum_{\ell,m,\ell',m'} \Pi^{-1}_\ell \Pi_{\ell'} (Y^{(1)})^m_\ell \left( \int_{S^2_q}^{(2)} [(Y^{(2)})^m_\ell]^* (Y^{(2)})^{m'}_{\ell'} \right) [(Y^{(3)})^{m'}_{\ell'}]^* \\
        &= \sum_{\ell,m,\ell',m'} \Pi^{-1}_\ell \Pi_{\ell'} \delta_{\ell,\ell'} \delta_{m,m'} (Y^{(1)})^m_\ell [(Y^{(3)})^{m'}_{\ell'}]^* \\
        &= \sum_{\ell,m} (Y^{(1)})^m_\ell [(Y^{(3)})^{m}_{\ell}]^*
        = \delta^{(1,3)}
    \end{align*}
    The other side is computed similarly.
\end{comp}

\paragraph{}
One further computes that
\begin{align*}
    \xLd{J^{(1)}} \left( \int^{(2)} \phi^{(2)} J^{(2)} + J^{(2)} \phi^{(2)} \right)
    &= \int^{(2)} \phi^{(2)} \delta^{(1,2)} + \delta^{(1,2)} \phi^{(2)} \\
    &= \int^{(2)} \phi^{(2)} (\nu^{-1} \actl \delta^{(2,1)}) + \phi^{(1)} \\
    &= (1 + \nu^{-1}) \actl \phi^{(1)},
\end{align*}
so that $n$-point functions are computed thanks to
\begin{align}
    \label{eq:ex_SUq2_fnpf}
\begin{aligned}
    \dlangle \phi^{(1)} \cdots \phi^{(n)} \drangle_0 
    &= \frac{1}{\mathcal{Z}_0(0)} \int \mathscr{D}\phi\, \phi^{(1)} \cdots \phi^{(n)} e^{-S_0(\phi) + \frac{1}{2} \int (\phi J + J \phi)} \\
    &= \left. \left( \left( \frac{1 + \nu^{-1}}{2} \right)^{-1} \actl \xLd{J^{(n)}} \right) \cdots \left( \left( \frac{1 + \nu^{-1}}{2} \right)^{-1} \actl \xLd{J^{(1)}} \right) \mathcal{Z}_0(J) \right|_{J=0}.
\end{aligned}
\end{align}

This leads to the free $2$-point function being given by a braided Green's function:
\begin{align}
    \label{eq:ex_SUq2_f2pf}
    \dlangle \phi^{(1)} \phi^{(2)} \drangle_0
    &= \frac{2}{1 + \nu^{-1}} \actl W^{(1,2)}
    = W_{\nu^{-1}}^{(1,2)}
\end{align}
\begin{comp}
The first and second derivatives of the generating functional are computed to be
\begin{align*}
    \xLd{J^{(1)}} \mathcal{Z}_0(J)
    &= \xLd{J^{(1)}} \left( \frac{1}{2} \int_{S^2_q}^{(2,3)} J^{(2)} W^{(2,3)} J^{(3)} \right) \mathcal{Z}_0(J) \\
    &= \left( \frac{1}{2} \int_{S^2_q}^{(2,3)} \delta^{(1,2)} W^{(2,3)} J^{(3)} + J^{(2)} W^{(2,3)} \delta^{(1,3)} \right) \mathcal{Z}_0(J) \\
    &= \frac{1}{2} \left( \int_{S^2_q}^{(3)} W^{(1,3)} J^{(3)} + \int_{S^2_q}^{(2,3)}J^{(2)} W^{(2,3)} (\nu^{-1} \actl \delta^{(3,1)}) \right) \mathcal{Z}_0(J) \\
    &= \frac{1}{2} \left( \int_{S^2_q}^{(3)} W^{(1,3)} J^{(3)} + J^{(3)} (\nu^{-1} \actl W^{(3,1)}) \right) \mathcal{Z}_0(J) \\
    &= \frac{1}{2} \left( \int_{S^2_q}^{(3)} W^{(1,3)} J^{(3)} + J^{(3)} W^{(1,3)} \right) \mathcal{Z}_0(J),
\end{align*}
and
\begin{align*}
    \left. \xLd{J^{(1)}} \xLd{J^{(2)}} \mathcal{Z}_0(J) \right|_{J=0}
    &= \frac{1}{2} \xLd{J^{(1)}} \left( \int_{S^2_q}^{(3)} W^{(2,3)} J^{(3)} + J^{(3)} W^{(2,3)} \right) \\
    &= \frac{1}{2} \int_{S^2_q}^{(3)} W^{(2,3)} \delta^{(1,3)} + \delta^{(1,3)} W^{(2,3)} \\
    &= \frac{1}{2} \int_{S^2_q}^{(3)} W^{(2,3)} (\nu^{-1} \actl \delta^{(3,1)}) + \delta^{(1,3)} (\nu^{-1} \actl W^{(3,2)}) \\
    &= \frac{1}{2} \big( \nu^{-1} \actl W^{(2,1)} + \nu^{-1} \actl W^{(1,2)} \big) \\
    &= \frac{1 + \nu^{-1}}{2} \actl W^{(1,2)}.
\end{align*}
\end{comp}
One should note that since the ribbon element $\nu$ only have a $\ell$-dependence and no $m$-dependence $W_{\nu^{-1}}$ is directly $SU_q(2)$-coinvariant because $W$ is. A doubtful reader can do the computation of $\coactl W_{\nu^{-1}}$ by following the same steps as the computation of $\coactl W$ above.

\paragraph{}
It follows from \eqref{eq:ex_SUq2_f2pf} that a braided Wick theorem holds. The $2n+1$-point functions vanish because $\mathcal{Z}_0$ is quadratic in the source, as in the commutative case. The $2n$-point function now reads
\begin{align}
    \label{eq:ex_SUq2_bWt}
\begin{aligned}
    \dlangle \phi^{(1)} \cdots \phi^{(2n)} \drangle_0
    &= \sum_{G \in \mathfrak{G}_{2n}} \prod_{\substack{e \in G \\ s(e) < t(e)}} W_{\nu^{-1}}^{(s(e),t(e))},
\end{aligned}
\end{align}
where $\mathfrak{G}_{2n}$ is the set of directed graphs with vertices labelled $1, \ldots, 2n$, such that each vertex is met by exactly one edge $e$. $s(e)$ and $t(e)$ denotes the source and the target of each edge $e$ respectively. We have called the Wick theorem \eqref{eq:ex_SUq2_bWt} ``braided'' because of the presence of the ribbon element $\nu$ which is intrinsicly linked to the braiding, but also because of elements of the form $W^{(1,3)} W^{(2,4)}$: the elements $(2)$ and $(3)$ needs to be ``reversed'' using the braiding.

Therefore, the free $4$-point function reads
\begin{align}
    \dlangle \phi^{(1)} \cdots \phi^{(4)} \drangle_0
    &= W_{\nu^{-1}}^{(1,2)} W_{\nu^{-1}}^{(3,4)} + W_{\nu^{-1}}^{(1,3)} W_{\nu^{-1}}^{(2,4)} + W_{\nu^{-1}}^{(1,4)} W_{\nu^{-1}}^{(2,3)}.
\end{align}
It is straightforwardly $SU_q(2)$-coinvariant as a sum of product of coinvariant elements (since $W_{\nu^{-1}}$ is coinvariant).

\paragraph{}
We consider now a $\phi^4$-interacting theory with interaction action defined by
\begin{align}
    \label{eq:ex_SUq2_int_act}
    S_{\text{int}}(\phi)
    &= \frac{\lambda}{4!} \int_{S^2_q} \phi^4,
\end{align}
where $\lambda$ is the coupling constant. The $2$-point function at one-loop order expresses as
\begin{align}
    \dlangle\phi^{(1)} \phi^{(2)} \drangle_4^{\text{1-loop}}
    &= - \frac{\lambda}{4!} \int_{S^2_q}^{(3)} \dlangle[\big] (\phi^{(3)})^4 \phi^{(1)} \phi^{(2)} \drangle[\big]_0.
\end{align}
From the computation of the free $n$-point function \eqref{eq:ex_SUq2_fnpf}, the previous formula amounts to compute a free $6$-point function and then integrate over it. After computation, one obtains
\begin{align}
    \dlangle\phi^{(1)} \phi^{(2)} \drangle_4^{\text{1-loop}}
    &= \frac{\lambda}{4!} \int_{S^2_q}^{(3)} 6\, W_{\nu^{-1}}^{(3,3)} \left( W_{\nu^{-1}}^{(1,3)} W_{\nu^{-1}}^{(2,3)} + W_{\nu^{-1}}^{(2,3)} W_{\nu^{-1}}^{(1,3)} \right) \\
    &= \frac{\lambda}{4} \left( \int_{S^2_q}^{(3)} W_{\nu^{-1}}^{(1,3)} W_{\nu^{-1}}^{(2,3)} + W_{\nu^{-1}}^{(2,3)} W_{\nu^{-1}}^{(1,3)} \right) \left( \sum_\ell \frac{2[2\ell+1]_q}{1 + q^{2\ell(\ell+1)}} \Pi_\ell \right)
\end{align}
where the loop contribution has been simplified to
\begin{align}
    \label{eq:ex_SUq2_loop}
    W_{\nu^{-1}}^{(a,a)}
    &= \sum_{\ell,m} \frac{2}{1 + q^{2\ell(\ell+1)}}\Pi_\ell (Y^{(a)})^m_\ell [(Y^{(a)})^m_\ell]^*
    = \sum_\ell \frac{2[2\ell+1]_q}{1 + q^{2\ell(\ell+1)}} \Pi_\ell,
\end{align}
since
\begin{align*}
    \sum_m Y^m_\ell (Y^m_\ell)^*
    &= \sum_m [2\ell+1]_q (-q)^{-2m} t^{(\ell)}_{-m0} (t^{(\ell)}_{-m0})^*
    = \sum_m [2\ell+1]_q (-q)^{-m} t^{(\ell)}_{-m0} t^{(\ell)}_{m0} \\
    &= \sum_m [2\ell+1]_q S(t^{(\ell)}_{0m}) t^{(\ell)}_{m0}
    = [2\ell+1]_q 1,
\end{align*}
where we used again $S(a_{(1)})a_{(2)} = \varepsilon(a)1$. Moreover, we can also simplify
\begin{align*}
    \int_{S^2_q}^{(3)} W_{\nu^{-1}}^{(a,3)} W_{\nu^{-1}}^{(b,3)}
    &= \int_{S^2_q}^{(3)} W_{\nu^{-1}}^{(a,3)} (\nu^{-1} \actl W_{\nu^{-1}}^{(3,b)}) \\
    &= \sum_{\ell,m,\ell',m'} \frac{2\Pi_\ell}{1 + q^{2\ell(\ell+1)}} \frac{2\Pi_{\ell'}}{1 + q^{2\ell'(\ell'+1)}} q^{2\ell'(\ell'+1)} (Y^{(a)})^m_\ell \left( \int_{S^2_q}^{(3)}  [(Y^{(3)})^m_\ell]^* (Y^{(3)})^{m'}_{\ell'} \right) [(Y^{(b)})^{m'}_{\ell'}]^* \\
    &= \sum_{\ell,m} \left( \frac{2\Pi_\ell}{1 + q^{2\ell(\ell+1)}} \right)^2 q^{2\ell(\ell+1)} (Y^{(a)})^m_\ell [(Y^{(b)})^{m}_{\ell}]^*,
\end{align*}
for any $a,b < 3$. Finally, the $2$-point function at one-loop reads
\begin{align}
    \label{eq:ex_SUq2_2pf1l}
    \dlangle\phi^{(1)} \phi^{(2)} \drangle_4^{\text{1-loop}}
    &= \frac{\lambda}{4} \sum_{\ell,m} \left( \frac{2\Pi_\ell}{1 + q^{2\ell(\ell+1)}} \right)^2 (1 + q^{2\ell(\ell+1)}) \, (Y^{(1)})^m_\ell [(Y^{(2)})^{m}_{\ell}]^*.
\end{align}
Several comments are in order.

\begin{itemize}
    \item \textbf{Invariance.}
    The expression \eqref{eq:ex_SUq2_2pf1l} is $SU_q(2)$-coinvariant. This is obtained thanks to a similar computation to~\eqref{eq:ex_SUq2_W_cov}: it consists in a term of the form $Y Y^*$ with a $m$-independent prefactor.
    \item \textbf{Regularisation.}
    Considering the usual $q$-deformed Laplacian of $S^2_q$, that is $\Pi_\ell^{-1} = [\ell]_q [\ell+1]_q + m^2$ (where $m$ is the mass of the scalar field), the loop contribution \eqref{eq:ex_SUq2_loop} is finite for $q > 1$. This is because
    \begin{align}
        \frac{2[2\ell+1]_q}{1 + q^{2\ell(\ell+1)}} \Pi_\ell
        &= \frac{2}{1 + q^{2\ell(\ell+1)}} \frac{[2\ell+1]_q}{[\ell]_q[\ell+1]_q + m^2}
        \underset{\ell \to +\infty}{\sim} \frac{2 q^{2\ell}}{q^{2\ell^2} q^\ell q^{\ell}}
        = 2 q^{-2\ell^2}.
    \end{align}
    The scalar theory is therefore $q$-regularised, as already discussed in \cite{Oeckl_2001}, and does not suffer from UV/IR mixing.
    \item \textbf{Diagramatics.}
    The $6+6$ splitting appears clearly in \eqref{eq:ex_SUq2_2pf1l}, where $6$ diagrams are of the form $Y^{(1)} [Y^{(2)}]^*$ and $6$ of the form $Y^{(2)} [Y^{(1)}]^*$. They therefore differ by a crossing of lines that corresponds analytically by a ribbon factor $q^{-2\ell(\ell+1)}$. This result was already found in \cite{Oeckl_2001}.
    \item \textbf{Comparison with Oeckl, and the origin of the prefactor.}
    Our $n$-point functions carry global factors $\tfrac{2}{1+q^{2\ell(\ell+1)}}$ where \cite{Oeckl_2001} carries $q^{-2\ell(\ell+1)} = \nu|_{V_\ell}$. Both prescriptions are covariant, share the same $6{+}6$ diagrammatics and the same $q$-regularised loop; they differ \emph{only} in this prefactor, and the difference is purely a convention in relating the field to the source. Oeckl relates them through the commutative $\gamma$-function (eq.~(21) of \cite{Oeckl_2001}), an \emph{asymmetric} prescription that produces a single contraction term and hence the bare factor $\nu$. Our generating functional uses the \emph{symmetric} coupling $\tfrac12\int(\phi J + J\phi)$ -- forced because $\phi$ and $J$ do not commute -- so the Leibniz rule produces two terms, with weights $1$ and $q^{2\ell(\ell+1)}=\nu^{-1}$, whose normalised average is exactly $\tfrac{2}{1+\nu^{-1}}$. The two results therefore coincide up to a global, $\ell$-dependent rescaling of the field -- the nonlinear redefinition of the kinetic operator already noted below \eqref{eq:ncpi_uxl_npf}. In short, we do not disagree with \cite{Oeckl_2001}: we adopt a different, symmetric normalisation, and the prefactor is the visible imprint of that choice.
\end{itemize}

\subsubsection{Explicit expressions of the ribbon elements}
\label{subsubapx:SUq2_ribbon}
Given the Hopf algebra of $U_q(\mathfrak{su}(2))$ generated by $\{X_+, X_-,q^{\frac{H}{2}}, q^{-\frac{H}{2}}\}$ (see Example 3.2.1 of \cite{Majid_1994})
\begin{subequations}
\begin{align}
    q^{\frac{H}{2}} X_\pm q^{-\frac{H}{2}}
    &= q^{\pm 1} X_\pm, &
    [X_+,X_-]
    &= \frac{q^H - q^{-H}}{q - q^{-1}}, &
    q^{-\frac{H}{2}} q^{\frac{H}{2}}
    &= q^{\frac{H}{2}} q^{-\frac{H}{2}} = 1, \\
    \Delta(q^{\pm \frac{H}{2}})
    &= q^{\pm \frac{H}{2}} \otimes q^{\pm \frac{H}{2}}, &
    \Delta X_\pm
    &= X_\pm \otimes q^{\frac{H}{2}} + q^{-\frac{H}{2}} \otimes X_\pm, &
    \varepsilon(q^{\pm \frac{H}{2}})
    &= 1, \\
    \varepsilon(X_\pm) 
    &= 0, &
    S(q^{\pm\frac{H}{2}}) 
    &= q^{\mp\frac{H}{2}}, &
    S(X_\pm)
    &= - q^{\pm1} X_\pm,
\end{align}
\end{subequations}
we compute the relations
\begin{align}
    \label{eq:suq2_square_antipode}
    S^2(q^{\pm\frac{H}{2}}) 
    &= q^{\pm \frac{H}{2}}, &
    S^2(X_\pm)
    &= q^{\pm 2} X_\pm
    = q^{H} X_\pm q^{-H},
\end{align}
where the last equality is obtained thanks to $q^\frac{H}{2} X_\pm = q^{\pm 1} X_\pm q^\frac{H}{2}$. The equation \eqref{eq:suq2_square_antipode} implies that the squared antipode $S^2$ corresponds to the adjoint action of $q^H$. However, by definition of $\mathfrak{u}$, we also have that $S^2$ is the adjoint action of $\mathfrak{u}$ (see Proposition 2.1.8 of \cite{Majid_1995}). Similarly, $S^{-2}$ can be computed to correspond both to the adjoint action of $\mathfrak{v}$ and of $q^{-H}$. By Schur's lemma,
\begin{align*}
    \mathfrak{u}|_{V_\ell} = \lambda_\ell q^H, &&
    \mathfrak{v}|_{V_\ell} = \mu_\ell q^{-H},
\end{align*}
where $V_\ell$ is the $U_q(\mathfrak{su}(2))$-modules\footnote{
    Recall that they are obtained through the Peter-Weyl decomposition.
}
of spin $\ell$ and $\lambda_\ell, \mu_\ell \in \mathbb{C}$ are constants. By applying these formula to the highest weight and lowest weight vectors of $V_\ell$, one computes that
\begin{align*}
    \mathfrak{u}|\ell, \ell\rangle
    = q^{-2\ell^2}|\ell,\ell\rangle
    = \lambda_\ell q^H|\ell,\ell\rangle
    = \lambda_\ell q^{2\ell}|\ell,\ell\rangle, &&
    \mathfrak{v}|\ell,-\ell\rangle 
    = q^{-2\ell^2}|\ell,-\ell\rangle
    = \mu_\ell q^{-H} |\ell,-\ell\rangle
    = \mu_\ell q^{2\ell}|\ell,-\ell\rangle,
\end{align*}
which implies that $\lambda_\ell = \mu_\ell = q^{-2\ell(\ell+1)}$. The values of $\mathfrak{u}|\ell,\ell\rangle$ and $\mathfrak{v}|\ell,-\ell\rangle$ can be computed easily since their expression involves $q$-exponential of nilpotent terms for which the $n = 0$ component of the exponential series is the only non-vanishing term. Explicitly, for $\mathfrak{u}$, one has the expression, deduced from the $\qRm$-matrix (see Proposition 2.1.8 of \cite{Majid_1995}),
\begin{align}
    \mathfrak{u} = \sum_{n=0}^{+\infty} \frac{(1-q^{-2})^n q^\frac{n(n-1)}{2}}{[n]!(-q)^n} (X_- q^\frac{H}{2})^n q^{-\frac{H^2}{2}} (q^\frac{H}{2} X_+)^n,
\end{align}
but since $X_+ |\ell,\ell\rangle = 0$, only the $n = 0$ term of the sum survives, so that
\begin{align*}
    \mathfrak{u}|\ell,\ell\rangle
    &= q^{-\frac{H^2}{2}} |\ell,\ell\rangle
    = q^{-2\ell^2}.
\end{align*}
The computation for $\mathfrak{v}$ is very similar, to the exception that the right hand side of $v$ is $X_-$. That is why the lowest weight vector since $X_-|\ell,-\ell\rangle = 0$ has been considered instead of the highest weight vector. Finally, we have shown that
\begin{align}
    \mathfrak{u}|_{V_\ell}
    &= q^{-2\ell(\ell+1)} q^H, &
    \mathfrak{v}|_{V_\ell}
    &= q^{-2\ell(\ell+1)} q^{-H}, &
    \nu|_{V_\ell}
    &= q^{-2\ell(\ell+1)},
\end{align}
where the result for the ribbon element $\nu^2 = \mathfrak{u}\mathfrak{v}$ is deduced from $\mathfrak{u}$ and $\mathfrak{v}$.

\section{Conclusions}
\paragraph{}
In this paper we have built the multi-particle algebra of a generic noncommutative deformation of Minkowski space as a braided tensor product algebra, the braiding encoding the deformed statistics dictated by the quantum Poincar\'{e} symmetry. This construction served a two-fold purpose: to give a well-defined, covariant notion of $n$-point function, and to keep track of the behaviour of these correlators under the quantum Poincar\'{e} symmetry. We then quantised a scalar field with quartic self-interaction in the path integral formalism, and showed that the procedure hinges on the choice of noncommutative functional derivative. We isolated two such choices, distinguished by the space in which the Leibniz rule remains undeformed: the $p$-Leibniz derivative, which carries the undeformed (bosonic) statistics, and the $x$-Leibniz derivative, which carries the braided one.

The two choices lead to markedly different theories. The $p$-Leibniz quantisation reproduces the standard noncommutative prescription: its $n$-point functions are not quantum Poincar\'{e} invariant and exhibit the UV/IR mixing of \cite{Minwalla_2000}, both features tracing to the same intertwining of external and loop momenta in the non-planar contributions. The $x$-Leibniz quantisation, which adapts the functional calculus to the braided algebra of fields along the lines of \cite{Oeckl_2001}, instead yields covariant $n$-point functions in which external momenta are never intertwined, and which are consequently free of this UV/IR mixing. The main conclusion is therefore that, in the models considered here, the UV/IR mixing of \cite{Minwalla_2000} is an artifact of a quantisation that breaks the deformed Poincar\'{e} symmetry, rather than a feature of noncommutativity itself: it disappears once covariance is imposed through a braided quantisation.

A second outcome of our analysis is a clean separation between two questions that are often conflated. Covariance is controlled by the quantisation scheme -- the braiding and the associated Leibniz rule -- whereas the finiteness of the theory is controlled by the propagator alone, through the convergence of its integral over momentum space. The two are independent: a covariant theory may still call for regularisation, and a finite theory may still break covariance. This brings the original motivation of \cite{Snyder_1947} back into focus, disentangled from the UV/IR mixing with which it is sometimes identified, as a property to be assessed model by model from the deformed propagator.

We have illustrated the formalism on three examples, collected in the appendices: the $T$-Minkowski models, the Euclidean $3$-dimensional quantum gravity (Ponzano--Regge) model, and scalar field theory on the quantum two-sphere $S^2_q$, the last reproducing the structure of Oeckl's original results \cite{Oeckl_2001}. We have also recast the canonical quantisation of the free theory in the present language, along the lines of \cite{Fabiano_2025}, where the choice between $p$- and $x$-Leibniz derivatives reappears as the choice between an undeformed and a braided oscillator algebra.

\section*{Acknowledgement}
The authors would like to thank R.~Oeckl for taking the time to come to Burgos and discuss his work with us. We would like to acknowledge M.~Dimitrijevi\'{c} \'{C}iri\'{c} for clarification on their braided BV quantisation. We also thank G.~Fabiano for thoughts and comments.

K.~H.~would like to thank J.~Barrett, J.~Burridge, S.~Koren, T.~Laird, V.~Maris, G.~Trojani, J.~C.~Wallet and F.~Wagner for discussions and giving their views on the subject.

This work has been supported by the grants CNS2023-143760, funded by NextGenerationEU and PID2023-148373NB-I00 funded by MCIN/AEI/10.13039/501100011033/FEDER – UE, and by the Q-CAYLE Project funded by the Regional Government of Castilla y León (Junta de Castilla y León) and by the Ministry of Science and Innovation MICIN through NextGenerationEU (PRTR C17.I1).

This work falls within the scopes of the COST Action CA23130 ``Bridging high and low energies in search of quantum gravity'' and the COST Action 21109 CaLISTA ``Cartan geometry, Lie, Integrable Systems, quantum group Theories for Applications'', from the European Cooperation in Science and Technology (COST).

\newpage
\appendix
\section{Hopf algebra symmetries and (co)quasitriangular structures}
\label{apx:Ha}
\paragraph{}
We review here some properties of quasitriangular Hopf algebras that one can find in textbooks, like in \cite{Majid_1995}. The appendix \ref{subapx:Ha_bc} also makes the link between the notion of braided derivations we are using in section \ref{sec:ncpi}, and the one already defined in the context of braided commutative spaces.

\subsection{Quasitriangular structures}
\label{subapx:Ha_qts}
\paragraph{}
A Hopf algebra of (infinitesimal) symmetries, called $\mathcal{H}$, is said to be quasitriangular if it exists an invertible element $\qRm = \qRm_{(1)} \otimes \qRm_{(2)} \in \mathcal{H \otimes H}$, called the $\qRm$-matrix, such that
\begin{subequations}
	\label{eq:Ha_Rmat_def}
\begin{align}
	(\Delta \otimes \id)(\qRm)
	&= \qRm_{13} \qRm_{23}, &
	(\id \otimes \Delta)(\qRm)
	&= \qRm_{13} \qRm_{12},
	\label{eq:Ha_Rmat_def_2c}
\end{align}
\ineqskip
\begin{align}
	\tau \circ \Delta(h)
	&= \qRm \Delta(h) \qRm^{-1},
	\label{eq:Ha_Rmat_def_dcop}
\end{align}
\end{subequations}
for any $h \in \mathcal{H}$, where $\tau$ is the flip map and
\begin{align*}
	\qRm_{12}
	&= \qRm_{(1)} \otimes \qRm_{(2)} \otimes 1, &
	\qRm_{13}
	&= \qRm_{(1)} \otimes 1 \otimes \qRm_{(2)}, &
	\qRm_{23}
	&= 1 \otimes \qRm_{(1)} \otimes \qRm_{(2)}.
\end{align*}
The equation \eqref{eq:Ha_Rmat_def_dcop} states that $\qRm$ deforms the cocommutativity condition of the coproduct $\Delta$ of $\mathcal{H}$. If one further requires that \begin{align}
	\qRm_{21} \qRm = 1 \otimes 1,
	\label{eq:Ha_Rmat_tri}
\end{align}
where $\qRm_{21} = \tau(\qRm) = \qRm_{(2)} \otimes \qRm_{(1)}$, then $\mathcal{H}$ is said to be triangular.

One can further show that
\begin{subequations}
	\label{eq:Ha_Rmat_prop}
\begin{align}
	(\varepsilon \otimes \id)(\qRm)
	&= (\id \otimes \varepsilon)(\qRm)
	= 1,
\end{align}
	\ineqskip
\begin{align}
	(S \otimes \id)(\qRm)
	&= \qRm^{-1}, &
	(\id \otimes S)(\qRm^{-1})
	&= \qRm,
\end{align}
	\ineqskip
\begin{align}
	\qRm_{12} \qRm_{13} \qRm_{23}
	= \qRm_{23} \qRm_{13} \qRm_{12}.
	\label{eq:Ha_Rmat_prop_qYB}
\end{align}
\end{subequations}
The equation \eqref{eq:Ha_Rmat_prop_qYB} is called the quantum Yang-Baxter equation.

\subsection{Coquasitriangular structures}
\label{subapx:Ha_cqts}
\paragraph{}
From a Hopf dual point of view, $\dual{\mathcal{H}}$ has a coquasitriangular structure, that we define here. Let us recall that $\dual{\mathcal{H}}$ corresponds here to the quantum deformation of the Lie group associated to the Lie algebra of which $\mathcal{H}$ is the deformation. In physical words, $\dual{\mathcal{H}}$ corresponds to the quantum group of symmetries.

The dual to the $\qRm$-matrix is called the $\qcRm$-matrix $\qcRm : \dual{\mathcal{H}} \otimes \dual{\mathcal{H}} \to \Cpx$ and is invertible.\footnote{
	Invertibility is here defined by the existence of $\qcRm^{-1}: \dual{\mathcal{H}} \otimes \dual{\mathcal{H}} \to \Cpx$ such that
	\begin{align}
		\qcRm^{-1}\big( X_{(1)} \otimes Y_{(1)} \big) \ \qcRm\big( X_{(2)} \otimes Y_{(2)} \big)
		&= \epsilon(T_1) \epsilon(T_2)
		= \qcRm\big( X_{(1)} \otimes Y_{(1)} \big) \ \qcRm^{-1}\big( X_{(2)} \otimes Y_{(2)} \big)
	\end{align}
}
It can be defined by
\begin{align}
	\qcRm(X \otimes Y)
	&= \langle \qRm, X \otimes Y \rangle,
	\label{eq:Ha_cRmat_dual}
\end{align}
for any $X, Y \in \dual{\mathcal{H}}$, where $\langle \cdot, \cdot \rangle: \mathcal{H} \times \dual{\mathcal{H}} \to \Cpx$ denotes the dual pairing. The definition of the $\qRm$-matrix \eqref{eq:Ha_Rmat_def}, now writes in term of the $\qcRm$-matrix
\begin{subequations}
	\label{eq:Ha_cRmat_def}
\begin{align}
	\begin{aligned}
		\qcRm(X Y \otimes Z)
		&= \qcRm(X \otimes Z_{(1)}) \ \qcRm(Y \otimes Z_{(2)}) \\
		\qcRm(X \otimes YZ)
		&= \qcRm(X_{(2)} \otimes Y) \ \qcRm(X_{(1)} \otimes Z)
	\end{aligned}
	\label{eq:Ha_cRmat_2c}\\
	\qcRm(X_{(2)} \otimes Y_{(2)}) Y_{(1)} X_{(1)}
	= \qcRm(X_{(1)} \otimes Y_{(1)}) X_{(2)} Y_{(2)}
	\label{eq:Ha_cRmat_dpdt}
\end{align}
\end{subequations}
for any $X, Y, Z \in \dual{\mathcal{H}}$, where $\Delta X = X_{(1)} \otimes X_{(2)}$ is the coproduct expressed in Sweedler's notation.

\begin{comp}
We here show that dualising \eqref{eq:Ha_Rmat_def} leads to \eqref{eq:Ha_cRmat_def}.
\begin{align*}
	\qcRm(X Y \otimes Z)
	&= \langle \qRm, XY \otimes Z \rangle
	= \langle (\Delta \otimes \id) \qRm, X \otimes Y \otimes Z \rangle \\
	&= \langle \qRm_{13} \qRm_{23}, X \otimes Y \otimes Z \rangle
	= \langle \qRm_{13}, X_{(1)} \otimes Y_{(1)} \otimes Z_{(1)} \rangle \ \langle \qRm_{23}, X_{(2)} \otimes Y_{(2)} \otimes Z_{(2)} \rangle \\
	&= \epsilon(Y_{(1)}) \langle \qRm, X_{(1)} \otimes Z_{(1)} \rangle \ \epsilon(X_{(2)}) \langle \qRm, Y_{(2)} \otimes Z_{(2)} \rangle \\
	&= \langle \qRm, X \otimes Z_{(1)} \rangle \ \langle \qRm, Y \otimes Z_{(2)} \rangle
	= \qcRm(X \otimes Z_{(1)}) \ \qcRm(Y \otimes Z_{(2)})
\end{align*}
	The proof for the other equality of \eqref{eq:Ha_cRmat_2c} is similar. For \eqref{eq:Ha_cRmat_dpdt}, let us consider $h \in \mathcal{H}$, then
\begin{align*}
	\qcRm(X_{(2)} \otimes Y_{(2)}) \ \langle h, Y_{(1)} X_{(1)} \rangle
	&= \langle \qRm, X_{(2)} \otimes Y_{(2)} \rangle \ \langle \tau \circ \Delta(h), X_{(1)} \otimes Y_{(1)} \rangle \\
	&= \langle \tau \circ \Delta(h) \qRm, X \otimes Y \rangle
	= \langle \qRm \Delta(h), X \otimes Y \rangle \\
	&= \langle \qRm, X_{(1)} \otimes Y_{(1)} \rangle \ \langle \Delta(h), X_{(2)} \otimes Y_{(2)} \rangle \\
	&= \qcRm( X_{(1)} \otimes Y_{(1)} ) \ \langle h, X_{(2)} Y_{(2)} \rangle,
\end{align*}
	which holds for any $h \in \mathcal{H}$.
\end{comp}

In a similar way as the $\qRm$-matrix deforms the coproduct of $\mathcal{H}$, the $\qcRm$-matrix deforms the product of $\dual{\mathcal{H}}$ through \eqref{eq:Ha_cRmat_dpdt}. Note that the relations \eqref{eq:Ha_cRmat_def} do not involve the $\qRm$-matrix and could be used to define a coquasitriangular Hopf algebra without referring to any quasitriangular structure.

\paragraph{}
The dual of the properties \eqref{eq:Ha_Rmat_prop} can be also directly computed from \eqref{eq:Ha_Rmat_def} and shown to be
\begin{subequations}
	\label{eq:Ha_cRmat_prop}
\begin{align}
	\qcRm(X \otimes 1)
	&= \epsilon(X)
	= \qcRm(1 \otimes X)
\end{align}
	\ineqskip
\begin{align}
	\qcRm(S(X) \otimes Y)
	&= \qcRm^{-1}(X \otimes Y), &
	\qcRm^{-1}(X \otimes S(Y))
	&= \qcRm(X \otimes Y),
\end{align}
	\ineqskip
\begin{align}
	\qcRm(X_{(1)} \otimes Y_{(1)}) \ \qcRm(X_{(2)} \otimes Z_{(1)}) \ \qcRm(Y_{(2)} \otimes Z_{(2)})
	&= \qcRm(Y_{(1)} \otimes Z_{(1)}) \ \qcRm(X_{(1)} \otimes Z_{(2)}) \ \qcRm(X_{(2)} \otimes Y_{(2)}).
	\label{eq:Ha_cRmat_prop_qYB}
\end{align}
\end{subequations}
We refer to Lemma 2.2.2 and 2.2.3 of \cite{Majid_1995} for a proof. Finally, we say that the $\qcRm$-matrix is triangular if, for any $X,Y \in \mathcal{H}$,
\begin{align}
    \label{eq:Ha_cRmat_triang}
    \qcRm(Y_{(1)} \otimes X_{(1)}) \qcRm(X_{(2)} \otimes Y_{(2)})
    &= \varepsilon(X) \varepsilon(Y).
\end{align}
The equation \eqref{eq:Ha_cRmat_triang} can be shown to be the dual of \eqref{eq:Ha_Rmat_tri}.

\subsection{Braided commutativity}
\label{subapx:Ha_bc}
\paragraph{}
We consider now our noncommutative space $\Malg$ to be a $\mathcal{H}$-module algebra, with $\mathcal{H}$ triangular. Then, the triangularity condition \eqref{eq:Ha_Rmat_tri} can be used to impose the relation
\begin{subequations}
	\label{eq:bc_def}
\begin{align}
	f g
	&= (\qRm_{(1)}^{-1} \actl g) (\qRm_{(2)}^{-1} \actl f)
	\label{eq:bc_def_nbra}
\end{align}
where $f, g \in \Malg$ and we used the product of $\Malg$. This relation is called the braided commutativity, and it can also be written under the form of a vanishing braided bracket
\begin{align}
	[f, g]_{\qRm}
	&= f g - (\qRm_{(1)}^{-1} \actl g) (\qRm_{(2)}^{-1} \actl f)
	= 0
	\label{eq:bc_def_bra}
\end{align}
\end{subequations}

This kind of commutativity takes its root in the context of Drinfel'd twist deformation of a classical differential calculus \cite{Majid_1999, Sitarz_2001}. This type of deformation has been adapted to deform $U(n)$ gauge theories (see for example \cite{Aschieri_2006a, Vassilevich_2006, Chaichian_2006}) and gravity theories (see \cite{Aschieri_2005, Aschieri_2006b}). The braided commutativity properties have rather been exploited in the context of ``braided geometry'' \cite{Aschieri_2020, Weber_2020b}, that corresponds to braided deformations of Riemannian geometries.

\paragraph{}
Let us consider the generating functional $\mathcal{Z}(J)$ defined in \eqref{eq:ncpi_gen_func}. By definition, a multiplication by $\phi$ corresponds to a derivation with respect to $J$ in the following sense
\begin{align*}
	\frac{\partial}{\partial J(x_1)} \mathcal{Z}(J)
	&= \frac{1}{\mathcal{Z}(0)} \int \tdl\phi \phi(x_1) \exp\left( -S(\phi) + \frac{1}{2} \int \big( \phi J + J \phi \big) \right).
\end{align*}
Due to the fact that the algebra of $\phi$'s is braided commutative, as in \eqref{eq:bc_def}, one can show that the derivations over $J$ forms a braided commutative Lie algebra with the bracket
\begin{align}
	\left[ \frac{\partial}{\partial J(x_1)}, \frac{\partial}{\partial J(x_2)} \right]_{\qRm}
	&= \frac{\partial}{\partial J(x_1)} \frac{\partial}{\partial J(x_2)} - \left( \qRm_{(1)}^{-1} \actl \frac{\partial}{\partial J(x_2)} \right) \left( \qRm_{(2)}^{-1} \actl \frac{\partial}{\partial J(x_1)} \right).
\end{align}
Indeed, using the braided commutativity \eqref{eq:bc_def_bra}, one computes
\begin{align*}
	\left[ \frac{\partial}{\partial J(x_1)}, \frac{\partial}{\partial J(x_2)} \right]_{\qRm} \mathcal{Z}(J)
	&= \frac{1}{\mathcal{Z}(0)} \int \tdl\phi [\phi(x_1), \phi(x_2)]_{\qRm} \exp\left( -S(\phi) + \frac{1}{2} \int \big( \phi J + J \phi \big) \right)
	= 0.
\end{align*}
In this respect, the work of \cite{Weber_2020b} showed (we refer to \cite{Weber_2020a} for the explicit computations) that the good object to study the braided differential calculus on $\Malg$ was not the derivations, but rather the braided derivations: $\frac{\partial}{\partial J}$ needs to be a braided derivation in order for it to close a (braided) Lie algebra for the bracket $[\cdot, \cdot]_{\qRm}$. Explicitly, one defines the derivative with respect to $J$ on a product as
\begin{align}
	\frac{\partial}{\partial J(x)} \big( U(J) V(J) \big)
	&= \frac{\partial U(J)}{\partial J(x)} V(J) + \big( \qRm_{(1)}^{-1} \actl U(J) \big) \left( \left( \qRm_{(2)}^{-1} \actl \frac{\partial}{\partial J(x)} \right) V(J) \right)
	\label{eq:bc_braid_der}
\end{align}
When the braiding $\Psi$ is inherited from a (co)triangular structure (see appendix \ref{subsubapx:mps_glb}) the braided Leibniz rule \eqref{eq:ncpi_bxlr} boils down to \eqref{eq:bc_braid_der}. Therefore, \eqref{eq:ncpi_bxlr} can be seen as a generalisation of \eqref{eq:bc_braid_der} for non-triangular deformations.

The fact that braided $x$-Leibniz rule, which corresponds to a $p$-Leibniz rule, is a key element in order to defined differential structure on braided commutative spaces therefore comes naturally. However, as detailed in this paper, the choice of a $x$-Leibniz rule might be preferable in order to preserve covariance under the quantum symmetries.

\section{Braided tensor algebras and their invariance}
\label{apx:mps}
\paragraph{}
This appendix is an extension to section \ref{sec:mps} and includes details of computations and explanations for the construction of the multi-particle algebra.

\subsection{The canonical product}
\label{subapx:mps_cp}
\paragraph{}
In this section, we briefly detail why the canonical product on tensor product of algebras is not suited for our construction of $\Malg_\ell^{\tpog n}$, which should satisfy the requirement \ref{it:mps_sp} and \ref{it:mps_act} of section \ref{sec:mps}. Explicitly, we construct the natural/canonical product on $\Malg_\ell^{\tpog 2}$ and show that it is not compatible with the action \eqref{eq:mps_gact}, \ie that the requirement \ref{it:mps_act} is not fulfilled.

\paragraph{}
Let us first consider the usual product $\star$ of tensor product algebra, which reads
\begin{align}
	(f_1 \tpog f_2) \star (g_1 \tpog g_2)
	&= f_1 g_1 \tpog f_2 g_2,
	\label{eq:mps_gpdt}
\end{align}
for any $f_1 \tpog f_2, g_1 \tpog g_2 \in \Malg_\ell^{\tpog 2}$. If we consider elements belonging to only one copy of the algebra, that is $f \tpog 1$ and $1 \tpog g$, then they commute in $\Malg_\ell^{\tpog 2}$ since
\begin{align*}
	(f \tpog 1) \star (1 \tpog g)
	= f \tpog g
	= (1 \tpog g) \star (f \tpog 1).
\end{align*}
By regarding physically $f \tpog 1$ and $1 \tpog g$ as two non-entangled different particles, then we have just defined a commutative multi-particle algebra. In other words, it imposes a bosonic statistic. Another way of considering this fact is that \eqref{eq:mps_gpdt} consists in the braided product \eqref{eq:mps_braid_pdt} with $\Psi = \tau$.

The product \eqref{eq:mps_gpdt} is associative thanks to the associativity of $\Malg_\ell$: the requirement \ref{it:mps_sp} is fulfilled.

Regarding \ref{it:mps_act}, that is the action compatibility with \eqref{eq:mps_gpdt}, one computes, for any $X \in \Poin_\ell$,
\begin{align*}
    X \actl \big( (f_1 \tpog f_2) \star (g_1 \tpog g_2) \big)
    &= X \actl (f_1 g_1 \tpog f_2 g_2) \\
    &= (X_{(1)} \actl f_1)(X_{(2)} \actl g_1) \tpog (X_{(3)} \actl f_2) (X_{(4)} \actl g_2) \\
    \big( X_{(1)} \actl (f_1 \tpog f_2) \big) \star \big( X_{(2)} \actl (g_1 \tpog g_2) \big)
    &= \big( (X_{(1)} \actl f_1) \tpog (X_{(2)} \actl f_2) \big) \star \big( (X_{(3)} \actl g_1) \tpog (X_{(4)} \actl g_2) \big) \\
    &= (X_{(1)} \actl f_1) (X_{(3)} \actl g_1) \tpog (X_{(2)} \actl f_2) (X_{(4)} \actl g_2)
\end{align*}
where we have used the Sweedler's notation $\Delta X = X_{(1)} \otimes X_{(2)}$. Note that both results are \textit{almost} equal up to $X_{(2)} \otimes X_{(3)} = X_{(3)} \otimes X_{(2)}$, which only holds if and only if $\Poin_\ell$ is cocommutative (since $f_2$ and $g_1$ are generic). By considering a quantum deformation of Poincar\'{e}, we explicitly make $\Poin_\ell$ noncocommutative: \ref{it:mps_act} is forced to fail. Therefore the product \eqref{eq:mps_gpdt} does not allow us to built a quantum Poincar\'{e} invariant multi-particle algebra.

\subsection{Braided tensor product}
\paragraph{}
In order to built the product of $\Malg_\ell^{\otimes n}$ in a covariant way, one idea \cite{Majid_1995} is to introduce a braiding \eqref{eq:mps_braid_def}, that corresponds to a generalised flip. This comes from the fact that multiplying elements with two components, say $f_1 \tpb f_2$ and $g_1 \tpb g_2$, requires to flip $f_2$ and $g_1$: diagrammatically, there is a crossing of lines which therefore corresponds to a braiding. This is pictured in Figure \ref{fig:braid_prod2}.

One can show that $\Malg_\ell^{\tpb n}$ equipped with the braided product \eqref{eq:mps_braid_pdt} forms a $\Poin_\ell$-module algebra if and only if
\begin{align}
	\Psi \circ \actl
	&= \actl \circ (\id_{\Poin_\ell} \otimes \Psi),
	\tag{\ref{eq:mps_braid_cov}}
\end{align}
where $\Psi$ is the braiding. A braiding is required to satisfy the property  \eqref{eq:mps_braid_cov} by definition since it is an (iso)morphism (see section 9.2 of \cite{Majid_1995}). This is what makes the braided formalism such a powerful tool to implement covariance. In this discussion, we say that a braiding $\Psi$ satisfying \eqref{eq:mps_braid_cov} is ``covariant'' with respect to $\actl$. Note that the covariance of $\Psi$ with respect to the coaction $\coactl$ of $\Poing_\ell$ reads
\begin{align}
	(\id_{\Poing_\ell} \otimes \Psi) \circ \coactl
	&= \coactl \circ \Psi,
	\label{eq:mps_braid_cov_coac}
\end{align}
and $\Malg_\ell^{\tpb n}$, equipped with the braided product \eqref{eq:mps_braid_pdt}, forms a $\Poing_\ell$-comodule algebra if and only if \eqref{eq:mps_braid_cov_coac} is satisfied.

In the following subsections, we study the product defined thanks to the quasitriangular braiding (section \ref{subsubapx:mps_qtb}) and the Hopf braiding (section \ref{subsubapx:mps_glb}) regarding their covariance and their associativity.

\subsubsection{Quasitriangular braiding}
\label{subsubapx:mps_qtb}
\paragraph{}
Let us suppose that $\Poin_\ell$ has a quasitriangular structure, \ie there exists an invertible element $\qRm \in \Poin_\ell \otimes \Poin_\ell$ called the $\qRm$-matrix that deforms the cocommutativity property (see appendix \ref{subapx:Ha_qts}). The $\qRm$-matrix has been defined for some deformations of Minkowski, especially those defined by a Drinfel'd twist deformation, as for example Moyal \cite{Aschieri_2005} or $\rho$-Minkowski \cite{Dimitrijevic_2018}. Moreover, all T-Minkowski models \cite{Mercati_2023} have a triangular structure. In most of these cases the $\qcRm$-matrix has not been explicitly defined, but it could be straightforwardly thanks to the defining relation \eqref{eq:Ha_cRmat_dual}.

In this case, one defines the braiding as being
\begin{align}
	\Psi(f \otimes g)
	&= (\qRm_{(2)} \actl f) \otimes (\qRm_{(1)} \actl g).
	\tag{\ref{eq:mps_qt_braid}}
\end{align}
This is a well-defined braiding as shown in Theorem 9.2.4 of \cite{Majid_1995}. For completeness, we detail here the proof of its covariance with respect to the coaction:
\begin{align*}
    \Psi \big( X \actl (f \otimes g) \big)
    &= \Psi \big( (X_{(1)} \actl f) \otimes (X_{(2)} \actl g) \big) \\
    &= (\qRm_{(2)} X_{(2)} \actl g) \otimes (\qRm_{(1)} X_{(1)} \actl f) \\
    &= (X_{(1)} \qRm_{(2)} \actl g) \otimes (X_{(2)} \qRm_{(1)} \actl f) \\
    &= X \actl \big( (\qRm_{(2)} \actl g) \otimes (\qRm_{(1)} \actl f) \big)\\
    &= X \actl \big( \Psi(f \otimes g) \big),
\end{align*}
where we used the deformed coproduct rule \eqref{eq:Ha_Rmat_def_dcop}, which writes here $X_{(2)} \qRm_{(1)} \otimes X_{(1)} \qRm_{(2)} = \qRm_{(1)} X_{(1)} \otimes \qRm_{(2)} X_{(2)}$ and composed the latter with $\tau$.

\paragraph{}
The braided product \eqref{eq:mps_braid_pdt} associated to the braiding \eqref{eq:mps_qt_braid} then writes
\begin{align}
	(f_1 \tpb f_2) \star_\Psi (g_1 \tpb g_2)
	&= f_1 (\qRm_{(2)} \actl g_1) \otimes (\qRm_{(1)} \actl f_2) g_2
	\tag{\ref{eq:mps_qtpdt}}
\end{align}
for any $f_1 \tpb f_2, g_1 \tpb g_2 \in \Malg_\ell^{\tpb 2}$. As compared to \eqref{eq:mps_gpdt}, this product is not ``bosonic'' anymore: by applying the definition \eqref{eq:mps_qtpdt} to single non-entangled particles, we have 
\begin{align*}
    (f \tpb 1) \star_\Psi (1 \tpb g)
    &= f \tpb g, &
    (1 \tpb f) \star_\Psi (g \tpb 1)
    &= (\qRm_{(2)} \actl g) \otimes (\qRm_{(1)} \actl f).
\end{align*}
Let us now focus on the requirements \ref{it:mps_sp} and \ref{it:mps_act} of section \ref{sec:mps}.

The condition \ref{it:mps_act} is a direct consequence of the covariance of the braiding \eqref{eq:mps_braid_cov}, together with the fact that $\Malg_\ell$ is a $\Poin_\ell$-module algebra. 
For completeness, one can still check explicitly the compatibility of the product \eqref{eq:mps_qtpdt} with respect to the action \eqref{eq:mps_gact}: 
\begin{align*}
	X \actl \big( (f_1 \tpb f_2) \star_\Psi (g_1 \tpb g_2) \big)
    &= X \actl \big( f_1(\qRm_{(2)} \actl g_1) \tpb (\qRm_{(1)} \actl f_2) g_2 \big) \\
    &= (X_{(1)} \actl f_1) (X_{(2)}\qRm_{(2)} \actl g_1) \tpb (X_{(3)} \qRm_{(1)} \actl f_2) (X_{(4)} \actl g_2) \\
    &= (X_{(1)} \actl f_1) (\qRm_{(2)} X_{(3)} \actl g_1) \tpb (\qRm_{(1)} X_{(2)} \actl f_2) (X_{(4)} \actl g_2) \\
    &= \big( (X_{(1)} \actl f_1) \tpb (X_{(2)} \actl f_2) \big) \star_\Psi \big( (X_{(3)} \actl g_1) \tpb (X_{(4)} \actl g_2) \big) \\
    &= \big( X_{(1)} \actl (f_1 \tpb f_2) \big) \star_\Psi \big( X_{(2)} \actl (g_1 \tpb g_2) \big)
\end{align*}
where we have used the fact that $\Malg_\ell$ is a $\Poin_\ell$-module algebra and the defining property of the $\qRm$-matrix \eqref{eq:Ha_Rmat_def_dcop}. Having the explicit proof is however enlightening on how the braiding solves the issue point out in appendix \ref{subapx:mps_cp}. Indeed, the compatibility of the canonical product \eqref{eq:mps_gpdt} with the coaction was shown to fail because $X_{(2)}$ and $X_{(3)}$ do not cocommute in general. However, in the braided setting, the $\qRm$-matrix compensate for the noncocommutativity of $X_{(2)}$ and $X_{(3)}$ through \eqref{eq:Ha_Rmat_def_dcop}. Indeed, the defining property of the $\qRm$-matrix precisely states that it controls the cocommutativity of $\Poin_\ell$.

As a side remark, let us phrase the condition \ref{it:mps_act} in the coaction formalism: the coaction of $\Poing_\ell$ on $\Malg_\ell$ extends to a coaction on $\Malg_\ell^{\tpb n}$ such that it is a $\Poing_\ell$-comodule algebra, that is (for $n=2$)
\begin{align}
    \label{eq:mps_An_comod_alg}
    \coactl \big( (f_1 \tpb f_2) \star_\Psi (g_1 \tpb g_2) \big)
    &= \big( \coactl (f_1 \tpb f_2) \big) \star_\Psi \big( \coactl(g_1 \tpb g_2) \big).
\end{align}

\paragraph{}
Finally, one needs to show that the product \eqref{eq:mps_qtpdt} is associative (requirement \ref{it:mps_sp}). This has already been proven in Corollary 9.2.13 of \cite{Majid_1995} but we perform this computation for completeness. Explicitly,
\begin{align*}
	\big((f_1 \tpb f_2) &\star_\Psi (g_1 \tpb g_2) \big) \star_\Psi (h_1 \tpb h_2)
	= \big( f_1 (\qRm_{(2)} \actl g_1) \tpb (\qRm_{(1)} \actl f_2) g_2 \big) \star_\Psi (h_1 \tpb h_2) \\
    &= f_1 (\qRm_{(2)} \actl g_1) (\qRm_{(2)} \actl h_1) \tpb \qRm_{(1)} \actl \big((\qRm_{(1)} \actl f_2) g_2 \big) h_2 \\
    &= f_1 (\qRm_{(2)} \actl g_1) (\qRm_{(2)} \qRm_{(2)} \actl h_1) \tpb (\qRm_{(1)}\qRm_{(1)} \actl f_2) (\qRm_{(1)} \actl g_2) h_2   \\
    &= f_1 \qRm_{(2)} \actl \big( g_1 (\qRm_{(2)} \actl h_1) \big) \tpb (\qRm_{(1)} \actl f_2) (\qRm_{(1)} \actl g_2) h_2 \\
    &= (f_1 \tpb f_2) \star_\Psi \big( g_1 (\qRm_{(2)} \actl h_1) \tpb (\qRm_{(1)} \actl g_2) h_2 \big) \\
	&= (f_1 \tpb f_2) \star_\Psi \big( (g_1 \tpb g_2) \star_\Psi (h_1 \tpb h_2) \big)
\end{align*}
where we have used the defining property of the $\qRm$-matrix \eqref{eq:Ha_Rmat_def_2c}. Note that the latter property implies the quantum Yang-Baxter equation \eqref{eq:Ha_Rmat_prop_qYB}, a property intrinsically linked to associativity. We have therefore built a quantum Poincar\'{e} covariant multi-particle algebra, fulfilling the requirements of section \ref{sec:mps}.

Notice that the non-covariance of \eqref{eq:mps_gpdt} and the establishment of a multi-particle algebra through \eqref{eq:mps_qtpdt} has already been done in the case of $\kappa$-Minkowski in \cite{Lizzi_2021}. But as remarked there, time-like $\kappa$-Poincar\'{e} do not fulfil the associativity property. Actually, $\kappa$-Poincar\'{e} does not even have an $\qRm$-matrix \cite{Maslanka_1993} in $4$ or higher dimensions.

\subsubsection{Hopf braiding}
\label{subsubapx:mps_glb}
\paragraph{}
In the case where $\Malg_\ell$ is a Hopf algebra (as in our case), $\Malg_\ell$ acts on itself via the adjoint action. This action can be used to define a braiding through
\begin{align}
	\Psi(f \otimes g)
	&= f_{(1)} g S(f_{(2)}) \tpb f_{(3)},
	\tag{\ref{eq:mps_glb}}
\end{align}
which is defined through the coproduct and antipode of $\Malg_\ell$. If one considers plane waves $f = e^{i p x}$ and $g = e^{i q x}$, then one obtains
\begin{align*}
	\Psi(e^{i p x} \otimes e^{i q x})
	&= e^{i (p \dplus q \dminus p) x} \tpb e^{i p x}
\end{align*}
and therefore $\Psi$ is the canonical braiding on the momentum group $G_\ell$. The product \eqref{eq:mps_braid_pdt} defined through the braiding \eqref{eq:mps_glb} is given by
\begin{align}
	(f_1 \tpb f_2) \star_\Psi (g_1 \tpb g_2)
	&= f_1 f_{2(1)} g_1 S(f_{2(2)}) \tpb f_{2(3)} g_2,
	\tag{\ref{eq:mps_glbpdt}}
\end{align}
and is ensured to be associative by its construction through a braiding (see Example 9.2.3 of \cite{Majid_1995}). However, one can derive it explicitly as
\begin{align*}
	\big( (f_1 \tpb f_2) \star_\Psi (g_1 \tpb g_2) \big) \star_\Psi (h_1 \tpb h_2)
	&= \big( f_1 f_{2(1)} g_1 S(f_{2(2)}) \tpb f_{2(3)} g_2 \big) \star_\Psi (h_1 \tpb h_2) \\
	&= f_1 f_{2(1)} g_1 S(f_{2(2)}) f_{2(3)} g_{2(1)} h_1 S(f_{2(4)} g_{2(2)}) \tpb f_{2(5)} g_{2(3)} h_2 \\
	&= f_1 f_{2(1)} g_1 \epsilon(f_{2(2)}) g_{2(1)} h_1 S(g_{2(2)}) S(f_{2(2)}) \tpb f_{2(3)} g_{2(3)} h_2 \\
	&= f_1 f_{2(1)} g_1 g_{2(1)} h_1 S(g_{2(2)}) S(f_{2(3)}) \tpb f_{2(4)} g_{2(3)} h_2 \\
	&= (f_1 \tpb f_2) \star_\Psi \big( g_1 g_{2(1)} h_1 S(g_{2(2)}) \tpb g_{2(3)} h_2 \big) \\
	&= (f_1 \tpb f_2) \star_\Psi \big( (g_1 \tpb g_2) \star_\Psi (h_1 \tpb h_2) \big),
\end{align*}
where we have used properties of the Hopf algebra: coassociativity, $f_{(1)} S(f_{(2)}) = \epsilon(f) 1$, $\epsilon(f_{(1)}) f_{(2)} = f$ and $S(fg) = S(g) S(f)$. The requirement \ref{it:mps_sp} therefore hold in this case.

\paragraph{}
Concerning the compatibility of the braided product \eqref{eq:mps_glb} with the action \eqref{eq:mps_gact} (requirement \ref{it:mps_act}), it can be shown that the Hopf braiding is not covariant\footnote{
    It should be underlined that the treatment we provide here differs of the one of \cite{Arzano_2023}. Indeed, they claim that the multi-particle algebra based on the product \eqref{eq:mps_glb} is ``covariant'' under the quantum Poincar\'{e} transformations, while we say it is not. This is because we do not agree on the definition of covariance. For us, covariance is related to the fact that $\Malg_\ell^{\tpb 2}$ forms a $\Poing_\ell$-comodule algebra (or a $\Poin_\ell$-module algebra), which we show is not the case for $\kappa$-Minkowski, because $[a,\Lambda] \neq 0$. Actually, the authors of \cite{Arzano_2023} agree with this statement: as they write it $(gh)' \neq g'h'$ (see equation (24) of \cite{Arzano_2023} and above). However, their notion of covariance rely on the fact that the total momentum of any element of $\Malg_\ell^{\tpb 2}$ should be equal to the deformed sum of momentum of individual elements, that is the total momentum of $\coactl( e^{ipx} \tpb e^{iqx})$ should be the one of $e^{ipx} \tpb e^{iqx}$, \ie $p \dplus q$. We agree with that statement since the product \eqref{eq:mps_glb} is shown to be covariant with respect to pure translations. 
}
\textit{in general}, and so that the tensor algebra $\Malg_\ell^{\tpb 2}$ does not form a $\Poin_\ell$-module algebra. Indeed,
\begin{align*}
	X \actl \big( (f_1 \tpb f_2) &\star_\Psi (g_1 \tpb g_2) \big)
    = X \actl \big( f_1 f_{2(1)} g_1 S(f_{2(2)}) \tpb f_{2(3)} g_2 \big) \\
    &= (X_{(1)} \actl f_1) (X_{(2)} \actl f_{2(1)}) (X_{(3)} \actl g_1) (X_{(4)} \actl S(f_{2(2)})) \tpb (X_{(5)} \actl f_{2(3)})(X_{(6)} \actl g_2) \\
    \big( X_{(1)} \actl (f_1 \tpb f_2) \big) & \star_\Psi \big( X_{(2)} \actl (g_1 \tpb g_2) \big) 
    = \big( (X_{(1)} \actl f_1) \tpb (X_{(2)} \actl f_2) \big) \star_\Psi \big( (X_{(3)} \actl g_1) \tpb (X_{(4)} \actl g_2) \big) \\
    &= (X_{(1)} \actl f_1) (X_{(2)} \actl f_2)_{(1)} (X_{(3)} \actl g_1) S\big( (X_{(2)} \actl f_2)_{(2)} \big) \tpb (X_{(2)} \actl f_2)_{(3)} (X_{(4)} \actl g_2)
\end{align*}
The non-covariance of the Hopf braiding can be understood as follows. The Hopf braiding \eqref{eq:mps_glb} is designed to make $\Malg_\ell^{\tpb 2}$ covariant under the self-coaction of $\Malg_\ell$ that is given by its coproduct $\Delta$. But $\Malg_\ell$ is defined as the Hopf subalgebra of translation of $\Poin_\ell$ and therefore shares the coproduct of $\Poing_\ell$ when restricted to pure translation:\footnote{
    We have written the ``$x$''-term as $a$ to match conventions of \eqref{eq:mps_x_fullcoac}, but $x$ and $a$ have the same nature: they are translations.
}
for a generating element $x^\mu \in \Malg_\ell$,
\begin{align} 
    \Delta x^\mu
    &= 1 \otimes x^\mu + a^\mu \otimes 1.
    \label{eq:mps_x_trcoac}
\end{align}
Yet, the full $\Poing_\ell$ coaction also involves deformed Lorentz rotations, which writes on generating elements
\begin{align}
    \label{eq:mps_x_fullcoac}
    \coactl x^\mu 
    &= \Lambda^\mu{}_\nu \otimes x^\nu + a^\mu \otimes 1,
\end{align}
where $\Lambda$ is the rotation and $a$ a translation. The self-coaction \eqref{eq:mps_x_trcoac} therefore corresponds to the full coaction \eqref{eq:mps_x_fullcoac} without rotation $\Lambda^\mu{}_\nu = \delta^\mu_\nu 1$. Put differently, one can define the coaction of pure rotations $\rho_{\mathrm{rot}}: \Malg_\ell \to \Poing_\ell \otimes \Malg_\ell$ and define the full $\Poing_\ell$ coaction as
\begin{align}
    \label{eq:mps_coac_compos_rot+tr}
    \coactl 
    &= (m_{\Poing_\ell} \otimes \id) \circ (\id \otimes \rho_{\mathrm{rot}}) \circ \Delta.
\end{align}
One can show that \eqref{eq:mps_x_trcoac} and \eqref{eq:mps_coac_compos_rot+tr} together with $\rho_{\mathrm{rot}}(1) = 1 \otimes 1$ and $\rho_{\mathrm{rot}}(x^\mu) = \Lambda^\mu{}_\nu \otimes x^\nu$ gives \eqref{eq:mps_x_fullcoac}. The non-covariance of $\coactl$ is then directly inherited from the non-covariance of $\rho_{\mathrm{rot}}$. This could be shown by a computation similar as above. However, a more telling counter-example is the one of generating elements. Explicitly, direct computations show the following properties
\begin{align*}
    \Psi(x^\mu \otimes x^\nu)
    &= [x^\mu, x^\nu] \tpb 1 + x^\nu \tpb x^\mu, \\
    \coactl [x^\mu, x^\nu]
    &= [\coactl x^\mu, \coactl x^\nu] \\
    &= [\Lambda^\mu{}_\rho, \Lambda^\nu{}_\lambda] \otimes x^\rho x^\lambda + [a^\mu, \Lambda^\nu{}_\lambda] \otimes x^\lambda + [\Lambda^\mu{}_\rho, a^\nu] \otimes x^\rho + [a^\mu, a^\nu] \otimes 1,
\end{align*}
which in turn allow us to compute
\begin{align}
    \label{eq:mps_x_nocov}
    \coactl(x_2^\mu \star_\Psi x_1^\nu) - (\coactl x_2^\mu) \star_\Psi(\coactl x_1^\nu)
    &= [\Lambda^\mu{}_\rho, \Lambda^\nu{}_\lambda] \otimes (x^\lambda x^\rho \tpb 1 + x^\lambda \tpb x^\rho) + [a^\nu, \Lambda^\mu{}_\rho] \otimes (1 \tpb x^\rho - x^\rho \tpb 1).
\end{align}
As expected, \eqref{eq:mps_x_nocov} vanishes for $\Lambda^\mu{}_\nu = \delta^\mu_\nu 1$ because the braiding \eqref{eq:mps_glb} is covariant with respect to the pure translation coaction \eqref{eq:mps_x_trcoac}. Moreover, we now see that the non-vanishing of $[\Lambda, \Lambda]$ and $[a, \Lambda]$ are responsible for the non-compatibility of the braiding \eqref{eq:mps_An_comod_alg}.

One should further note that this braiding is covariant in particular cases, as for example in $(2+1)$-dimensional $\kappa$-deformation (studied in section \ref{subsec:ex_3qg}). Quite remarkably, in this case, the quasitriangular braiding \eqref{eq:mps_cqt_braid} and the Hopf braiding \eqref{eq:mps_glb} coincide so that both are covariant.

\subsection{Generalisation to \tops{$\Malg_\ell^{\tpb n}$}{A\^{}n}}
\label{subapx:mps_An}
\paragraph{}
In this section, we prove that the product \eqref{eq:mps_braid_pdt_An} of $\Malg_\ell^{\tpb n}$ introduced in section \ref{subsec:mps_An} is well-defined and associative.

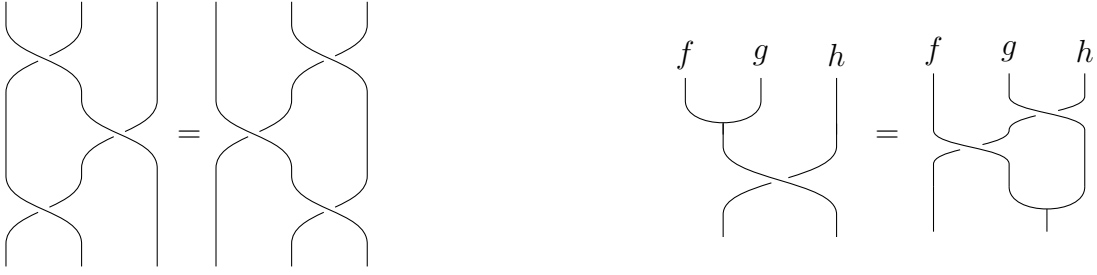
\begin{figure}
    \centering
    %\iffalse % Begin commenting out
    $\vcenter{\hbox{%
    \begin{tikzpicture}[braid/.cd,]
        \pic[braid/number of strands = 3,]
            (b) {
                braid={a_1 a_2 a_1}
        };
    \end{tikzpicture}
    }}
    =\,
    \vcenter{\hbox{%
    \begin{tikzpicture}[braid/.cd,]
        \pic[braid/number of strands = 3,]
            (b) {
                braid={a_2 a_1 a_2}
        };
    \end{tikzpicture}
    }}$%
    \hspace{.2\textwidth}%
    $\vcenter{\hbox{%
    \begin{tikzpicture}[braid/.cd,]
        \draw (0,0) node[yshift=.3cm] {$f$}
            to (0,-.3)
            to[out=-90, in=-90] node[above] (p1){} (1,-.3) 
            to (1,0) node[yshift=.3cm] {$g$};
        \draw (p1) -- +(0,-.3) node[above] (p2){};

        \node[at=(p2), xshift=1.5cm] (p3) {};
    
        \draw (2,0) node[yshift=.3cm] {$h$} to (p3.south);
    
        \pic[braid/width=1.5cm]
            at (p2) {
                braid={a_1}
        };
    \end{tikzpicture}
    }}
    =\,
    \vcenter{\hbox{%
    \begin{tikzpicture}[braid/.cd,]
        \pic[braid/number of strands = 3,]
            (b) {
                braid={a_{1-3}}
        };

        \node[at=(b-1-s), yshift=.3cm] {$f$};
        \node[at=(b-2-s), yshift=.3cm] {$g$};
        \node[at=(b-3-s), yshift=.3cm] {$h$};

        \draw (b-rev-2-e) to[out=-90, in=-90] node (p1){} (b-rev-3-e);
        \draw (p1.center) -- +(0,-.3) node[xshift=-1.5cm] (p2) {};
        \draw (b-rev-1-e) -- (p2.center);
    \end{tikzpicture}
    }}$
    %\fi % End commenting out
    \caption{Diagrammatic representation of the coherence theorem \eqref{eq:mps_braid_ct} (left) and of the compatibility between the product of $\Malg_\ell$ and the braiding \eqref{eq:mps_braid_pdt_comp} (right).}
    \label{fig:braid_ct+pdtc}
\end{figure}

\paragraph{}
Before starting, we need some properties of the braidings $\Psi_j$ (defined in section \ref{subsec:mps_An}), notably the one induced by the coherence theorem. The coherence theorem expresses how the defining property of braidings \eqref{eq:mps_braid_defprop} behaves for higher number of spaces and higher number of braidings. Given three spaces $U$, $V$ and $W$, the lowest degree equality writes
\begin{align}
    \label{eq:mps_braid_ct}
    (\Psi_{U,V} \otimes \id_W) \circ (\id_V \otimes \Psi_{U,W}) \circ (\Psi_{W,V} \otimes \id_U)
    = (\id_U \otimes \Psi_{V,W}) \circ (\Psi_{U,W} \otimes \id_V) \circ (\id_W \otimes \Psi_{U,V})
\end{align}
that both go from $U \otimes V \otimes W$ to $W \otimes V \otimes U$ (see equation (9.7) of \cite{Majid_1994}). This theorem is represented on the braided diagram\footnote{
    We refer the reader to section \ref{subsec:mps_bd} on how to read braided diagrams.
}
of Figure \ref{fig:braid_ct+pdtc}. Considering $U = V = W = \Malg_\ell$, \eqref{eq:mps_braid_ct} writes
\begin{align*}
    \Psi_1 \circ \Psi_2 \circ \Psi_1
    &= \Psi_2 \circ \Psi_1 \circ \Psi_2,
\end{align*}
and corresponds to the Yang-Baxter equation for $\Psi$. This translates to the Yang-Baxter equation for the $\qcRm$-matrix in the case of the coquasitriangular braiding of $SU_q(2)$ \eqref{eq:mps_cqt_braid}. By straightforward application of \eqref{eq:mps_braid_ct}, we have more generally, for any $j \in \{1,\ldots, n-1\}$,
\begin{align}
    \label{eq:mps_braid_ct_A}
    \Psi_j \circ \Psi_{j+1} \circ \Psi_j
    &= \Psi_{j+1} \circ \Psi_j \circ \Psi_{j+1}.
\end{align}

Note also that the braiding of elements that are distant of more then two positions can be done independently, or more explicitly
\begin{align}
    \label{eq:mps_braid_dist+2}
    \Psi_{j} \circ \Psi_{j+m} = \Psi_{j+m} \circ \Psi_j,
\end{align}
for any $m > 1$. As an example for $n = 4$, $\Psi_1$ and $\Psi_3$ can be applied independently, since
\begin{align*}
    \Psi_1 \circ \Psi_3 (f_1 \tpb f_2 \tpb f_3 \tpb f_4)
    &= \Psi(f_1 \tpb f_2) \tpb \Psi(f_3 \tpb f_4)
    = \Psi_3 \circ \Psi_1 (f_1 \tpb f_2 \tpb f_3 \tpb f_4),
\end{align*}
for any $f_1, f_2, f_3, f_4 \in \Malg_\ell$. Properties \eqref{eq:mps_braid_ct_A} and \eqref{eq:mps_braid_dist+2} actually tells us that the $\Psi_j$'s forms a braid group and justifies why we can use braided diagrams in this context.

Finally, the braiding is an (iso)morphism and therefore has a compatibility relation with the product of $\Malg_\ell$. Explicitly, performing a product and braiding with a third element is the same as braiding first the two elements and then performing the product. In terms of algebraic relation this writes
\begin{align}
    \label{eq:mps_braid_pdt_comp}
    \Psi( f_1 f_2 \otimes f_3)
    &= (\id \otimes m_{\Malg_\ell}) \circ \Psi_1 \circ \Psi_2(f_1 \tpb f_2 \tpb f_3).
\end{align}
This equality is represented on the braided diagram of Figure \ref{fig:braid_ct+pdtc}. 
\begin{comp}
    The proof of \eqref{eq:mps_braid_pdt_comp}, is actually based on the property of the braiding \eqref{eq:mps_braid_defprop}, which defines the braiding for higher order of the tensor product. We already saw that \eqref{eq:mps_braid_defprop} implies $\Psi_{\Malg_\ell \tpb \Malg_\ell, \Malg_\ell} = \Psi_1 \circ \Psi_2$ so that
    \begin{align*}
        \Psi \circ (m_{\Malg_\ell} \otimes \id) 
        &= (m_{\Malg_\ell} \otimes \id) \circ \Psi_{\Malg_\ell \tpb \Malg_\ell, \Malg_\ell}
        = (m_{\Malg_\ell} \otimes \id) \circ \Psi_1 \circ \Psi_2.
    \end{align*}
\end{comp}

\paragraph{}
The definition \eqref{eq:mps_braid_pdt_An} of the product of $\Malg_\ell^{\tpb n}$, say $(f_1 \tpb \cdots \tpb f_n) \star_\Psi (g_1 \tpb \cdots \tpb g_n)$, is made by sending $f_n$ to the $(2n-1)^\text{th}$ position (through $\Psi_{2n-2} \circ \cdots \circ \Psi_{n}$), then $f_{n-1}$ to the $(2n-3)^\text{th}$ position (though $\Psi_{2n-4} \circ \cdots \circ \Psi_{n-1}$) and so on, and finally taking the product of elements two by two. One could have equivalently define the same product in different ways by moving the $g$'s instead of the $f$'s, or by taking some of the products before some braidings. All the different formulations sparked by these changes can be shown to be all equal thanks to \eqref{eq:mps_braid_ct_A}, \eqref{eq:mps_braid_dist+2} and \eqref{eq:mps_braid_pdt_comp}. In other words, the braidings $\Psi_j$ and the product $m_{\Malg_\ell}$ can all be reordered in coherent ways thanks to these three equations. In that sense, this product is well-defined.

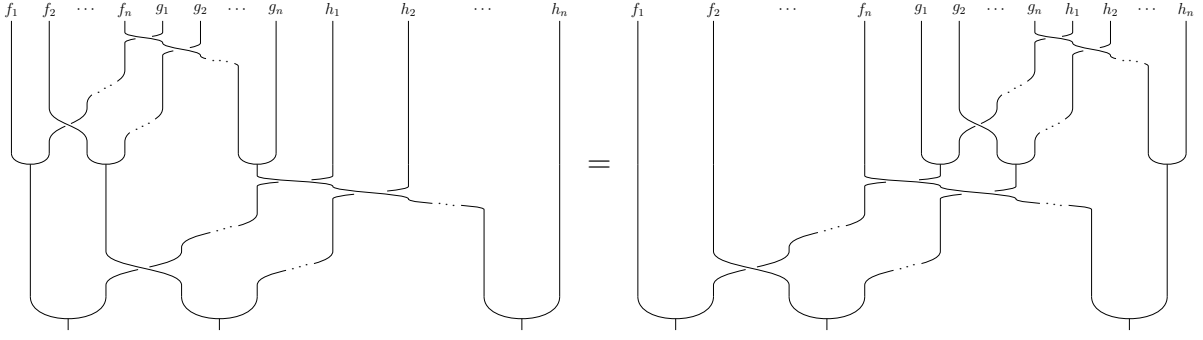
\begin{figure}
    \centering
    %\iffalse %
    $\vcenter{\hbox{\scalebox{.5}{%
    \begin{tikzpicture}[braid/.cd,]
        \pic[braid/number of strands = 8,
            braid/width = 1cm,
            braid/crossing convention=under,
            braid/set symbols=under,
            braid/strand 3/.style={white},
            braid/strand 7/.style={white},]
            (b1) {
                braid={s_{4-7} s_3  |s_2-s_4| }
        };
        \node[fill=white, rotate=-8, xshift=-.5cm, yshift=.12cm] at (b1-4-1) {$\ldots$};
        \node[fill=white, rotate=28, xshift=-.25cm, yshift=.65cm] at (b1-3-2) {$\ldots$};
        \node[fill=white, rotate=28, xshift=-.25cm, yshift=.65cm] at (b1-3-3) {$\ldots$};
    
        \draw (b1-rev-1-e) to[out=-90, in=-90] node (p11){} (b1-rev-2-e);
        \draw (b1-rev-3-e) to[out=-90, in=-90] node (p12){} (b1-rev-4-e);
        \draw (b1-rev-7-e) to[out=-90, in=-90] node (p13){} (b1-rev-8-e);

        \pic[at = (p11.center),
            braid/number of strands = 8,
            braid/width = 2cm,
            braid/crossing convention=under,
            braid/set symbols=under,
            braid/strand 3/.style={white},
            braid/strand 7/.style={white},]
            (b2) {
                braid={s_{4-7} s_3  |s_2-s_4| }
        };
        \node[fill=white, rotate=-5, xshift=-1cm, yshift=.1cm] at (b2-4-1) {$\ldots$};
        \node[fill=white, rotate=15, xshift=-.7cm, yshift=.75cm] at (b2-3-2) {$\ldots$};
        \node[fill=white, rotate=15, xshift=-.7cm, yshift=.75cm] at (b2-3-3) {$\ldots$};

        \draw (b2-rev-1-e) to[out=-90, in=-90] node (p21){} (b2-rev-2-e);
        \draw (b2-rev-3-e) to[out=-90, in=-90] node (p22){} (b2-rev-4-e);
        \draw (b2-rev-7-e) to[out=-90, in=-90] node (p23){} (b2-rev-8-e);

        \foreach \p in {1,2,3}
            \draw (p2\p.center) -- +(0,-.3);

        \foreach \p in {5,6,8}
            \draw (b2-\p-s) -- (b2-\p-s |-, |- b1-1-s) node (p3\p){};
        \draw[transparent] (b2-7-s) -- (b2-7-s |-, |- b1-1-s) node (p37){};

        \node[at=(b1-1-s), yshift=.3cm] {$f_1$};
        \node[at=(b1-2-s), yshift=.3cm] {$f_2$};
        \node[at=(b1-3-s), yshift=.3cm] {$\cdots$};
        \node[at=(b1-4-s), yshift=.3cm] {$f_n$};
        \node[at=(b1-5-s), yshift=.3cm] {$g_1$};
        \node[at=(b1-6-s), yshift=.3cm] {$g_2$};
        \node[at=(b1-7-s), yshift=.3cm] {$\cdots$};
        \node[at=(b1-8-s), yshift=.3cm] {$g_n$};
        \node[at=(p35.center), yshift=.3cm] {$h_1$};
        \node[at=(p36.center), yshift=.3cm] {$h_2$};
        \node[at=(p37.center), yshift=.3cm] {$\cdots$};
        \node[at=(p38.center), yshift=.3cm] {$h_n$};
    \end{tikzpicture}
    }}}
    =\,
    \vcenter{\hbox{\scalebox{.5}{%
    \begin{tikzpicture}[braid/.cd,]
        \pic[braid/number of strands = 8,
            braid/width = 1cm,
            braid/crossing convention=under,
            braid/set symbols=under,
            braid/strand 3/.style={white},
            braid/strand 7/.style={white},]
            (b1) {
                braid={s_{4-7} s_3  |s_2-s_4| }
        };
        \node[fill=white, rotate=-8, xshift=-.5cm, yshift=.12cm] at (b1-4-1) {$\ldots$};
        \node[fill=white, rotate=28, xshift=-.25cm, yshift=.65cm] at (b1-3-2) {$\ldots$};
        \node[fill=white, rotate=28, xshift=-.25cm, yshift=.65cm] at (b1-3-3) {$\ldots$};
    
        \draw (b1-rev-1-e) to[out=-90, in=-90] node (p11){} (b1-rev-2-e);
        \draw (b1-rev-3-e) to[out=-90, in=-90] node (p12){} (b1-rev-4-e);
        \draw (b1-rev-7-e) to[out=-90, in=-90] node (p13){} (b1-rev-8-e);

        \pic[at = {(-7.5 , |- p11.center)},
            braid/number of strands = 8,
            braid/width = 2cm,
            braid/crossing convention=under,
            braid/set symbols=under,
            braid/strand 3/.style={white},
            braid/strand 7/.style={white},]
            (b2) {
                braid={s_{4-7} s_3  |s_2-s_4| }
        };
        \node[fill=white, rotate=-5, xshift=-1cm, yshift=.1cm] at (b2-4-1) {$\ldots$};
        \node[fill=white, rotate=15, xshift=-.7cm, yshift=.75cm] at (b2-3-2) {$\ldots$};
        \node[fill=white, rotate=15, xshift=-.7cm, yshift=.75cm] at (b2-3-3) {$\ldots$};

        \draw (b2-rev-1-e) to[out=-90, in=-90] node (p21){} (b2-rev-2-e);
        \draw (b2-rev-3-e) to[out=-90, in=-90] node (p22){} (b2-rev-4-e);
        \draw (b2-rev-7-e) to[out=-90, in=-90] node (p23){} (b2-rev-8-e);

        \foreach \p in {1,2,3}
            \draw (p2\p.center) -- +(0,-.3);

        \foreach \p in {1,2,4}
            \draw (b2-\p-s) -- (b2-\p-s |-, |- b1-1-s) node (p3\p){};
        \draw[transparent] (b2-3-s) -- (b2-3-s |-, |- b1-1-s) node (p33){};

        \node[at=(p31.center), yshift=.3cm] {$f_1$};
        \node[at=(p32.center), yshift=.3cm] {$f_2$};
        \node[at=(p33.center), yshift=.3cm] {$\cdots$};
        \node[at=(p34.center), yshift=.3cm] {$f_n$};
        \node[at=(b1-1-s), yshift=.3cm] {$g_1$};
        \node[at=(b1-2-s), yshift=.3cm] {$g_2$};
        \node[at=(b1-3-s), yshift=.3cm] {$\cdots$};
        \node[at=(b1-4-s), yshift=.3cm] {$g_n$};
        \node[at=(b1-5-s), yshift=.3cm] {$h_1$};
        \node[at=(b1-6-s), yshift=.3cm] {$h_2$};
        \node[at=(b1-7-s), yshift=.3cm] {$\cdots$};
        \node[at=(b1-8-s), yshift=.3cm] {$h_n$};
\end{tikzpicture}
}}}$
    %\fi %
    \caption{Diagram of the associativity property of the product \eqref{eq:mps_braid_pdt_An}.}
    \label{fig:pdt_An_assos}
\end{figure}

\paragraph{}
The associativity property of the product \eqref{eq:mps_braid_pdt_An} is again more easily proved thanks to braid diagrams. It is displayed in Figure \ref{fig:pdt_An_assos}. 
In order to prove this associativity property, we first need the diagram of the associativity of the product of $\Malg_\ell$, \ie
\begin{align}
    \label{eq:mps_assos_prod_A}
    m_{\Malg_\ell} \circ (\id \otimes m_{\Malg_\ell})
    &= m_{\Malg_\ell} \circ (m_{\Malg_\ell} \otimes \id )
    & \Longleftrightarrow &&
    \vcenter{\hbox{%
    \begin{tikzpicture}
        \draw (0,0) to (0,-.3)
            to[out=-90, in=-90] node (p1){} (1,-.3) 
            to (1,0);
        \draw (p1.center) -- +(0,-.3) node (p2){};
        \node[at=(p2.center), xshift=1.5cm] (p3) {};
        \draw (2,0) to (p3.center)
            to[out=-90, in=-90] node (p4){} (p2.center);
        \draw (p4.center) -- +(0,-.3);
    \end{tikzpicture}
    }}%
    =\,
    \vcenter{\hbox{%
    \begin{tikzpicture}
        \draw (1,0) to (1,-.3)
            to[out=-90, in=-90] node (p1){} (2,-.3) 
            to (2,0);
        \draw (p1.center) -- +(0,-.3) node (p2){};
        \node[at=(p2.center), xshift=-1.5cm] (p3) {};
        \draw (0,0) to (p3.center)
            to[out=-90, in=-90] node (p4){} (p2.center);
        \draw (p4.center) -- +(0,-.3);
    \end{tikzpicture}
    }}.
\end{align}
The proof for general $n$ goes as follows: one needs first to ``bring down'' all the products so that they are all performed after the braidings thanks to \eqref{eq:mps_braid_pdt_comp}, then apply the associativity property \eqref{eq:mps_assos_prod_A}, and finally move back up some products with again \eqref{eq:mps_braid_pdt_comp}. For simplicity, we only perform here the proof for $n=2$, but it directly generalises to any $n$.

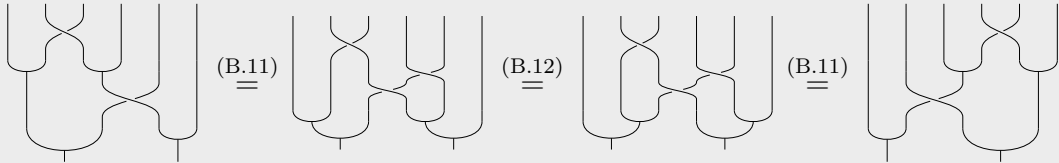
\begin{comp}
    \centering
    %\iffalse
$\vcenter{\hbox{\scalebox{.5}{%
\begin{tikzpicture}[braid/.cd,]
    \pic[braid/number of strands = 6,
        braid/width = 1cm,]
        (b1) {
            braid={a_2}
    };

    \draw (b1-rev-1-e) to[out=-90, in=-90] node (p1){} (b1-rev-2-e);
    \draw (b1-rev-3-e) to[out=-90, in=-90] node (p2){} (b1-rev-4-e);

    \pic[
        braid/number of strands = 2,
        at = (p2.center),
        braid/width = 1.5cm,]
        (b2) {
            braid={a_1}
        };

    \draw (p1.center) -- (p1.center |- , |- b2-rev-1-e) node (p4){};
    \draw (b2-rev-1-e) to[out=-90, in=-90] node (p5){} (p4.center);
    \draw (p5.center) -- +(0,-.3) node (bot){};

    \draw (b1-rev-5-e) to (b2-2-s);
    \draw (b1-rev-6-e) to (b1-rev-6-e |- , |- b2-rev-2-e) node (p6){};
    \draw (b2-rev-2-e) to[out=-90, in=-90] node (p7){} (p6.center);
    \draw (p7.center) to (p7.center |- , |- bot);
\end{tikzpicture}
}}}
\overset{\eqref{eq:mps_braid_pdt_comp}}{=} \,
\vcenter{\hbox{\scalebox{.5}{%
\begin{tikzpicture}[braid/.cd,]
    \pic[braid/number of strands = 6,
        braid/width = 1cm,]
        (b1) {
            braid={a_2 a_{3-5}}
    };

    \draw (b1-rev-1-e) to[out=-90, in=-90] node (p1){} (b1-rev-2-e);
    \draw (b1-rev-4-e) to[out=-90, in=-90] node (p2){} (b1-rev-5-e);

    \draw (b1-rev-3-e) -- (b1-rev-3-e |- , |- p1.center) node (p11){};
    \draw (p11.center) to[out=-90, in=-90] node (p12){} (p1.center);
    \draw (p12.center) -- +(0,-.3) node (bot){};

    \draw (b1-rev-6-e) -- (b1-rev-6-e |- , |- p2.center) node (p21){};
    \draw (p21.center) to[out=-90, in=-90] node (p22){} (p2.center);
    \draw (p22.center) -- (p22.center |- , |- bot);
\end{tikzpicture}
}}}
\overset{\eqref{eq:mps_assos_prod_A}}{=}\,
\vcenter{\hbox{\scalebox{.5}{%
\begin{tikzpicture}[braid/.cd,]
    \pic[braid/number of strands = 6,
        braid/width = 1cm,]
        (b1) {
            braid={a_2 a_{3-5}}
    };

    \draw (b1-rev-2-e) to[out=-90, in=-90] node (p1){} (b1-rev-3-e);
    \draw (b1-rev-5-e) to[out=-90, in=-90] node (p2){} (b1-rev-6-e);

    \draw (b1-rev-1-e) -- (b1-rev-1-e |- , |- p1.center) node (p11){};
    \draw (p11.center) to[out=-90, in=-90] node (p12){} (p1.center);
    \draw (p12.center) -- +(0,-.3) node (bot){};

    \draw (b1-rev-4-e) -- (b1-rev-4-e |- , |- p2.center) node (p21){};
    \draw (p21.center) to[out=-90, in=-90] node (p22){} (p2.center);
    \draw (p22.center) -- (p22.center |- , |- bot);
\end{tikzpicture}
}}}
\overset{\eqref{eq:mps_braid_pdt_comp}}{=}\,
\vcenter{\hbox{\scalebox{.5}{%
\begin{tikzpicture}[braid/.cd,]
    \pic[braid/number of strands = 6,
        braid/width = 1cm,]
        (b1) {
            braid={a_4}
    };

    \draw (b1-rev-3-e) to[out=-90, in=-90] node (p1){} (b1-rev-4-e);
    \draw (b1-rev-5-e) to[out=-90, in=-90] node (p2){} (b1-rev-6-e);

    \pic[
        braid/number of strands = 2,
        at = {(b1-rev-2-s |- , |- p1.center)},
        braid/width = 1.5cm,]
        (b2) {
            braid={a_1}
        };

    \draw (p2.center) -- (p2.center |- , |- b2-rev-2-e) node (p4){};
    \draw (b2-rev-2-e) to[out=-90, in=-90] node (p5){} (p4.center);
    \draw (p5.center) -- +(0,-.3) node (bot){};

    \draw (b1-rev-2-e) to (b2-1-s);
    \draw (b1-rev-1-e) to (b1-rev-1-e |- , |- b2-rev-2-e) node (p6){};
    \draw (b2-rev-1-e) to[out=-90, in=-90] node (p7){} (p6.center);
    \draw (p7.center) to (p7.center |- , |- bot);
\end{tikzpicture}
}}}$
    %\fi
\end{comp}

\section{Noncommutative canonical field quantisation}
\label{apx:cq}
\paragraph{}
In this appendix, we reproduce the results obtained with path integral quantisation in section \ref{sec:ncpi} using canonical quantisation of fields. The choice of functional derivatives ($p$-Leibniz or $x$-Leibniz) turns, in canonical quantisation, into a choice of oscillator algebra: the usual brackets give non-$\Poin_\ell$-invariant $n$-point functions, while braided brackets give $\Poin_\ell$-invariant functions. This result has already been obtained in \cite{Fabiano_2025}, in the context of $T$-deformations, and we summarise their argument within our formalism. Since we deal with canonical quantisation, we stick with the free theory only in this appendix.

\paragraph{}
Let us first consider that the field $\phi$ expresses thanks to annihilation and creation operators as
\begin{align}
	\phi(x)
	&= \int \Haar{p} \big( \mathscr{I}(p) a_p e^{i(\dminus p)x} + \mathscr{I}(\dminus p) a_p^\dagger e^{i p x} \big)
	\label{eq:cq_ca_dec}
\end{align}
where $a_p, a_p^\dagger \in \mathcal{O}(\mathscr{H})$ are operators on the Fock space $\mathscr{H}$. Thus, we consider our quantised field to be an element of $\mathcal{O}(\mathscr{H}) \otimes \Malg_\ell$. The quantisation procedure consists in specifying a (noncommutative) product on $\mathcal{O}(\mathscr{H})$ and therefore on $\mathcal{O}(\mathscr{H}) \otimes \Malg_\ell$.

\subsection{Undeformed oscillator algebra}
\paragraph{}
We detail here the case of an undeformed oscillator algebra. Explicitly, we suppose that
\begin{align}
	[a_p, a^\dagger_q] = \delta(p - q) \Pi(p), &&
	[a_p, a_q] = 0, &&
	[a^\dagger_p, a^\dagger_q] = 0,
	\label{eq:cq_uoa_def}
\end{align}
together with the existence of a vacuum state $|0\rangle$ such that $a_p|0\rangle = 0$, $\langle0| a^\dagger_p = 0$ and $\langle 0 | 0 \rangle = 1$. Here $\Pi(p)$ is a scalar function standing for the propagator term in the free action \eqref{eq:ncpi_free_action}. Indeed, under these assumptions 
\begin{align}
	\label{eq:cq_uoa_2pf}
\begin{aligned}
	\langle 0| \phi(x_1) \phi(x_2) |0\rangle
	&= \int \Haar{p} \Haar{q} \mathscr{I}(p \dminus q) \langle0| a_{p} e^{i(\dminus p) x_1} a^\dagger_{q} e^{i q x_2} |0\rangle
	= \int \Haar{p} \Pi(p) e^{i (\dminus p) x_1} e^{i p x_2} \\
	&= W(x_1, x_2)
\end{aligned}
\end{align}

For convenience, we introduce the following notations
\begin{align}
	\phi_+(x_1) = \int \Haar{p} \mathscr{I}(p) a_p e^{i (\dminus p) x_1}, &&
	\phi_-(x_2) = \int \Haar{p} \mathscr{I}(\dminus p) a_p^\dagger e^{i p x_2}.
\end{align}
From the general formula
\begin{align}
	[a \otimes b, c \otimes d]
	= ac \otimes bd - ca \otimes db
	= [a, c] \otimes bd + ca \otimes [b, d],
\end{align}
we compute
\begin{align}
\begin{aligned}[]
	[\phi_+(x_1), \phi_-(x_2)]
	&= \int \Haar{p} \Haar{q} \mathscr{I}(p\dminus q) [a_p, a^\dagger_q] e^{i (\dminus p) x_1} e^{i q x_2} + \mathscr{I}(p\dminus q) a_q^\dagger a_p [e^{i (\dminus p) x_1}, e^{i q x_2}] \\
	&= \idu \otimes W(x_1, x_2) + \int \Haar{p} \Haar{q} \mathscr{I}(p\dminus q) a_q^\dagger a_p [e^{i (\dminus p) x_1}, e^{i q x_2}].
\end{aligned}
\end{align}
To write it under a more useful form, one has
\begin{align}
	\phi_+(x_1) \phi_-(x_2)
	&= \idu \otimes W(x_1, x_2) + \int \Haar{p} \Haar{q} \mathscr{I}(p\dminus q) a_q^\dagger a_p e^{i (\dminus p) x_1} e^{i q x_2}.
	\label{eq:cq_uoa_phi_invert}
\end{align}
The last term is not exactly $\phi_-(x_2)\phi_+(x_1)$ since the exponentials are in the reverse order. We can write this term as $(\idu \otimes \Psi_\tau) \big( \phi_-(x_2)\phi_+(x_1) \big)$, where $\idu \otimes \Psi_\tau : \mathcal{O(H)} \otimes \Malg_\ell^{\otimes 2} \to \mathcal{O(H)} \otimes \Malg_\ell^{\otimes 2}$. Therefore,
\begin{align}
	\label{eq:cq_uoa_bfa}
\begin{aligned}[]
	\idu \otimes W(x_1,x_2)
	&= \phi_+(x_1) \phi_-(x_2) - (\idu \otimes \Psi_\tau) \big(\phi_-(x_2) \phi_+(x_1)\big) \\
	&= [ \phi_+(x_1), \phi_-(x_2)]_{\Psi_\tau}.
\end{aligned}
\end{align}
We have just shown that an undeformed oscillator algebra is equivalent to a braided field algebra in the sense of \eqref{eq:cq_uoa_bfa}. Following section \ref{subsec:ncpi_fd}, we represent this observation through the diagram
\begin{center}
	\begin{tikzpicture}[every node/.style={align=center, outer sep=.5cm}]
		\node (uoa) {Undeformed\\ oscillator algebra};
		\node[right= 3cm of uoa] (bfa) {Braided\\ field algebra};
		
		\draw[{Implies[]}-{Implies[]}, double distance=3pt] (uoa.east) to (bfa.west);
	\end{tikzpicture}
\end{center}

\paragraph{}
Using \eqref{eq:cq_uoa_phi_invert}, one can compute directly the (free) $4$-point function to be
\begin{align}
    \label{eq:cq_uoa_4pf}
    \begin{aligned}
        \langle 0 | & \phi(x_1) \cdots \phi(x_4) |0\rangle \\
        &= \int \Haar{p} \Haar{q} \Pi(p) \Pi(q) \Big( e^{i(\dminus p) x_1} e^{ip x_2} e^{i(\dminus q) x_3} e^{i q x_4}
		+ e^{i(\dminus p) x_1} e^{i(\dminus q) x_2} e^{i q x_3} e^{i p x_4}
		+ e^{i(\dminus p) x_1} e^{i(\dminus q) x_2} e^{i p x_3} e^{i q x_4} \Big)
    \end{aligned}
\end{align}
One can see that the expressions \eqref{eq:ncpi_upl_4pf} and \eqref{eq:cq_uoa_4pf} are similar, up to the modular factors, and so that this expression is not $\Poin_\ell$-invariant, mainly because it is not translation invariant.

\begin{comp}[ams align*]
		\langle 0 | & \phi(x_1) \cdots \phi(x_4) |0\rangle \\
		&= \int \Haar{p_1} \cdots \Haar{p_4} 
		\langle 0| \mathscr{I}(p_1) a_{p_1} e^{i(\dminus p_1)x_1} \big( \mathscr{I}(p_2) a_{p_2} e^{i(\dminus p_2)x_2} + \mathscr{I}(\dminus p_2) a_{p_2}^\dagger e^{i p_2 x_2} \big) \\
		&\phantom{=} \times \big( \mathscr{I}(p_3) a_{p_3} e^{i(\dminus p_3)x_3} + \mathscr{I}(\dminus p_3) a_{p_3}^\dagger e^{i p_3 x_3} \big) \mathscr{I}(\dminus p_4) a^\dagger_{p_4} e^{i p_4 x_4} |0\rangle \\
		&= \int \Haar{p_1} \cdots \Haar{p_4} \Big(
		\cancel{\mathscr{I}(\cdots) \langle 0| a_{p_1} a_{p_2} a_{p_3} a^\dagger_{p_4} |0\rangle} e^{\cdots}
		+ \bcancel{\mathscr{I}(\cdots) \langle 0| a_{p_1} a^\dagger_{p_2} a^\dagger_{p_3} a^\dagger_{p_4} |0\rangle} e^{\cdots} \\
		&\phantom{=} + \mathscr{I}(p_1 \dminus p_2 \dplus p_3 \dminus p_4) \langle 0| a_{p_1} a^\dagger_{p_2} a_{p_3} a^\dagger_{p_4} |0\rangle e^{i(\dminus p_1) x_1} e^{ip_2 x_2} e^{i(\dminus p_3) x_3} e^{i p_4 x_4} \\
		&\phantom{=} + \mathscr{I}(p_1 \dplus p_2 \dminus p_3 \dminus p_4) \langle 0| a_{p_1} a_{p_2} a^\dagger_{p_3} a^\dagger_{p_4} |0\rangle e^{i(\dminus p_1) x_1} e^{i(\dminus p_2) x_2} e^{i p_3 x_3} e^{i p_4 x_4} \Big) \\
		&= \int \Haar{p_1} \cdots \Haar{p_4} \Big(
		\mathscr{I}(p_1 \dminus p_2 \dplus p_3 \dminus p_4) \delta(p_1 - p_2) \Pi(p_1) \langle0| a_{p_3} a^\dagger_{p_4} |0\rangle e^{i(\dminus p_1) x_1} e^{ip_2 x_2} e^{i(\dminus p_3) x_3} e^{i p_4 x_4} \\
		&\phantom{=} + \mathscr{I}(p_1 \dplus p_2 \dminus p_3 \dminus p_4) \delta(p_2 - p_3) \Pi(p_2) \langle 0| a_{p_1} a^\dagger_{p_4} |0 \rangle e^{i(\dminus p_1) x_1} e^{i(\dminus p_2) x_2} e^{i p_3 x_3} e^{i p_4 x_4} \\
		&\phantom{=} + \mathscr{I}(p_1 \dplus p_2 \dminus p_3 \dminus p_4) \langle0| a_{p_1} a^\dagger_{p_3} a_{p_2} a^\dagger_{p_4} |0\rangle e^{i(\dminus p_1) x_1} e^{i(\dminus p_2) x_2} e^{i p_3 x_3} e^{i p_4 x_4} \Big) \\
		&= \! \int \! \Haar{p_1}\! \cdots \Haar{p_4}\! \Big( \! \mathscr{I}(p_1 \dminus p_2 \dplus p_3 \dminus p_4) \delta(p_1 - p_2) \delta(p_3 - p_4) \Pi(p_1) \Pi(p_3) e^{i(\dminus p_1) x_1} e^{ip_2 x_2} e^{i(\dminus p_3) x_3} e^{i p_4 x_4} \\
		&\phantom{=} + \mathscr{I}(p_1 \dplus p_2 \dminus p_3 \dminus p_4) \delta(p_2 - p_3) \delta(p_1 - p_4) \Pi(p_1) \Pi(p_2) e^{i(\dminus p_1) x_1} e^{i(\dminus p_2) x_2} e^{i p_3 x_3} e^{i p_4 x_4} \\
		&\phantom{=} + \mathscr{I}(p_1 \dplus p_2 \dminus p_3 \dminus p_4) \delta(p_1 - p_3) \delta(p_2 - p_4) \Pi(p_1) \Pi(p_2) e^{i(\dminus p_1) x_1} e^{i(\dminus p_2) x_2} e^{i p_3 x_3} e^{i p_4 x_4} \Big) \\
		&= \! \int \! \Haar{p} \Haar{q} \Pi(p) \Pi(q) \Big(\! e^{i(\dminus p) x_1} e^{ip x_2} e^{i(\dminus q) x_3} e^{i q x_4}
		+ e^{i(\dminus p) x_1} e^{i(\dminus q) x_2} e^{i q x_3} e^{i p x_4}
		+ e^{i(\dminus p) x_1} e^{i(\dminus q) x_2} e^{i p x_3} e^{i q x_4} \!\Big)
\end{comp}

\subsection{Braided oscillator algebra}
\paragraph{}
Following the main idea of \cite{Fabiano_2025}, we want to impose that $[\phi_+(x_1), \phi_-(x_2)] = \idu \otimes W(x_1, x_2)$, instead of \eqref{eq:cq_uoa_bfa}. In order to do so, we suppose that the hypothesis \eqref{eq:mps_braid_mom} now holds. One can then compute directly $[\phi(x_1), \phi(x_2)]$ thanks to \eqref{eq:cq_ca_dec}, and perform a change of variable of the form $\tilde{p}, \tilde{q} \to \Psi_\tau^{-1}(p, q)$ in the momentum integral. We then have

\begin{align*}
	[\phi(x_1), &\ \phi(x_2)]
	= \int \Haar{p} \Haar{q} \big[ a_p e^{i(\dminus p)x_1} + a_p^\dagger e^{ipx_1}, a_q e^{i(\dminus q)x_2} + a_q^\dagger e^{iqx_2} \big] \\
	&= \int \Haar{p} \Haar{q} \Big( a_p a_q e^{i(\dminus p)x_1} e^{i(\dminus q)x_2} - a_q a_p e^{i(\dminus q)x_2} e^{i(\dminus p)x_1}
	+ a_p^\dagger a_q e^{ipx_1} e^{i(\dminus q)x_2} - a_q a_p^\dagger e^{i(\dminus q)x_2} e^{ipx_1} \\
	&\phantom{=} + a_p a_q^\dagger e^{i(\dminus p)x_1} e^{iqx_2} - a_q^\dagger a_p e^{iqx_2} e^{i(\dminus p)x_1}
	+ a_p^\dagger a_q^\dagger e^{ipx_1} e^{iqx_2} - a_q^\dagger a_p^\dagger e^{iqx_2} e^{ipx_1} \Big) \\
	&= \int \Haar{p} \Haar{q} \Big( 
	\big( a_p a_q - J_{\Psi_\tau}(\dminus q, \dminus p) a_{\dminus \Psi^{-1}_{\tau(2)}(\dminus q, \dminus p)} a_{\dminus \Psi^{-1}_{\tau(1)}(\dminus q, \dminus p)} \big) e^{i(\dminus p)x_1} e^{i(\dminus q)x_2} \\
	&\phantom{=} + \big( a_p^\dagger a_q - J_{\Psi_\tau}(\dminus q, p) a_{\dminus \Psi^{-1}_{\tau(2)}(\dminus q, p)} a_{\Psi^{-1}_{\tau(1)}(\dminus q, p)}^\dagger \big) e^{ipx_1} e^{i(\dminus q)x_2} \\
	&\phantom{=} + \big( a_p a_q^\dagger - J_{\Psi_\tau}(q, \dminus p) a_{\Psi^{-1}_{\tau(2)}(q, \dminus p)}^\dagger a_{\dminus \Psi^{-1}_{\tau(1)}(q, \dminus p)} \big) e^{i(\dminus p)x_1} e^{i q x_2} \\
	&\phantom{=}+ \big( a_p^\dagger a_q^\dagger - J_{\Psi_\tau}(q,p) a_{\Psi^{-1}_{\tau(2)}(q,p)}^\dagger a_{\Psi^{-1}_{\tau(1)}(q,p)}^\dagger \big) e^{ipx_1} e^{iqx_2} \Big),
\end{align*}
where $J_{\Psi_\tau}(p,q)$ denotes the Jacobian of the change of variable. If we impose that the previous expression corresponds to $\idu \otimes W(x_1, x_2)$, one needs to have a braided oscillator algebra with a structure of the form
\begin{align}
	[a^\dagger_p, a_q]_{\Psi_\tau}
	&= \delta(p-q) \Pi(p), &
	[a_p, a_q]_{\Psi_\tau}
	&= 0, &
	[a^\dagger_p, a^\dagger_q]_{\Psi_\tau}
	&= 0,
	\label{eq:bcpi_ca_balg}
\end{align}
together with $a_p|0\rangle = 0$ and $\langle 0| a_p^\dagger = 0$. The explicit expressions of the braided brackets is to be read from the previous computation. Therefore, if one wants a trivial unbraided commutator of $\phi$'s, then one needs to have a braided oscillator algebra. These result can be summarised by the following diagram, where the * indicates the fact that the hypothesis \eqref{eq:mps_braid_mom} has been used.
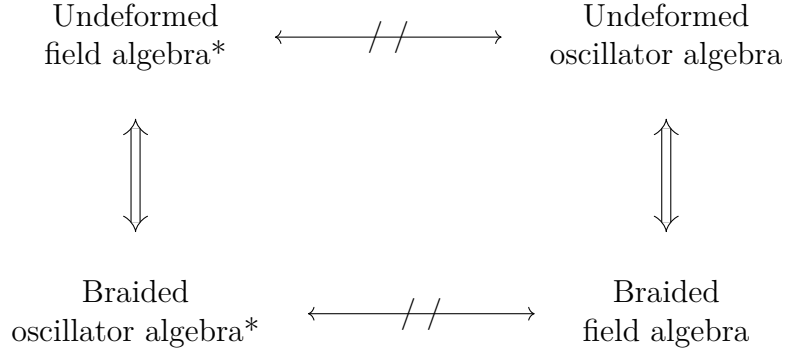
\begin{center}
	\begin{tikzpicture}[every node/.style={align=center, outer sep=.5cm}]
		\node (ufa) {Undeformed\\ field algebra*};
		\node[right= 3cm of ufa] (uoa) {Undeformed\\ oscillator algebra};
		\node[below= 1.5cm of ufa] (boa) {Braided\\ oscillator algebra*};
		\node[right= 3cm of boa] (bfa) {Braided\\ field algebra};	
		
		\draw[{To}-{To}] (ufa.east) to node{/ /} (uoa.west);
		\draw[{To}-{To}] (boa.east) to node{/ /} (bfa.west);
		\draw[{Implies[]}-{Implies[]}, double distance=3pt] (ufa.south) to (boa.north);
		\draw[{Implies[]}-{Implies[]}, double distance=3pt] (uoa.south) to (bfa.north);
	\end{tikzpicture}
\end{center}
The similarity of this diagram with respect to the on of section \ref{subsec:ncpi_fd} underlines that the canonical quantisation scheme undergoes the same disjunction as the path integral quantisation: the undeformed algebras match in the commutative case but differ in the noncommutative case.

\paragraph{}
Finally, we need to assess the  $\Poin_\ell$-invariance of the (free) $4$-point function. The later is computed as follows
\begin{align}
	\begin{aligned}
	\langle 0| & \phi(x_1) \cdots \phi(x_4) |0\rangle
	= \langle 0| \phi_+(x_1) \phi_-(x_2) \phi_+(x_3) \phi_-(x_4) |0 \rangle
	+ \langle 0| \phi_+(x_1) \phi_+(x_2) \phi_-(x_3) \phi_-(x_4) |0\rangle \\
	&= W(x_1,x_2) W(x_3,x_4)
	+ \langle 0| \phi_+(x_1) W(x_2, x_3) \phi_-(x_4) |0 \rangle
	+ \langle 0| \phi_+(x_1) \phi_-(x_3) \phi_+(x_2) \phi_-(x_4) |0\rangle \\
	&= \int \Haar{p} \Haar{q} \Pi(p) \Pi(q) \Big( e^{i(\dminus p) x_1} e^{i p x_2} e^{i(\dminus q) x_3} e^{i q x_4}
	+ e^{i (\dminus p) x_1} e^{i (\dminus q) x_2} e^{i q x_3} e^{i p x_4} 
	+ e^{i(\dminus p) x_1} e^{i p x_3} e^{i (\dminus q) x_2} e^{i q x_4} \Big).
	\end{aligned}
	\label{eq:cq_boa_4pf}
\end{align}
Since \eqref{eq:cq_boa_4pf} expresses as sums of products of $2$-point functions of the form \eqref{eq:mps_inv_2pf}, it is indeed $\Poin_\ell$-invariant.

\end{document}